\documentclass[12pt,epsf]{article}

\usepackage{times}
\usepackage{graphicx}
\usepackage{amsfonts}
\usepackage{amsmath}
\usepackage{amssymb}
\usepackage{amstext}
\usepackage{amsthm}
\usepackage{slashed}

\usepackage{epsfig}
\usepackage{color}
\psfigdriver{dvips}
\usepackage{latexsym}
\usepackage[matrix,curve]{xypic}
\usepackage{rotating}
\rotdriver{dvips}

\begin{document}

\baselineskip 6mm
\renewcommand{\thefootnote}{\fnsymbol{footnote}}

%------------ Hyun Seok's macro's, etc  -----------

\newcommand{\nc}{\newcommand}
\newcommand{\rnc}{\renewcommand}

%\headheight=0truein
%\headsep=0truein
%\topmargin=0truein
%\oddsidemargin=0truein
%\evensidemargin=0truein
%\textheight=9truein
%\textwidth=6.5truein

\rnc{\baselinestretch}{1.24}    % 1.5 spacing btwn text lines
\setlength{\jot}{6pt}       % spacing btwn the rows of an eqnarray
\rnc{\arraystretch}{1.24}   % spacing btwn the rows of a non-eqn array

%%%%%%%%%%%%%%%%%%%%%% Equation Numbering %%%%%%%%%%%%%%%%%%%%%%%
\makeatletter
\rnc{\theequation}{\thesection.\arabic{equation}}
\@addtoreset{equation}{section}
\makeatother

%%%%%%%%%%%%%%%%%%%%%%%%%%%%%%%%%%%%%%%%%%%%%%%%%%%%%%%%%%%%%%%%%
%                                                               %
%                NEW COMMANDS AND MACROS                        %
%                                                               %
%%%%%%%%%%%%%%%%%%%%%%%%%%%%%%%%%%%%%%%%%%%%%%%%%%%%%%%%%%%%%%%%%

%%%%% Simplify some frequently used LaTeX commands %%%%%

\nc{\be}{\begin{equation}}

\nc{\ee}{\end{equation}}

\nc{\bea}{\begin{eqnarray}}

\nc{\eea}{\end{eqnarray}}

\nc{\xx}{\nonumber\\}

\nc{\ct}{\cite}

\nc{\la}{\label}

\nc{\eq}[1]{(\ref{#1})}

\nc{\newcaption}[1]{\centerline{\parbox{6in}{\caption{#1}}}}

\nc{\fig}[3]{

\begin{figure}
\centerline{\epsfxsize=#1\epsfbox{#2.eps}}
\newcaption{#3. \label{#2}}
\end{figure}
}

%%% Caligraphic letters %%%%

\def\CA{{\cal A}}
\def\CC{{\cal C}}
\def\CD{{\cal D}}
\def\CE{{\cal E}}
\def\CF{{\cal F}}
\def\CG{{\cal G}}
\def\CH{{\cal H}}
\def\CK{{\cal K}}
\def\CL{{\cal L}}
\def\CM{{\cal M}}
\def\CN{{\cal N}}
\def\CO{{\cal O}}
\def\CP{{\cal P}}
\def\CR{{\cal R}}
\def\CS{{\cal S}}
\def\CU{{\cal U}}
\def\CV{{\cal V}}
\def\CW{{\cal W}}
\def\CY{{\cal Y}}
\def\CZ{{\cal Z}}

%%% Double line letters %%%

\def\IB{{\hbox{{\rm I}\kern-.2em\hbox{\rm B}}}}
\def\IC{\,\,{\hbox{{\rm I}\kern-.50em\hbox{\bf C}}}}
\def\ID{{\hbox{{\rm I}\kern-.2em\hbox{\rm D}}}}
\def\IF{{\hbox{{\rm I}\kern-.2em\hbox{\rm F}}}}
\def\IH{{\hbox{{\rm I}\kern-.2em\hbox{\rm H}}}}
\def\IN{{\hbox{{\rm I}\kern-.2em\hbox{\rm N}}}}
\def\IP{{\hbox{{\rm I}\kern-.2em\hbox{\rm P}}}}
\def\IR{{\hbox{{\rm I}\kern-.2em\hbox{\rm R}}}}
\def\IZ{{\hbox{{\rm Z}\kern-.4em\hbox{\rm Z}}}}

%%% Greek letters %%%

\def\a{\alpha}
\def\b{\beta}
\def\d{\delta}
\def\ep{\epsilon}
\def\ga{\gamma}
\def\k{\kappa}
\def\l{\lambda}
\def\t{\theta}
\def\w{\omega}
\def\G{\Gamma}

%%%%% Mathematical Symbols

\def\half{\frac{1}{2}}
\def\dint#1#2{\int\limits_{#1}^{#2}}
\def\goto{\rightarrow}
\def\para{\parallel}
\def\brac#1{\langle #1 \rangle}
\def\curl{\nabla\times}
\def\div{\nabla\cdot}
\def\p{\partial}

%%%%% Roman pont in math

\def\Tr{{\rm Tr}\,}
\def\det{{\rm det}}

%%%%% Special Letters

\def\vare{\varepsilon}
\def\zbar{\bar{z}}
\def\wbar{\bar{w}}
\def\what#1{\widehat{#1}}

%%%%% For this paper only

\def\ad{\dot{a}}
\def\bd{\dot{b}}
\def\cd{\dot{c}}
\def\dd{\dot{d}}
\def\so{SO(4)}
\def\bfr{{\bf R}}
\def\bfc{{\bf C}}
\def\bfz{{\bf Z}}

\begin{titlepage}

%---------------- preprint number ---------------

\hfill\parbox{3.7cm} {{\tt arXiv:1609.00753}}

\vspace{15mm}

\begin{center}
%------------------------ title ------------------------
{\Large \bf Matrix models from localization of five-dimensional supersymmetric noncommutative U(1) gauge theory}

\vspace{10mm}
%---------------- authors and addresses ----------------

Bum-Hoon Lee ${}^{a,b,c}$\footnote{bhlee@sogang.ac.kr}, Daeho Ro ${}^b$\footnote{dhro@sogang.ac.kr}
and Hyun Seok Yang ${}^c$\footnote{hsyang@sogang.ac.kr}
\\[10mm]

${}^a$ {\sl Department of Physics, Sogang University, Seoul 04107, Korea}

${}^b$ {\sl Asia Pacific Center for Theoretical Physics, POSTECH, Pohang, Gyeongbuk 37673, Korea}

${}^c$ {\sl Center for Quantum Spacetime, Sogang University, Seoul 04107, Korea}

\end{center}

\thispagestyle{empty}

\vskip1cm

%----------------------- abstract ----------------------

\centerline{\bf ABSTRACT}
\vskip 4mm
\noindent

We study localization of five-dimensional supersymmetric $U(1)$ gauge theory
on $\mathbb{S}^3 \times \mathbb{R}_{\theta}^{2}$ where $\mathbb{R}_{\theta}^{2}$ is a noncommutative
(NC) plane. The theory can be isomorphically mapped to three-dimensional supersymmetric $U(N \to \infty)$
gauge theory on $\mathbb{S}^3$ using the matrix representation on a separable Hilbert space on which
NC fields linearly act. Therefore the NC space $\mathbb{R}_{\theta}^{2}$ allows for a flexible path
to derive matrix models via localization from a higher-dimensional supersymmetric NC $U(1)$ gauge theory.
The result shows a rich duality between NC $U(1)$ gauge theories and large $N$ matrix models
in various dimensions.
\\

%PACS numbers: 11.10.Nx, 02.40.Gh, 11.25.Tq


\vspace{1cm}

\today

\end{titlepage}

\renewcommand{\thefootnote}{\arabic{footnote}}
\setcounter{footnote}{0}

\section{Introduction}

The existence of gravity introduces the gravitational constant $G_N$ into a physical theory.
It is well-known that the gravitational constant $G_N$ leads to a certain scale
known as the Planck length $L_P = \sqrt{\frac{G_N \hbar}{c^3}} = 1.6 \times 10^{-33} \mathrm{cm}$
in which spacetime coordinates become noncommutative (NC) operators obeying the commutation relation
\begin{equation}\label{nc-space}
    [y^a, y^b] = i \theta^{ab}, \qquad (a, b = 1, \cdots, 2n).
\end{equation}
We are interested in the NC space with a constant symplectic matrix
$(\theta^{ab}) = \alpha' (\mathbb{I}_n \otimes i \sigma^2)$ which is isomorphic to the Heisenberg algebra
of an $n$-dimensional harmonic oscillator. The NC space \eq{nc-space} will be denoted
by $\mathbb{R}^{2n}_\theta$ and $l_s \equiv \sqrt{\alpha'}$ is a typical length scale for
the noncommutativity. Thus the NC space \eq{nc-space} is similar to the NC phase space in
quantum mechanics obeying the commutation relation given by
\begin{equation}\label{nc-phase}
[x^i, p_j] = i \hbar \delta^i_j, \qquad (i,j =1, \cdots,n).
\end{equation}
We will get an important insight from this similarity to understand the NC spacetime correctly.
As we have learned from quantum mechanics, the NC phase space \eq{nc-phase} introduces a complex vector space
called the Hilbert space. This is also true for the NC space \eq{nc-space} since its mathematical structure
is essentially the same as quantum mechanics.
Therefore the NC space \eq{nc-space} admits a separable Hilbert space $\mathcal{H}$ on which
any object $\mathcal{O}$ defined on $\mathbb{R}^{2n}_\theta$ linearly acts.
In particular, NC fields become linear operators acting on the Hilbert space $\mathcal{H}$.
Since the NC space $\mathbb{R}^{2n}_\theta$ is isomorphic to the Heisenberg algebra
of an $n$-dimensional harmonic oscillator, its Hilbert space is given by the Fock space and
so has a countable basis, the representation of NC fields on the Hilbert space $\mathcal{H}$
is given by $N \times N$ matrices where $N = \mathrm{dim}(\mathcal{H}) \to \infty$.
Consequently, the NC space \eq{nc-space} brings about an interesting equivalence between
a lower-dimensional large $N$ gauge theory and a higher-dimensional NC $U(1)$
gauge theory \cite{hsy-ijmp09,hsy-jhep09,q-emg}.

To illuminate the remarkable equivalence, let us consider a five-dimensional field theory
defined on the NC space $\mathbb{R}^{3} \times \mathbb{R}^2_{\theta}$ with coordinates $X^M = (x^m, y^a)$
where $M = 1, 2, \cdots, 5, \; m = 1, 2, 3$ and $a = 4, 5$.
Here $\mathbb{R}^3$ is a usual commutative space while $\mathbb{R}^2_{\theta}$ is a NC plane whose
coordinates obey the commutation relation
\begin{equation}\label{nc2-space}
    [y^4, y^5] = i \alpha'
\end{equation}
where $\alpha' = l_s^2$ is a constant parameter measuring the noncommutativity of
the space $\mathbb{R}^2_{\theta}$.
If we define annihilation and creation operators as
\begin{equation}\label{ancr}
    a = \frac{y^{4} + i y^{5}}{\sqrt{2\alpha'}}, \qquad
    a^\dagger = \frac{y^{4} - i y^{5}}{\sqrt{2\alpha'}},
\end{equation}
the NC algebra \eq{nc2-space} reduces to the Heisenberg algebra
of harmonic oscillator, i.e.,
\begin{equation}\label{haho}
    [a, a^\dagger]= 1.
\end{equation}
The representation space of the Heisenberg algebra (\ref{haho}) is thus given by the Fock space
\begin{equation}\label{fock-space}
    \mathcal{H} = \{| n \rangle | \; n \in \mathbb{Z}_{\geq 0} \},
\end{equation}
which is orthonormal, i.e., $\langle n| m \rangle = \delta_{nm}$ and
complete, i.e., $\sum_{n = 0}^{\infty} | n \rangle \langle n | = \mathbb{I}_{\mathcal{H}}$,
as is well-known from quantum mechanics.

The field theory we will consider is defined by
dynamical fields on $\mathbb{R}^{3} \times \mathbb{R}^2_{\theta}$ which are elements of
$\mathcal{A}_\theta (\mathbb{R}^{3}) \equiv C^\infty(\mathbb{R}^{3}) \otimes \mathcal{A}_\theta$
where $\mathcal{A}_\theta$ is a NC $\star$-algebra
generated by the NC space \eq{nc2-space}.
Consider two arbitrary fields $\widehat{\Phi}_1(X)$ and $\widehat{\Phi}_2(X)$
on $\mathbb{R}^{3} \times \mathbb{R}^2_{\theta}$ whose multiplication is defined by the star product
\begin{equation}\label{star-prod}
    (\widehat{\Phi}_1 \star \widehat{\Phi}_2) (X) = e^{\frac{i}{2} \theta^{ab} \frac{\partial}{\partial y^a}
    \otimes \frac{\partial}{\partial z^b}} \widehat{\Phi}_1 (x,y) \widehat{\Phi}_2 (x, z)|_{y=z}.
\end{equation}
See \cite{ncft-rev,rev-mncg} for a review of NC field theories and matrix models.
In quantum mechanics, physical observables are considered as operators acting on a Hilbert space.
Similarly the dynamical variables $\widehat{\Phi}_1(X)$ and $\widehat{\Phi}_2(X)$ can be regarded as
operators acting on the Hilbert space $\mathcal{H}$. Thus one can represent the operators acting
on the Fock space (\ref{fock-space}) as $N \times N$ matrices in $\mathrm{End}(\mathcal{H})
\equiv \mathcal{A}_N (\mathbb{R}^3)$ where $N = \mathrm{dim}(\mathcal{H}) \to \infty$:
\begin{eqnarray}\label{matrix-rep}
     && \widehat{\Phi}_1(X) = \sum_{n,m=0}^\infty | n \rangle \langle n| \widehat{\Phi}_1 (x, y)
     | m \rangle \langle m| := \sum_{n,m=0}^\infty (\Phi_1)_{nm} (x) | n \rangle \langle m|, \nonumber \\
     && \widehat{\Phi}_2(X) = \sum_{n,m=0}^\infty | n \rangle \langle n| \widehat{\Phi}_2 (x,y)
     | m \rangle \langle m| := \sum_{n,m=0}^\infty (\Phi_2)_{nm}(x) | n \rangle \langle m|,
\end{eqnarray}
where $\Phi_1 (x)$ and $\Phi_2 (x)$ are $N \times N$ matrices on $\mathbb{R}^{3}$.\footnote{Note that
the eigenvalue $n$ corresponds to the radius $r^2 = y_4^2 + y_5^2$ of the plane since the radial operator
is given by $\widehat{r}^2 = 2 \alpha' (a^\dagger a + \frac{1}{2})$ and $\widehat{r}^2 | n \rangle
= 2 \alpha' (n + \frac{1}{2}) | n \rangle $. Therefore, for a well-localized NC field which rapidly
decays at asymptotic regions, one may truncate the matrix representation of the NC field
to a finite-dimensional space, i.e.,  $N = \mathrm{dim}(\mathcal{H}) < \infty$.}
Then one gets a natural composition rule for the products
\begin{eqnarray}\label{matrix-comp}
 (\widehat{\Phi}_1 \star \widehat{\Phi}_2) (X) &=& \sum_{n,l,m=0}^\infty | n \rangle \langle n|
 \widehat{\Phi}_1 (x,y) | l \rangle \langle l| \widehat{\Phi}_2(x,y) | m \rangle \langle m| \nonumber \\
      &=& \sum_{n,l,m=0}^\infty (\Phi_1)_{nl} (x) (\Phi_2)_{lm} (x) | n \rangle \langle m|.
\end{eqnarray}
The above composition rule implies that the ordering in the NC algebra $\mathcal{A}_\theta (\mathbb{R}^3)$
is compatible with the ordering in the matrix algebra $\mathcal{A}_N (\mathbb{R}^3)$ and
so it is straightforward to translate multiplications of NC fields in $\mathcal{A}_\theta (\mathbb{R}^3)$
into those of matrices in $\mathcal{A}_N (\mathbb{R}^3)$ using the matrix representation (\ref{matrix-rep})
without any ordering ambiguity.

To formulate a gauge theory on $\mathbb{R}^{3} \times \mathbb{R}^2_{\theta}$,
it is necessary to dictate the gauge covariance under the NC star product \eq{star-prod}.
The covariant field strength of NC $U(1)$ gauge fields $\widehat{A}_M (X)
= (\widehat{A}_m, \widehat{A}_a)(x,y)$ is then given by
\begin{equation}\label{d-ncfs}
    \widehat{F}_{MN}(X) = i[\widehat{D}_M, \widehat{D}_N]_\star
    = \partial_M \widehat{A}_N (X) - \partial_N \widehat{A}_M (X)
    - i[\widehat{A}_M, \widehat{A}_N]_\star (X)
\end{equation}
where the covariant derivative is defined by $\widehat{D}_M (X) = \partial_M
- i \widehat{A}_M (X)$. Note that the covariant derivative along $\mathbb{R}^2_{\theta}$
can be written as an inner derivation, i.e., $\widehat{D}_a (X) = - i [\widehat{\phi}_a (x, y),
\, \cdot \,]_\star$ where $\widehat{\phi}_a (x, y) \equiv p_a + \widehat{A}_a (x,y)$
with $B_{ab} = (\theta^{-1} )_{ab}$ and $p_a = B_{ab} y^b$.
Now we will apply the matrix representation \eq{matrix-rep} to a five-dimensional
NC $U(1)$ gauge theory whose action is given by\footnote{\label{w-sign}Note that the kinetic term for the $\widehat{\sigma}$ field has the unusual sign. It is to follow the convention in \cite{pestun}.
(See also the footnote 3 in \cite{mnz13}.) A motivation for the wrong sign is to consider Euclidean
Yang-Mills theory and yet work with physical fermions. This can be accomplished by making $\widehat{\sigma}$ time-like. As a result, the $\mathcal{N} = 2$ gauge theory on $\mathbb{R}^{3} \times \mathbb{R}^2_{\theta}$,
that we will consider later, can be reduced from a six-dimensional super Yang-Mills theory
on $\mathbb{R}^{3,1} \times \mathbb{R}^2_{\theta}$.
The scalar field $\widehat{\sigma}$ in the five-dimensional theory corresponds to the gauge field
component compactified along the time-like direction.}
\begin{equation} \label{5d-action}
S = \frac{1}{g_5^2} \int d^5 X \Big( \dfrac{1}{4} \widehat{\mathcal{F}}_{MN}
\widehat{\mathcal{F}}^{MN} - \dfrac{1}{2} \widehat{D}_{M} \widehat{\sigma}
\widehat{D}^{M} \widehat{\sigma} \Big),
\end{equation}
where $\widehat{\mathcal{F}}_{MN} = \widehat{F}_{MN} - B_{MN}$ and
\begin{equation}\label{back-b}
B_{MN} = \left(
                  \begin{array}{cc}
                    0 & 0 \\
                  0 & B_{ab} \\
                  \end{array}
                \right).
\end{equation}
Using the relations,
\begin{equation}\label{nc-inner}
    \begin{array}{l}
     \widehat{\mathcal{F}}_{ab} = -i [\widehat{\phi}_a, \widehat{\phi}_b ]_\star \\
      \widehat{\mathcal{F}}_{ma} = \partial_m \widehat{\phi}_a
     - i [\widehat{A}_m, \widehat{\phi}_a ]_\star = \widehat{D}_m \widehat{\phi}_a,  \\
     \widehat{D}_a \widehat{\sigma} = -i [\widehat{\phi}_a, \widehat{\sigma}]_\star,
    \end{array}
\end{equation}
and the matrix representation \eq{matrix-rep}, the above five-dimensional NC $U(1)$ gauge theory
is exactly mapped to the three-dimensional $U(N \to \infty)$ gauge theory on $\mathbb{R}^3$
with a scalar triplet $\Phi_A = (\sigma, \phi_4, \phi_5), \; A=0,4,5$:
\begin{eqnarray} \label{equiv-ncu1}
 S &=& \frac{1}{g_5^2} \int d^5 X \Big( \dfrac{1}{4} \widehat{\mathcal{F}}_{MN}
\widehat{\mathcal{F}}^{MN} - \dfrac{1}{2} \widehat{D}_{M} \widehat{\sigma}
\widehat{D}^{M} \widehat{\sigma} \Big), \\
 \label{equiv-un}
   &=& \frac{1}{g_3^2} \int d^3 x \mathrm{Tr} \Bigl( \frac{1}{4} F_{mn}F^{mn}
   + \frac{1}{2} \eta^{AB} D_m \Phi_A D^m \Phi_B
   - \frac{1}{4} \eta^{AC}\eta^{BD} [\Phi_A, \Phi_B] [\Phi_C, \Phi_D] \Bigr)
\end{eqnarray}
where $g_5^2 = (2 \pi \alpha') g_3^2$ and $\eta^{AB} = \mathrm{diag} (-1, 1, 1)$.
Now all the dynamical fields in the action \eq{equiv-un} are $N \times N$ matrices
in the adjoint representation of $U(N \to \infty)$.
The action \eq{equiv-un} respects the $SO(3) \times SO(2,1)$ global symmetry where $SO(3)$ is
the Lorentz symmetry group acting on $\mathbb{R}^3$ and $SO(2,1)$ is the R-symmetry group
acting on $(x^0, y^4, y^5)$ (see footnote \ref{w-sign}).

Let us summarize the isomorphic map from a five-dimensional NC $U(1)$ gauge theory taking values
in $\mathcal{A}_\theta (\mathbb{R}^3) = C^\infty (\mathbb{R}^3) \otimes \mathcal{A}_\theta$
to a three-dimensional large $N$ gauge theory taking values in $\mathcal{A}_N (\mathbb{R}^3)
= C^\infty (\mathbb{R}^3) \otimes \mathcal{A}_N$ \cite{hsy-ijmp09,hsy-jhep09,q-emg}:
\begin{eqnarray} \label{imap-53}
  && \mathcal{A}_\theta (\mathbb{R}^3) \to \mathcal{A}_N (\mathbb{R}^3):
\widehat{\Phi} (x,y) \mapsto \Phi (x), \xx
  && \int \frac{d^2 y}{2 \pi \alpha'} \to \mathrm{Tr}_{\mathcal{H}} = \mathrm{Tr},
\qquad g_5 \to g_3 = \frac{g_5}{\sqrt{2 \pi \alpha'}}.
\end{eqnarray}

The conventional Coulomb branch of the large $N$ gauge theory \eq{equiv-un} is defined by
\begin{equation}\label{cc-vacuum}
    [\Phi_a, \Phi_b]|_{\mathrm{vac}} = 0 \quad \Rightarrow \quad
    \langle \Phi_a \rangle_{\mathrm{vac}} = \mathrm{diag} \big( (\alpha_a)_1, (\alpha_a)_2,
    \cdots, (\alpha_a)_N \big)
\end{equation}
for $a=4,5$. In this case the $U(N)$ gauge symmetry is broken to $U(1)^N$.
It is important to perceive that, in the limit $N \to \infty$, we have a new phase of
the Coulomb branch in addition to the conventional Coulomb branch \eq{cc-vacuum} \cite{q-emg,hea}.
The new vacuum is called the NC Coulomb branch and it is defined by
\begin{equation}\label{nc-vacuum}
   \langle \Phi_a \rangle_{\mathrm{vac}} = p_a, \qquad [p_a, p_b] = - i B_{ab}.
\end{equation}
Note that the NC Coulomb branch \eq{nc-vacuum} saves the NC nature of matrices
while the conventional vacuum \eq{cc-vacuum} dismisses the property. We emphasize that
the NC space $\mathbb{R}^2_{\theta}$ in Eq. \eq{nc-vacuum} arises as a vacuum solution of
the large $N$ gauge theory \eq{equiv-un} when we take the limit $N \to \infty$.
Consequently, the three-dimensional large $N$ gauge theory \eq{equiv-un}
in the NC Coulomb branch is exactly mapped to the five-dimensional NC $U(1)$ gauge
theory \eq{equiv-ncu1} and thus we verify their equivalence in a reverse way.
If the conventional vacuum \eq{cc-vacuum} were chosen, we would have failed to realize
the equivalence. Indeed it turns out \cite{q-emg,hea} that the NC Coulomb branch \eq{nc-vacuum} is
crucial to realize the large $N$ duality which implies the emergent gravity from matrix models
or large $N$ gauge theories.

Recently a localization technique using fixed point theorems provides us a very powerful tool for
the exact computation of the path integral both for topologically twisted supersymmetric theories
and for more general rigid supersymmetric theories defined on curved spaces.
See Refs. \cite{local-qft1}-\cite{local-qft18} for the collection of reviews of this subject.
The power of localization is to reduce the dimensionality of the path integral using supersymmetries
such that the path integral receives contribution from the locus of fixed points of supersymmetry.
We will put a supersymmetric quantum field theory on $\mathbb{S}^{3} \times \mathbb{R}^2_{\theta}$
so that the path integral on $\mathbb{S}^{3}$ is free of infrared divergences. Our aim is to exactly
compute the expectation value $\langle \mathcal{O} \rangle$ of a BPS observable in the quantum theory,
which is defined by
\begin{equation}\label{bps-eval}
   \langle \mathcal{O} \rangle = \int \mathcal{D} \Phi \, \mathcal{O} \exp(-S[\Phi]),
\end{equation}
where $\Phi$ is the set of fields in the action. The usual partition function $Z$ corresponds to
the expectation value of the identity operator, i.e., $Z = \langle \mathbb{I} \rangle$.
We are interested in supersymmetric field theories with a supercharge $Q$ which obeys $Q^2 = B$
with $B$ a linear combination of bosonic charges conserved by the theory. We will assume that
the BPS observable $\mathcal{O}$ as well as the action $S[\Phi]$ is preserved by the supercharge $Q$, i.e.,
\begin{equation}\label{sup-inv-qs}
    Q \mathcal{O} = Q S[\Phi] = 0,
\end{equation}
and the fermionic symmetry generated by $Q$ is free of anomaly.
Then we can use the freedom to deform the path integral of a supersymmetric quantum field theory
by adding a $Q$-exact term to the classical action because
\begin{eqnarray} \label{q-exact}
  \frac{d}{dt} \int \mathcal{D} \Phi \, \mathcal{O} \exp \big(-S[\Phi] - t Q P[\Phi] \big)  &=&
   - \int \mathcal{D} \Phi \, \mathcal{O} Q P[\Phi] \, \exp \big(-S[\Phi] - t Q P[\Phi] \big) \nonumber \\
   &=& \mp \int \mathcal{D} \Phi \, Q \Big\{ \mathcal{O} P[\Phi] \, \exp \big(-S[\Phi]
   - t Q P[\Phi] \big) \Big\} \nonumber \\
   &=& 0,
\end{eqnarray}
where $t$ is a non-negative real parameter and $P[\Phi]$ is a fermionic functional invariant under $B$.
This means that
\begin{equation} \label{t-ind}
   \langle \mathcal{O} \rangle = \int \mathcal{D} \Phi \, \mathcal{O} \exp(-S[\Phi]) =
 \int \mathcal{D} \Phi \, \mathcal{O} \exp \big(-S[\Phi] - t Q P[\Phi] \big).
\end{equation}
Since the equality \eq{t-ind} is valid for any $t \in \mathbb{R}_{\geq 0}$, we can then calculate
the right-hand side of \eq{t-ind} by taking $t \to + \infty$.
In practice, we can choose the functional $P[\Phi]$ properly such that the bosonic part of the deformation
term $QP[\Phi]$ is positive-definite. In this case, the integrand is dominated by the saddle point of
the localizing action
\begin{equation}\label{loc-action}
    S_{\mathrm{loc}} [\Phi] := QP[\Phi].
\end{equation}

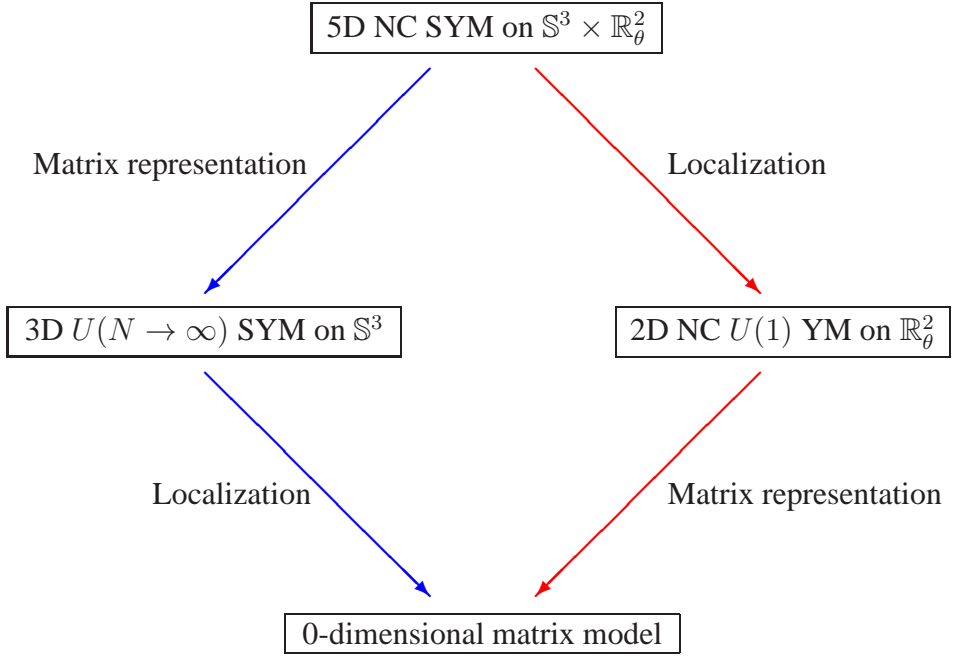
\begin{figure}
\centering
\begin{picture}(400,250)
 \label{flowchart}
%%%%%%%%%%%%%%%%%%%%%%%%%%%%%%%%%%%%%%%%%%%%%%%%%%%%%%%%%%%%%%%%%%%%%%%%%%
% \put(initial point of x,y coord. where the origin is SW corner){Text}  %
% \put(coord.){\vector(x,y direction of vector){Length}}                 %
%%%%%%%%%%%%%%%%%%%%%%%%%%%%%%%%%%%%%%%%%%%%%%%%%%%%%%%%%%%%%%%%%%%%%%%%%%

% Top
\put(135,242){\framebox[1.1\width]{5D NC SYM on $\mathbb{S}^3 \times \mathbb{R}_{\theta}^2$}}
% Left
\put(20,127){\framebox[1.1\width]{3D $U(N \to \infty)$ SYM on $\mathbb{S}^{3}$}}
% Right
\put(250,127){\framebox[1.1\width]{2D NC $U(1)$ YM on $\mathbb{R}_{\theta}^{2}$}}
% Bottom
\put(125,12){\framebox[1.1\width]{0-dimensional matrix model}}

% NE
\put(270,190){Localization}
\thicklines
\textcolor[rgb]{0.00,0.00,1.00}{\put(180,230){\vector(-1,-1){85}}}

% NW
\put(30,190){Matrix representation}
\thicklines
\textcolor[rgb]{0.98,0.00,0.00}{\put(220,230){\vector(1,-1){85}}}

% SE
\put(270,65){Matrix representation}
\thicklines
\textcolor[rgb]{0.00,0.00,1.00}{\put(95,115){\vector(1,-1){85}}}

% SW
\put(75,65){Localization}
\thicklines
\textcolor[rgb]{0.98,0.00,0.00}{\put(305,115){\vector(-1,-1){85}}}
\end{picture}
\caption{Flowchart for zero-dimensional matrix model}
\end{figure}

In the end, the path integral \eq{bps-eval} is localized to the locus $\mathfrak{F}_Q = \{ \Phi|
S_{\mathrm{loc}} [\Phi] = 0 \}$ which is BPS field configurations annihilated by the supercharge $Q$.
Depending on the spacetime dependence of the field configuration in the localization locus
$\mathfrak{F}_Q$, we may be left with the path integral of lower-dimensional field theory or,
in favorable cases, $\mathfrak{F}_Q$ consists of constant field configurations with a finite-dimensional
integral of a zero-dimensional quantum field theory such as matrix models \cite{local-qft1}-\cite{local-qft18}.
Since we will consider a supersymmetric field theory on $\mathbb{S}^{3} \times \mathbb{R}^2_{\theta}$,
we will have the supersymmetric version of the equality in Eqs. \eq{equiv-ncu1} and \eq{equiv-un}.
Although two theories are defined in different dimensions with different gauge groups,
they are mathematically equivalent to each other.
Therefore, we can apply the localization to either a five-dimensional
supersymmetric NC $U(1)$ gauge theory or a three-dimensional supersymmetric $U(N \to \infty)$ gauge theory.
On the one hand, we can first apply the localization to the five-dimensional theory to obtain
a two-dimensional NC gauge theory and then consider the matrix representation of the resulting NC gauge theory
to yield a zero-dimensional matrix model, as depicted in Fig. \ref{flowchart}.
On the other hand, we can first implement the matrix representation to the five-dimensional theory to get
a three-dimensional large $N$ gauge theory and then apply the localization to the large $N$ gauge theory on $\mathbb{S}^{3}$ to derive a zero-dimensional matrix model. Both routes should end in an identical
zero-dimensional matrix model. The aim of this paper is to verify the flowchart outlined
in Fig. \ref{flowchart} using the localization technique and the matrix representation of NC field theories.

This paper is organized as follows.
In section 2, we construct a five-dimensional $\mathcal{N}=2$ supersymmetric NC $U(1)$ gauge theory
on $\mathbb{S}^3 \times \mathbb{R}_\theta^2$.
Using the matrix representation \eq{matrix-rep}, the five-dimensional supersymmetric NC $U(1)$
gauge theory is isomorphically mapped to a three-dimensional supersymmetric $U(N \to \infty)$
gauge theory on $\mathbb{S}^3$.

In section 3 we perform the localization of the five-dimensional supersymmetric NC $U(1)$ gauge theory
on $\mathbb{S}^3 \times \mathbb{R}_\theta^2$, which results in a two-dimensional
NC $U(1)$ gauge theory. The matrix representation of the two-dimensional
NC $U(1)$ gauge theory leads to a zero-dimensional matrix model at the localization locus.
We thus explore the red arrows in Fig. \ref{flowchart} to derive a zero-dimensional matrix model
via the localization of a five-dimensional NC $U(1)$ gauge theory.

In section 4 we follow the blue arrows in Fig. \ref{flowchart} to get the zero-dimensional matrix model
via the localization of a three-dimensional $U(N \to \infty)$ gauge theory.
Using the fact that a large $N$ gauge theory admits a NC Coulomb branch in the limit $N \to \infty$
and any $N \times N$ (Hermitian) matrix can be regarded as the matrix representation of
a higher-dimensional NC field, the localization of the three-dimensional large $N$ gauge theory
can be easily done by mapping the problem to the one of a five-dimensional NC $U(1)$ gauge theory.

In section 5 we appeal to the mathematical identity depicted in Fig. \ref{flowchart},
in particular, the equivalence between a higher-dimensional NC $U(1)$ gauge theory and
a lower-dimensional large $N$ gauge theory. This implies \cite{q-emg,hea} that
the five-dimensional NC $U(1)$ gauge theory describes a five-dimensional gravity
according to the large $N$ duality or gauge/gravity duality.
We discuss a physical implication of the localization in Fig. \ref{flowchart}
from the point of view of the emergent gravity.

We include five appendices, containing our notation and conventions, some details on the supersymmetric
transformations, the harmonic analysis on $\mathbb{S}^3$, and the Clebsch-Gordan coefficients
for the irreducible representation of the tensor products $j \otimes 1$ and $j \otimes \frac{1}{2}$.

\section{Three-dimensional large $N$ gauge theory from five-dimensional NC $U(1)$ gauge theory}

The vector multiplet in the five-dimensional $\mathcal{N}=2$ supersymmetric Yang-Mills (SYM) theory
consists of an $\mathcal{N}=1$ vector multiplet and an $\mathcal{N}=1$ hypermultiplet in the adjoint
representation of gauge group $\mathfrak{g}$. We start with the $\mathcal{N}=2$ SYM theory on the NC space
$\mathbb{R}^{3} \times \mathbb{R}^2_{\theta}$ with the gauge group $\mathfrak{g} = U(1)_\star$.
We will follow Refs. \cite{kama12,fukama13,fukama15} for the supersymmetric actions with minor modifications.
Later we will put the theory on $\mathbb{S}^{3} \times \mathbb{R}^2_{\theta}$ to carry out
the localization, which will require some additional terms in the action and a modification of
supersymmetry transformations.
For a notational simplicity, we will omit the hat symbol to indicate five-dimensional NC fields
and implicitly assume the star product \eq{star-prod} for the multiplication between NC fields.
We hope it does not cause much confusion with three-dimensional large $N$ matrices.
The vector multiplet in the five-dimensional $\mathcal{N}=1$ SYM theory consists of a gauge field
$A_M$, a real scalar field $\sigma$ and a doublet of spinor fields $\Psi^{\dot{\alpha}}$
where the doublet of $SU(2)$ R-symmetry is labeled by the indices $\dot{\alpha}, \dot{\beta} = 1, 2$.
The spinor field obeys the symplectic Majorana condition
\begin{equation}\label{sym-majorana}
    (\Psi^{\dot{\alpha}})^\dagger = (\Psi^{\dot{\beta}})^T C_5 \varepsilon_{\dot{\beta}\dot{\alpha}}
    \equiv  \overline{\Psi}_{\dot{\alpha}}.
\end{equation}
To realize an off-shell supersymmetry, we also introduce an auxiliary field $D^{\dot{\alpha}}_{~\dot{\beta}}$
in the adjoint representation of $SU(2)_R$ R-symmetry:
\begin{equation}\label{aux-d}
D^{\dot{\alpha}}_{~\dot{\beta}} = ( D^{\dot{\beta}}_{~\dot{\alpha}})^\dagger,
\qquad  D^{\dot{\alpha}}_{~\dot{\alpha}} = 0.
\end{equation}
Thus the auxiliary field may be represented by $D^{\dot{\alpha}}_{~\dot{\beta}} = D_m
(\sigma^m)^{\dot{\alpha}}_{~\dot{\beta}}$ where $D_m \, (m=1,2,3)$ is the triplet of real scalar fields.
The $\mathcal{N}=1$ vector multiplet describes the supersymmetric version of the
five-dimensional NC $U(1)$ gauge theory \eq{5d-action}. The $\mathcal{N}=1$ hypermultiplet
consists of complex scalar fields $H_{\dot{\alpha}}$, a spinor field $\Theta = \Theta_1 + i \Theta_2$
and two auxiliary fields $F_{\alpha} \, (\alpha=1,2)$. They are also in the adjoint representation of
the gauge group $\mathfrak{g}$ so that the hypermultiplet is combined with the $\mathcal{N}=1$ vector
multiplet to form a vector multiplet in the $\mathcal{N}=2$ theory.

The action for the $\mathcal{N}=1$ vector multiplet is given by
\begin{equation} \label{vector-action}
S_{5V} = \int d^5 X \bigg[ \dfrac{1}{4g_5^{2}} \mathcal{F}_{MN} \mathcal{F}^{MN}
- \dfrac{1}{2g_5^{2}} D_{M} \sigma D^{M} \sigma
- i \overline{\Psi}_{\dot{\alpha}} \Gamma^{M} D_{M} \Psi^{\dot{\alpha}} - \overline{\Psi}_{\dot{\alpha}}
[\sigma, \Psi^{\dot{\alpha}} ] - \dfrac{1}{4} D^{\dot{\alpha}}_{~\dot{\beta}} D^{\dot{\beta}}_{~\dot{\alpha}} \bigg],
\end{equation}
where $\mathcal{F}_{MN} = F_{MN} - B_{MN}$.
The above action is invariant under the following supersymmetric transformations
\begin{eqnarray} \label{susytr}
\delta A_{M} &=& i g_5 \overline{\Sigma}_{\dot{\alpha}} \Gamma_{M} \Psi^{\dot{\alpha}},
\qquad \delta\sigma = -i g_5 \overline{\Sigma}_{\dot{\alpha}} \Psi^{\dot{\alpha}}, \nonumber\\
\delta \Psi^{\dot{\alpha}}
&=& \frac{1}{2} \left( \frac{1}{2g_5} \mathcal{F}_{MN} \Gamma^{MN} \Sigma^{\dot{\alpha}}
+ \frac{1}{g_5}  D_{M}\sigma \Gamma^{M} \Sigma^{\dot{\alpha}} - i D_{~\dot{\beta}}^{\dot{\alpha}}
\Sigma^{\dot{\beta}} \right), \\
\delta D_{~\dot{\beta}}^{\dot{\alpha}} &=&
D_{M} \overline{\Psi}_{\dot{\beta}} \Gamma^{M} \Sigma^{\dot{\alpha}}
+ \overline{\Sigma}_{\dot{\beta}} \Gamma^{M} D_{M} \Psi^{\dot{\alpha}}
- i \left( [\sigma, \overline{\Psi}_{\dot{\beta}}] \Sigma^{\dot{\alpha}}
+ \overline{\Sigma}_{\dot{\beta}} [\sigma, \Psi^{\dot{\alpha}}] \right), \nonumber
\end{eqnarray}
where the transformation parameters $\Sigma^{\dot{\alpha}}$ are the $SU(2)_R$ doublet of symplectic Majorana spinors. Using the definition of symplectic Majorana spinor,
one can deduce the supersymmetric transformation for $\overline{\Psi}_{\dot{\alpha}}$:
\begin{equation}
\delta \overline{\Psi}_{\dot{\alpha}} = \frac{1}{2} \left( - \frac{1}{2g_5} \overline{\Sigma}_{\dot{\alpha}}
\Gamma^{MN} \mathcal{F}_{MN} + \frac{1}{g_5} \overline{\Sigma}_{\dot{\alpha}} \Gamma^{M} D_{M}\sigma
+ i \overline{\Sigma}_{\dot{\beta}} D_{~\dot{\alpha}}^{\dot{\beta}}  \right). \\
\end{equation}
It is straightforward to check the supersymmetric invariance of the action \eq{vector-action}
if the cyclic permutation under the integral such as Eqs. \eq{com-int} and \eq{cyc-int}
is carefully used. As in the commutative case,
after cancellation of all the quadratic fermion terms, we are left with the cubic terms coming from
the supersymmetric transformations
\begin{equation}\label{f-cubic-term}
- \overline{\Psi}_{\dot{\alpha}} \Gamma^{M} [\delta A_{M}, \Psi^{\dot{\alpha}}]
- \overline{\Psi}_{\dot{\alpha}} [\delta\sigma, \Psi^{\dot{\alpha}}].
\end{equation}
One can show by applying the Fierz identity \eq{fierz-id} that these terms cancel each other.
The details show up in appendix B.

We can apply the matrix representation \eq{matrix-rep} to the supersymmetric action \eq{vector-action}.
The bosonic part was already done in \eq{equiv-un} except the auxiliary term.
The matrix representation for the five-dimensional spinors $\Psi^{\dot{\alpha}}(x,y)$ is
basically the same as the bosonic fields and it is denoted by the same symbol $\Psi^{\dot{\alpha}}(x)$
that are now $N \times N$ matrices over $\mathbb{R}^3$. But, in order to get a three-dimensional
picture after the matrix representation, it is convenient to represent the symplectic Majorana spinors
$\Psi^{\dot{\alpha}}(x)$ in terms of three-dimensional spinors:\footnote{The multiplication factor
$\sqrt{2\pi \theta}$ is to match the physical mass dimension of five- and three-dimensional spinors
since $[\Psi] = M^2$ and $[\lambda] = [\psi] = M$. Then this factor will compensate the one in
the measure \eq{int=tr}.}
\begin{equation} \label{sdec-5d}
\begin{array}{l}
 \sqrt{2\pi \theta} \Psi^{1} (x) = \lambda (x) \otimes \zeta_{+} + \psi (x) \otimes \zeta_{-}, \\
 \sqrt{2\pi \theta} \Psi^{2} (x) = C_{3}^{-1} \psi^{*} (x) \otimes \zeta_{+}
 + C_{3}^{-1} \lambda^{*} (x) \otimes \zeta_{-}
\end{array}
\end{equation}
where two-dimensional Weyl spinors
\begin{equation}\label{2d-weyl}
    \zeta_\pm = \frac{1}{\sqrt{2}} \left(
                                    \begin{array}{c}
                                      1 \\
                                      \pm i \\
                                    \end{array}
                                  \right)
\end{equation}
are the eigenvectors of $i\Gamma^4 \Gamma^5$, i.e., $i\Gamma^4 \Gamma^5 \zeta_\pm = \pm \zeta_\pm$.
Then the conjugate spinors are given by
\begin{equation} \label{csdec-5d}
\begin{array}{l}
 \sqrt{2\pi \theta} \overline{\Psi}_{1} (x) = \overline{\lambda} (x) \otimes \overline{\zeta}_{+}
 + \overline{\psi} (x) \otimes \overline{\zeta}_{-}, \\
 \sqrt{2\pi \theta} \overline{\Psi}_{2} (x) = \psi^{T} (x) C_{3} \otimes \overline{\zeta}_{+}
 + \lambda^{T} (x) C_{3} \otimes \overline{\zeta}_{-}
\end{array}
\end{equation}
where the barred spinors are defined by $\overline{\lambda} \equiv \lambda^\dagger = (\lambda^*)^T$, etc.
Using this result, we can get the three-dimensional supersymmetric large $N$ gauge theory
whose action takes the form
\begin{eqnarray} \label{3d-sym}
 S_{3V} &=& \frac{1}{g_3^2} \int d^3 x \mathrm{Tr} \Bigl( \frac{1}{4} F_{mn}F^{mn}
   + \frac{1}{2} \eta^{AB} D_m \Phi_A D^m \Phi_B
   - \frac{1}{4} \eta^{AC}\eta^{BD} [\Phi_A, \Phi_B] [\Phi_C, \Phi_D] \Bigr) \xx
   && + 2 \int d^3 x \mathrm{Tr} \Bigl( i \overline{\psi} \gamma^m D_m \psi
   - i \overline{\lambda} \gamma^m D_m \lambda - 2i \big(\overline{\psi} [\phi_z, \lambda] -
   \overline{\lambda} [\phi_{\overline{z}}, \psi] \big)  \Bigr.\xx
 && \hspace{2.3cm} - \overline{\psi} [\sigma, \psi] - \overline{\lambda} [\sigma, \lambda]
 - \dfrac{1}{8} d^{\dot{\alpha}}_{~\dot{\beta}} d^{\dot{\beta}}_{~\dot{\alpha}} \Bigr),
\end{eqnarray}
where we have introduced the complex coordinates for $\mathbb{R}^2 \cong \mathbb{C}$
given by\footnote{In terms of complex coordinates, the commutator for the star product \eq{star-prod}
is then given by $[f, g] = f \star g - g \star f =
2 \alpha' \big(\frac{\partial f}{\partial z} \frac{\partial g}{\partial \overline{z}}
- \frac{\partial g}{\partial z} \frac{\partial f}{\partial \overline{z}} \big) + \mathcal{O} ({\alpha'}^3)$
for any two NC fields $f, g \in C^\infty (\mathbb{R}^3) \otimes \mathcal{A}_\theta$.}
\begin{equation}\label{comp-coord}
    z = y^4 + i y^5, \qquad \overline{z} = y^4 - i y^5
\end{equation}
and the complex scalar field defined by
\begin{equation}\label{comp-scalar}
    \phi_z = \frac{1}{2}(\phi_4 - i \phi_5), \qquad
    \phi_{\overline{z}} = \frac{1}{2}(\phi_4 + i \phi_5).
\end{equation}
The auxiliary fields $D^{\dot{\alpha}}_{~\dot{\beta}} (x,y)$ are also represented by matrices  $d^{\dot{\alpha}}_{~\dot{\beta}} (x) \equiv \sqrt{2\pi \theta} D^{\dot{\alpha}}_{~\dot{\beta}} (x)$.
Since the chirality condition, $i\Gamma^4 \Gamma^5 \zeta_\pm = \pm \zeta_\pm$, has been imposed on
the spinor $\zeta_\pm$, the $SO(2,1)$ global symmetry now reduces to $U(1)$.

Since the three-dimensional large $N$ gauge theory \eq{3d-sym} has been obtained from the matrix
representation of the five-dimensional $\mathcal{N}=1$ vector multiplet without any loss of supersymmetry,
the theory \eq{3d-sym} must preserve $\mathcal{N}=2$ supersymmetries in three dimensions.
Using the matrix representation \eq{matrix-rep} again, it should be straightforward to
derive the supersymmetry transformations for the three-dimensional large $N$ gauge theory \eq{3d-sym}
from the five-dimensional ones in Eq. \eq{susytr}
by taking the supersymmetry transformation parameters
\begin{equation}\label{susypa-n=2}
    \Sigma^1 = \epsilon_1 \otimes \zeta_+ + \epsilon_2 \otimes \zeta_- ,
    \qquad \Sigma^2 = C_3^{-1} \epsilon_2^* \otimes \zeta_+ + C_3^{-1} \epsilon_1^* \otimes \zeta_-.
\end{equation}
For localization, we will eventually put the $\mathcal{N}=2$ theory on $\mathbb{S}^3$,
which is a maximally supersymmetric background, i.e., preserves four supercharges \cite{susy-3dim}.
However we must select the supercharges that will be used to localize.
Therefore we want to focus on the $\mathcal{N}=1$ sector in the $\mathcal{N}=2$
supersymmetries which will be identified with a localizing supercharge on $\mathbb{S}^3$.
For that reason, we consider the supersymmetry transformation parameters given by
\begin{equation}\label{susypa-53}
    \Sigma^1 = \epsilon \otimes \zeta_+, \qquad \Sigma^2 = C_3^{-1} \epsilon^* \otimes \zeta_-.
\end{equation}
Then the supersymmetry transformations generated by the above parameter are given by \cite{fukama13}
\begin{eqnarray}
&& \delta A_m = i g_3 (\overline{\epsilon} \gamma_m \lambda - \overline{\lambda} \gamma_m \epsilon ), \xx
&& \delta \phi_z = g_3 \overline{\epsilon}  \psi, \xx
&& \delta \sigma = - i g_3 (\overline{\epsilon} \lambda - \overline{\lambda} \epsilon ), \xx
\label{susytr-3d}
&& \delta \lambda = \frac{1}{2} \Big( \frac{1}{2g_3} F_{mn} \gamma^{mn} + \frac{1}{g_3}
D_m \sigma \gamma^m - i D \Big) \epsilon, \\
&& \delta \psi = - \frac{i}{g_3} D_m \phi_z \gamma^m \epsilon + \frac{1}{g_3} [\phi_z, \sigma] \epsilon
- i F C_3^{-1} \epsilon^*, \xx
&& \delta F = - \epsilon^T C_3 \big( \gamma^m D_m \psi + i [\sigma, \psi] - 2 [\phi_z, \lambda] \big), \xx
&& \delta D =  D_m \overline{\lambda} \gamma^m \epsilon + \overline{\epsilon} \gamma^m D_m \lambda
- i \big( [\sigma, \overline{\lambda}] \epsilon + \overline{\epsilon} [\sigma, \lambda] \big), \nonumber
\end{eqnarray}
where
\begin{equation}\label{fd}
    F \equiv \frac{1}{2} {d^1}_2, \qquad D \equiv {d^1}_1 + \frac{2}{g_3} [\phi_{\overline{z}}, \phi_z].
\end{equation}

The $\mathcal{N}=1$ hypermultiplet is described by the action given by
\begin{eqnarray} \label{hyper-action}
S_{5H} &=& \int d^5 X \bigg[ \frac{1}{g_5^{2}} D^{M} \overline{H}^{\dot{\alpha}}
D_{M} H_{\dot{\alpha}} + i \overline{\Theta} \Gamma^{M} D_{M} \Theta - \overline{F}^\alpha F_\alpha
+ \frac{1}{g_5^{2}} [\sigma, \overline{H}^{\dot{\alpha}}] [\sigma, H_{\dot{\alpha}} ] \nonumber \\
&& \hspace{1.5cm} + \frac{1}{g_5} D^{\dot{\alpha}}_{~\dot{\beta}} [\overline{H}^{\dot{\beta}}, H_{\dot{\alpha}}]
- \overline{\Theta} [\sigma, \Theta] - 2 \overline{\Theta}[ H_{\dot{\alpha}}, \Psi^{\dot{\alpha}}]
+ 2 [ \overline{H}^{\dot{\alpha}}, \overline{\Psi}_{\dot{\alpha}}] \Theta \bigg]
\end{eqnarray}
where $\overline{H}^{\dot{\alpha}}$ is the complex conjugate of $H_{\dot{\alpha}}$ and
$\overline{\Theta} = \Theta^\dagger$.
It is invariant under the following supersymmetry transformations
\begin{eqnarray}
  && \delta H_{\dot{\alpha}} = i g_5 \overline{\Sigma}_{\dot{\alpha}} \Theta, \xx
\label{hsusytr}
  && \delta \Theta = - \frac{1}{g_5} \Big( \Gamma^M D_M  H_{\dot{\alpha}}
  - i [\sigma, H_{\dot{\alpha}} ] \Big) \Sigma^{\dot{\alpha}}  + F_\alpha \Lambda^\alpha, \\
  && \delta F_\alpha = \overline{\Lambda}_\alpha \Big( i \Gamma^{M} D_{M} \Theta
  -   [\sigma, \Theta ]  - 2 [ H_{\dot{\beta}}, \Psi^{\dot{\beta}}]  \Big), \nonumber
\end{eqnarray}
where $\Lambda^\alpha$ are symplectic Majorana spinors and $\overline{\Lambda}_\alpha
= (\Lambda^\beta)^T C_5 \varepsilon_{\beta\alpha}$.
Like the vector multiplet, the action \eq{hyper-action} is invariant under the above supersymmetric
transformations but it is not necessary to use the Fierz identity for fermionic cubic terms.

We can similarly apply the matrix representation \eq{matrix-rep} to the hypermultiplet.
For this purpose, let us represent the complex scalar fields $H_{\dot{\alpha}} (x,y)$
and spinor $\Theta (x, y)$ in terms of three-dimensional fields in the adjoint representation
of $U(N \to \infty)$:\footnote{\label{japan-ref}Our representation \eq{spin-theta}
for the spinor $\Theta (x, y)$ is different from that in Ref. \cite{fukama13} taking
the form $\sqrt{2\pi \theta}  \Theta (x, y) \mapsto \eta (x) \otimes \zeta_+
+ C_3^{-1} \chi(x)^* \otimes \zeta_-.$}
\begin{eqnarray} \label{matrix-h}
&& H_{\dot{\alpha}} (x,y) \mapsto h_{\dot{\alpha}} (x), \\
\label{spin-theta}
&& \sqrt{2\pi \theta}  \Theta (x, y) \mapsto \eta (x) \otimes \zeta_+ + \chi(x) \otimes \zeta_-.
\end{eqnarray}
We also denote the spinors in Eq. \eq{sdec-5d} with a compact notation:
\begin{equation} \label{matrix-psi}
\sqrt{2\pi \theta}  \Psi^{\dot{\alpha}} (x) \equiv \xi_+^{\dot{\alpha}} (x) \otimes \zeta_+
+ \xi_-^{\dot{\alpha}} (x) \otimes \zeta_-.
\end{equation}
Using this matrix representation, it is straightforward to get the three-dimensional action
for the hypermultiplet given by
\begin{eqnarray} \label{mhyper-action}
S_{3H} &=& \int d^3 x \mathrm{Tr} \bigg[ \frac{1}{g_3^{2}} D^{m} \overline{h}^{\dot{\alpha}}
D_{m} h_{\dot{\alpha}} - \frac{1}{g_3^{2}} \eta^{AB} [\Phi_A, \overline{h}^{\dot{\alpha}}]
[\Phi_B, h_{\dot{\alpha}}]  + \frac{1}{g_3} d^{\dot{\alpha}}_{~\dot{\beta}}
[\overline{h}^{\dot{\beta}}, h_{\dot{\alpha}} ] - \overline{f}^\alpha f_\alpha \nonumber \\
&& \hspace{1.5cm} + i \overline{\eta} \gamma^{m} D_{m} \eta - i \overline{\chi} \gamma^{m} D_{m} \chi
+ 2i \big( \overline{\chi} [\phi_z, \eta] - \overline{\eta} [\phi_{\overline{z}}, \chi] \big)
- \overline{\eta} [\sigma, \eta] - \overline{\chi} [\sigma, \chi] \nonumber \\
&&  \hspace{1.5cm} - 2 \Big( \overline{\eta}[ h_{\dot{\alpha}}, \xi_+^{\dot{\alpha}}]
+ \overline{\chi} [ h_{\dot{\alpha}}, \xi_-^{\dot{\alpha}}]
- [ \overline{h}^{\dot{\alpha}}, \overline{\xi}_{+ \dot{\alpha}}] \eta
- [ \overline{h}^{\dot{\alpha}}, \overline{\xi}_{- \dot{\alpha}}] \chi \Big) \bigg],
\end{eqnarray}
where $f_{\alpha} (x) \equiv \sqrt{2\pi \theta}  F_{\alpha} (x)$ is the matrix representation of
the auxiliary fields $F_{\alpha} (x, y)$.

The supersymmetry transformations for the three-dimensional hypermultiplet will be obtained by
the matrix representation of the five-dimensional ones in Eq. \eq{hsusytr}.
Having in mind a localizing supersymmetry on $\mathbb{S}^3$,
let us consider the supersymmetry transformation parameters given by
\begin{equation}\label{susypa-h3}
    \Lambda^1 = \epsilon \otimes \zeta_-, \qquad \Lambda^2 = C_3^{-1} \epsilon^* \otimes \zeta_+
\end{equation}
which are related to the spinors $\Sigma^{\dot{\alpha}}$ in Eq. \eq{susypa-53} by
$\Lambda^{\dot{\alpha}}=\Gamma^5 \Sigma^{\dot{\alpha}}$. For this reason, we will use
the spinor index $\dot{\alpha}$ for the spinors in Eq. \eq{susypa-h3}.
The supersymmetry transformations generated by the spinors $\Sigma^{\dot{\alpha}}$
and $\Lambda^{\dot{\alpha}}$ are easily deduced from Eq. \eq{hsusytr},
which are given by
\begin{eqnarray}
&& \delta h_1 = i g_3 \overline{\epsilon} \eta, \qquad
\delta h_2 = i g_3 \epsilon^T C_3 \chi, \xx
&& \delta \eta = - \frac{1}{g_3} \Big( \gamma^m \epsilon D_m h_1 - 2 C^{-1}_3 \epsilon^*
[\phi_{\overline{z}}, h_2] - i \epsilon [\sigma, h_1] \Big) + f_2 C^{-1}_3 \epsilon^*, \xx
\label{3hsusy}
&& \delta \chi = \frac{1}{g_3} \Big( \gamma^m C^{-1}_3 \epsilon^* D_m h_2 - 2 \epsilon
[\phi_{z}, h_1] + i C^{-1}_3 \epsilon^* [\sigma, h_2] \Big) + f_1 \epsilon, \\
&& \delta f_1 = - \overline{\epsilon} \Big( i \gamma^m D_m \chi - 2 i
[\phi_{z}, \eta] +  [\sigma, \chi] + 2 [h_{\dot{\alpha}}, \xi_-^{\dot{\alpha}}] \Big), \xx
&& \delta f_2 = - \epsilon^T C_3 \Big( - i \gamma^m D_m \eta + 2 i [\phi_{\overline{z}}, \chi]
+ [\sigma, \eta] + 2 [h_{\dot{\alpha}}, \xi_+^{\dot{\alpha}}] \Big). \nonumber
\end{eqnarray}

Using the matrix representation \eq{matrix-rep}, we have obtained two mathematically equivalent theories,
that are defined in different dimensions with different gauge groups.
Although the two theories superficially look quite different, they should be physically equivalent
to each other. To explore the physical implications of the equivalence,
one may apply localization techniques to each of them to compute, for example, partition functions
and some correlators exactly. On the one hand, one can first apply the localization to the five-dimensional
theory to obtain a two-dimensional NC gauge theory and then take the matrix representation to
the two-dimensional NC gauge theory to yield a zero-dimensional matrix model,
as depicted in Fig. \ref{flowchart}.
On the other hand, one can first apply the matrix representation to the five-dimensional theory
to have a three-dimensional large $N$ gauge theory and then consider the localization of
the large $N$ gauge theory to get a zero-dimensional matrix model.
Both routes should end up with an identical zero-dimensional matrix model.
In the end, the localization will verify a rich duality between NC $U(1)$ gauge theories and
large $N$ matrix models in various dimensions as outlined in Fig. \ref{flowchart}.

To carry out the localization, we put the theory on $\mathbb{S}^3 \times \mathbb{R}_{\theta}^2$.
Let $r$ be the radius of $\mathbb{S}^3$.
For this purpose, it is convenient to represent five-dimensional fields as the form of
three-dimensional fields in which extra coordinates $y^a$ are regarded as parameters
living in $\mathbb{R}_{\theta}^2$.
For the vector multiplet on $\mathbb{S}^3 \times \mathbb{R}_{\theta}^2$,
we take the following representation \cite{kama12,fukama13}:\footnote{From now on,
we distinguish the curved space indices,
$\mu, \nu, \cdots = 1,2,3$ from the flat space ones $m, n, \cdots = 1,2,3$.
The dreibein on $\mathbb{S}^3$ is denoted by $e^m = e^m_\mu dx^\mu$ and obeys the structure
equation $de^m = \frac{1}{r} \varepsilon^{mnp} e^n \wedge e^p$.
The metric and spin connections are given by $ds^2 = e^m \otimes e^m = e^m_\mu e^m_\nu
dx^\mu \otimes dx^\nu$ and $\omega_\mu = \frac{1}{4} \omega_\mu^{mn} \gamma_{mn}
= \frac{i}{2r} \gamma_\mu$, respectively.}
\begin{equation}\label{3-split}
  \begin{array}{l}
    A_\mu (x, y) \; \; (\mu=1,2,3), \qquad A_a (x, y) = \phi_a (x,y) - B_{ab} y^b
    \;\; (a = 4,5), \\
    F (x, y) =  \frac{1}{2} {D^1}_2 (x,y), \qquad D (x,y) = {D^1}_1 (x, y)
    + \frac{2}{g_5} [\phi_{\overline{z}}, \phi_z] (x, y), \\
\Psi^{1} (x, y) = \lambda (x, y) \otimes \zeta_{+} + \psi (x, y) \otimes \zeta_{-}, \\
\Psi^{2} (x, y) = C_{3}^{-1} \psi^{*} (x, y) \otimes \zeta_{+}
+ C_{3}^{-1} \lambda^{*} (x, y) \otimes \zeta_{-}.
\end{array}
\end{equation}
One may notice that the pattern of the above decomposition is equal to the three-dimensional
large $N$ matrices in the action \eq{3d-sym}. This replica is not accidental because
the matrix representation of the NC fields in Eq. \eq{3-split} will give rise to
the three-dimensional large $N$ gauge theory on $\mathbb{S}^3$ whose action
is precisely equal to Eq. \eq{3d-sym}. Indeed the corresponding five-dimensional action
after the decomposition \eq{3-split} can be written as the same form as Eq. \eq{3d-sym} with
simple replacements $g_3 \to g_5, \; d^{\dot{\alpha}}_{~\dot{\beta}} \to D^{\dot{\alpha}}_{~\dot{\beta}}$
and $\int d^3x \Tr \to \int d\upsilon \equiv \int \frac{d^2 y}{2 \pi \alpha'} d^3 x \sqrt{g}$.

Let us consider the supersymmetry transformation parameters as Eq. \eq{susypa-53}
and take $\epsilon$ to be a Killing spinor on $\mathbb{S}^3$ obeying the following equation
\begin{equation}\label{kill-spinor}
    \nabla_\mu \epsilon = \frac{i}{2r} \gamma_\mu \epsilon,
\end{equation}
where the covariant derivative acting on a spinor is given by
\begin{equation} \label{spin-cov}
\nabla_\mu = \partial_\mu + \omega_\mu = \partial_\mu + \frac{i}{2r} \gamma_\mu.
\end{equation}
Then the covariant derivative acting on a spinor $\Psi$ in the adjoint
representation of $\mathfrak{g} = U(1)_\star$ is defined by
\begin{equation} \label{gspin-cov}
D_\mu \Psi = \nabla_\mu \Psi - i [A_\mu, \Psi].
\end{equation}
For a gauge singlet which does not depend on $(y^4, y^5)$, it reduces to Eq. \eq{spin-cov}.
The supersymmetry transformations generated by the Killing spinor $\epsilon$ obeying
Eq. \eq{kill-spinor} are also given by Eq. \eq{susytr-3d} with the replacement
$g_3 \to g_5$. However, after imposing the condition \eq{kill-spinor},
the supersymmetry transformations will no longer be closed
because the covariant derivative now acts on the spinor $\epsilon$ nontrivially.
Fortunately, to achieve a closed algebra, it is enough to modify the transformation law only
for the auxiliary fields by adding \begin{equation}\label{mod-susy}
\delta' D = \frac{i}{2r} (\overline{\epsilon} \lambda - \overline{\lambda} \epsilon),
\qquad   \delta' F = - \frac{i}{2r} \epsilon^T C_3 \psi.
\end{equation}
We verify the closed algebra in appendix C. The result
is essentially the same as the one in Refs. \cite{kama12,fukama13} although two-dimensional
surface in our case is a NC space.

Using the matrix representation of the NC fields in Eq. \eq{3-split},
the five-dimensional $U(1)$ gauge theory on $\mathbb{S}^3 \times \mathbb{R}_{\theta}^2$ can
easily be transformed to a three-dimensional large $N$ gauge theory on $\mathbb{S}^3$.
The three-dimensional action on $\mathbb{S}^3$ is obtained from the result in Eq. \eq{3d-sym}
on $\mathbb{R}^3$ with an obvious replacement, $\int d^3x \to \int d^3 x \sqrt{g}$.
The supersymmetry transformations of large $N$ matrices on $\mathbb{S}^3$ can easily be deduced
from the five-dimensional ones using the matrix representation in a similar way.
Moreover, the closedness of the three-dimensional supersymmetric algebra simply results from
the five-dimensional one.

The modified supersymmetry transformations generated by the spinor $\Sigma^{\dot{\alpha}}$
obeying Eq. \eq{kill-spinor} will be denoted by $\Delta_\epsilon = \delta
+ \delta'$ where $\delta'$-transformation is given by Eq. \eq{mod-susy}.
Since the supersymmetry transformation parameter $\epsilon$ obeys the nontrivial Killing spinor
equation \eq{kill-spinor}, the action \eq{vector-action} is no longer invariant under the
$\Delta_\epsilon$-transformations. Indeed its variation reduces to
\begin{equation}\label{mod-variation}
    \Delta_\epsilon S_{5V} = \delta S_{5V} + \delta' S_{5V}
\end{equation}
where
\begin{equation}\label{s3-vvar}
\delta S_{5V} = \frac{i}{g_5}  \int d \upsilon \Big(  \nabla_N \overline{\Sigma}_{\dot{\alpha}}
\Gamma^M \Gamma^N \Psi^{\dot{\alpha}} D_M \sigma - \frac{1}{2} \nabla_L \overline{\Sigma}_{\dot{\alpha}}
\Gamma^{MN} \Gamma^L \Psi^{\dot{\alpha}} F_{MN} \Big)
\end{equation}
and
\begin{equation}\label{mod-vvar}
\delta' S_{5V} = - \frac{i}{r} \int d \upsilon \Big( \frac{D}{2}(\overline{\epsilon} \lambda
- \overline{\lambda} \epsilon) - \frac{1}{g_5} [\phi_{\overline{z}}, \phi_z]
(\overline{\epsilon} \lambda - \overline{\lambda} \epsilon) + \overline{\psi} C_3^{-1} \epsilon^* F
- \overline{F} \epsilon^T C_3 \psi \Big).
\end{equation}
In order to preserve the $\Delta_\epsilon$-supersymmetry, it is necessary to add an extra action
such that its supersymmetric transformation cancels the variation \eq{mod-variation}.
It turns out \cite{kama12,fukama13} that the extra action is given by
\begin{equation}\label{add-5v}
    S'_{5V} = S_{5M} + S_{CS}
\end{equation}
where
\begin{equation}\label{mass-action}
    S_{5M} = - \frac{1}{r} \int d\upsilon \Big(\overline{\psi} \psi + \overline{\lambda} \lambda
    + \frac{1}{g_5^2 r} \sigma^2 + \frac{1}{g_5} \sigma \big( D - \frac{4}{g_5}
    [\phi_{\overline{z}}, \phi_z] \big) \Big)
\end{equation}
and
\begin{equation}\label{cs-term}
   S_{CS} = - \frac{1}{2g_5^2} \int \big( A \wedge F + \frac{i}{3} A \wedge A \wedge A \big) \wedge \varpi
\end{equation}
with $\varpi = \frac{1}{r} dy^4 \wedge dy^5$. Then the total action $S_{5V} + S'_{5V}$ is invariant
under the supersymmetry transformations $\Delta_\epsilon$.

We also put the hypermultiplet on $\mathbb{S}^3 \times \mathbb{R}_{\theta}$ for the localization.
The strategy is the same as the vector multiplet.
In order to achieve a closed algebra of the supersymmetry on $\mathbb{S}^3 \times \mathbb{R}_{\theta}$
generated by the Killing spinor obeying the condition \eq{kill-spinor}, it is necessary to modify
the supersymmetry transformations in Eq. \eq{hsusytr} by simply adding additional transformations
\begin{equation}\label{add-hsusy}
    \delta' \Theta = \frac{1}{g_5 r} H_{\dot{\alpha}}\Gamma^{45} \Sigma^{\dot{\alpha}},
    \qquad \delta' F_{\dot{\alpha}} = \frac{i}{2r} \overline{\Lambda}_{\dot{\alpha}} \Gamma^{45} \Theta,
\end{equation}
where the spinors $\Sigma^{\dot{\alpha}}$ and $\Lambda^{\dot{\alpha}}$ are given by Eqs. \eq{susypa-53}
and \eq{susypa-h3}, respectively. It is straightforward (though a bit tedious) to verify that
the modified supersymmetry transformation $\Delta_\Sigma = \delta + \delta'$ leads to
a closed supersymmetry algebra on the fields in the hypermultiplet. For example, one can show that
\begin{eqnarray}\label{h-closusy}
   [ \Delta_{\Sigma_1}, \Delta_{\Sigma_2}] F_{\dot{\alpha}} &=&
i \big( \overline{\Sigma}_{2\dot{\alpha}}\Gamma^\mu \Sigma_1^{\dot{\beta}}
- \overline{\Sigma}_{1\dot{\alpha}}\Gamma^\mu \Sigma_2^{\dot{\beta}} \big) D_\mu F_{\dot{\beta}}
- \big( \overline{\Sigma}_{2\dot{\alpha}} \Sigma_1^{\dot{\beta}}
- \overline{\Sigma}_{1\dot{\alpha}} \Sigma_2^{\dot{\beta}} \big) [\sigma, F_{\dot{\beta}}] \xx
&& - \frac{2i}{r} \big( \overline{\Sigma}_{2\dot{\alpha}}\Gamma^{45} \Sigma_1^{\dot{\beta}}
- \overline{\Sigma}_{1\dot{\alpha}}\Gamma^{45} \Sigma_2^{\dot{\beta}} \big) F_{\dot{\beta}}.
\end{eqnarray}
Note that the adjoint scalar field $\sigma$ in Eq. \eq{h-closusy} can be absorbed into
a local gauge transformation parameter \cite{hst-5dsusy}.
Since the supersymmetry transformations are now generated by the Killing spinor $\epsilon$
obeying Eq. \eq{kill-spinor}, there are extra contributions from the derivative
of the Killing spinor given by
\begin{eqnarray} \label{5h-svar}
\delta S_{5H} &=& - \frac{1}{g_5} \int d \upsilon \Big(
i \big( \overline{\Theta} \Gamma^M \Gamma^N \nabla_M \Sigma^{\dot{\alpha}} D_N H_{\dot{\alpha}} -
\nabla_N \overline{\Sigma}_{\dot{\alpha}} \Gamma^M \Gamma^N \Theta D_M \overline{H}^{\dot{\alpha}} \big) \xx
&& \qquad  - \nabla_M \overline{\Sigma}_{\dot{\alpha}} \Gamma^M \Theta [\sigma, \overline{H}^{\dot{\alpha}}]
+ \overline{\Theta} \Gamma^M \nabla_M \Sigma^{\dot{\alpha}} [\sigma, H_{\dot{\alpha}}] \xx
&& \qquad + \big( \nabla_M \overline{\Sigma}_{\dot{\beta}} \Gamma^M \Psi^{\dot{\alpha}}
+ \overline{\Psi}_{\dot{\beta}} \Gamma^M \nabla_M \Sigma^{\dot{\alpha}} \big)
[\overline{H}^{\dot{\beta}}, H_{\dot{\alpha}}] \Big),
\end{eqnarray}
and the modified transformations introduced in Eqs. \eq{mod-susy} and \eq{add-hsusy}.
They are combined to get the total variation of the action \eq{hyper-action} generated
by the supersymmetry transformation $\Delta_\epsilon = \delta + \delta'$
and denoted by $\Delta_\epsilon S_{5H} = \delta S_{5H} + \delta'  S_{5H}$,
which turns out to be non-vanishing.
Therefore, as in the vector multiplet, it is necessary to add a compensating action given by
\begin{equation}\label{hyper-mass}
     S_{5H}^M = \frac{1}{r} \int d\upsilon \Big( \frac{i}{2} \overline{\Theta}\Gamma^{45} \Theta
     + \frac{1}{g_5^2 r} \overline{H}^{\dot{\alpha}} H_{\dot{\alpha}} \Big).
\end{equation}
Then the total action is invariant under the supersymmetry transformations, i.e.,
$\Delta_\epsilon (S_{5H} + S^M_{5H}) = 0$.

\section{Localization of five-dimensional NC $U(1)$ gauge theory}

In this section we will compute the partition function of five-dimensional $\mathcal{N}=2$ supersymmetric
NC $U(1)$ gauge theory on $\mathbb{S}^3 \times \mathbb{R}^2_{\theta}$ by using the localization method.
For the localization of five-dimensional quantum field theories, see Refs. \cite{local-qft15,local-qft16}
and references therein.
As we pointed out in footnote \ref{w-sign}, the adjoint scalar field $\sigma$ in the vector multiplet has
a wrong sign. In order to define the path integral properly, it needs to be analytically continued by
replacing $\sigma$ by $i\sigma$.

To carry out the localization procedure, we need to identify the Grassmann-odd symmetry,
denoted by $\delta_Q$. It is required that $\delta_Q$ is a symmetry of path integral (i.e. $\delta_Q$
is not anomalous and preserves the action) and obey $\delta_Q^2 = \mathcal{L}_B$ where
$\mathcal{L}_B$ is a Grassmann-even symmetry that could be a combination of Lorentz, R-symmetry, and
gauge transformations. We apply the same twisting procedure as \cite{witten-tft} by considering
a global symmetry group $H = SU(2)_1 \times SU(2)_2 \times SU(2)_R$ where $SU(2)_1 \times SU(2)_2$ is
the rotational symmetry on $\mathbb{S}^3$ and $SU(2)_R$ is the global symmetry of the $\mathcal{N}=2$
theory. We embed the Lorentz group $K$ into $H$ as $K = SU(2)_1 \times SU(2)'_2$ where
$SU(2)'_2$ is a diagonal subgroup of $SU(2)_2 \times SU(2)_R$. After the twisting,
we get a scalar supercharge $Q$ which is thus Lorentz invariant (in the $K$ sense).
It is enough to have one scalar supercharge $Q$ for localization and $Q$ is regarded as a BRST operator.

\subsection{Localization of vector multiplet}

According to our twisting, we will define the BRST transformation $\delta_Q \equiv \epsilon Q$ by
setting $\overline{\epsilon}$ to zero and replacing the Grassmann-odd parameter $\epsilon$
by a Grassmann-even parameter in the supersymmetric transformations \eq{susytr-3d}.
Then $\delta_Q$ is anticommuting with twisted spinors although all fields have integer spins with respect
to $K$. The corresponding BRST transformations for the vector multiplet are then given by
\begin{eqnarray}\label{twvsusy-5d}
&& \delta_Q A_\mu = i g_5 \overline{\lambda} \gamma_\mu \epsilon, \quad
\delta_Q \sigma = - g_5 \overline{\lambda} \epsilon, \quad \delta_Q \phi_z = 0,
\quad \delta_Q \phi_{\overline{z}} = - g_5 \overline{\psi} \epsilon, \xx
&& \delta_Q \lambda = \frac{1}{2} \Big( \frac{1}{2g_5} F_{\mu\nu} \gamma^{\mu\nu} + \frac{i}{g_5}
\gamma^\mu D_\mu \sigma  - i D \Big) \epsilon, \quad \delta_Q \overline{\lambda} = 0, \xx
&& \delta_Q \psi = - \frac{i}{g_5} D_\mu \phi_z \gamma^\mu \epsilon + \frac{i}{g_5} [\phi_z, \sigma] \epsilon,
\quad \delta_Q \overline{\psi} = i \overline{F} \epsilon^T C_3, \\
&& \delta_Q F = - \epsilon^T C_3 \big( \gamma^\mu D_\mu \psi - [\sigma, \psi] - 2 [\phi_z, \lambda]
+ \frac{i}{2r} \psi \big), \quad  \delta_Q \overline{F} = 0, \xx
&& \delta_Q D =  - D_\mu \overline{\lambda} \gamma^\mu \epsilon
- [\sigma, \overline{\lambda}] \epsilon + \frac{i}{2r} \overline{\lambda}\epsilon. \nonumber
\end{eqnarray}
It is straightforward to check that the above BRST transformations are nilpotent, i.e.,
$\delta^2_Q = 0$.\footnote{It may be useful to use the Fierz identity: $\delta_{\alpha\gamma} \delta_{\delta\beta}
= \frac{1}{2} \delta_{\alpha\beta} \delta_{\delta\gamma}
+ \frac{1}{2} (\gamma_\mu)_{\alpha\beta} (\gamma^\mu)_{\delta\gamma}$.}
The BRST invariant action on $\mathbb{S}^3 \times \mathbb{R}^2_{\theta}$ is given by
\begin{eqnarray}\label{tw-vaction-5d}
    S^{(\mathrm{inv})}_{5V} &=& \int d\upsilon \left[ \frac{1}{g_5^2} \Big( \frac{1}{4} F_{\mu\nu} F^{\mu\nu}
    + \frac{1}{2} D_\mu \sigma D^\mu \sigma + 2 D_\mu \phi_{\overline{z}} D^\mu \phi_z
    - 2 [\phi_{\overline{z}}, \sigma] [\phi_z, \sigma] \Big) \right. \xx
    && \quad - \frac{1}{2} D \big( D + \frac{4}{g_5} [\phi_{\overline{z}}, \phi_z] \big)
    - 2 \overline{F} F - 2i \Big( \overline{\lambda} \gamma^\mu D_\mu \lambda
    - \overline{\psi} \gamma^\mu D_\mu \psi + 2 \big( \overline{\psi} [\phi_z, \lambda]
    - \overline{\lambda} [\phi_{\overline{z}}, \psi] \big) \xx
    && \left. \quad + \overline{\lambda} [\sigma, \lambda] + \overline{\psi} [\sigma, \psi] \Big)
    - \frac{1}{r} \Big( \overline{\lambda} \lambda + \overline{\psi} \psi - \frac{1}{g_5^2 r} \sigma^2
    - \frac{i}{g_5} \sigma \big(  D - \frac{4}{g_5} [\phi_{\overline{z}}, \phi_z]  \big) \Big) \right] \xx
    && \quad - \frac{1}{2 g_5^2} \int \big( A \wedge F + \frac{i}{3} A \wedge A \wedge A \big)
    \wedge \varpi.
\end{eqnarray}

We deform the action \eq{tw-vaction-5d} by adding a BRST $Q$-exact term
\begin{equation}\label{brst-term}
    S_{5V}^Q =  2  \int d\upsilon \delta_Q \Big[ (\delta_Q \lambda)^\dagger \lambda
    + (\delta_Q \psi)^\dagger \psi + \overline{\psi} (\delta_Q \overline{\psi})^\dagger \Big]
\end{equation}
such that the total classical action is given by
\begin{equation}\label{tv-action}
    \widetilde{S}_{5V} \equiv  S^{(\mathrm{inv})}_{5V} + t S_{5V}^Q
\end{equation}
where $t$ is a non-negative real parameter.
The explicit form for the $Q$-exact Lagrangian is given by
\begin{eqnarray}\label{q-action-5d}
    \mathcal{L}_{5V}^Q &=& \frac{1}{g_5^2} \Big( \frac{1}{4} F_{\mu\nu} F^{\mu\nu}
    + \frac{1}{2} D_\mu \sigma D^\mu \sigma + 2 D_\mu \phi_{\overline{z}} D^\mu \phi_z
    - 2 [\phi_{\overline{z}}, \sigma] [\phi_z, \sigma] \Big) - \frac{1}{2} D^2 - 2 \overline{F} F \xx
    && +  \frac{2k_\mu}{g_5^2} \Big( i \varepsilon^{\mu\nu\rho} D_\nu \phi_{\overline{z}} D_\rho \phi_z
    - D^\mu \phi_{\overline{z}} [\phi_z, \sigma] + [\phi_{\overline{z}}, \sigma] D^\mu \phi_z \Big) \xx
&&- 2i \Big( \overline{\lambda} \gamma^\mu D_\mu \lambda
    - \overline{\psi} \gamma^\mu D_\mu \psi + \overline{\psi} [\phi_z, (1 - k_\mu \gamma^\mu) \lambda]
    - 2 \overline{\lambda} [\phi_{\overline{z}}, \psi]
    + \overline{\lambda} [\sigma, \lambda] + \overline{\psi} [\sigma, \psi] \Big) \xx
    && + \frac{1}{r} \big( \overline{\lambda} \lambda + \overline{\psi} \psi
    + 2 k_\mu \overline{\psi} \gamma^\mu \psi \big),
\end{eqnarray}
where $k_\mu = \overline{\epsilon} \gamma_\mu \epsilon$ and $\overline{\epsilon} \epsilon = 1$.
Note that the second line of $\mathcal{L}_{5V}^Q$ can be recast as
\begin{equation}\label{bps-top}
    \frac{2k_\mu}{g_5^2} \Big( \frac{1}{2} \varepsilon^{\mu\nu\rho} F_{\nu\rho}
    + D^\mu \sigma \Big) [\phi_{\overline{z}}, \phi_z]
   + \frac{4ik_\mu}{g_5^2 r} \phi_{\overline{z}} D^\mu \phi_z
\end{equation}
up to total derivatives.

Since the Lagrangian \eq{q-action-5d} is BRST-exact, the modified action \eq{tv-action} with a parameter $t$
leads to the same partition function as the undeformed one as was explained in Eq. \eq{q-exact}.
To be precise, the partition function $Z(t)$ for the modified action is $t$-independent, i.e.,
$\frac{d Z(t)}{dt} = 0$. Therefore we can calculate the partition function in the large $t$ limit.
In this limit, especially $ t \to \infty$, the fixed point is given by a solution obeying
\begin{equation}\label{q-fixed}
\delta_Q \lambda = 0, \qquad \delta_Q \psi = 0, \qquad \delta_Q \overline{\psi} = 0
\end{equation}
and fermions $=0$. Since the deformation term \eq{q-action-5d} is positive semi-definite,
the solution of Eq. \eq{q-fixed} constitutes the localization locus $\mathfrak{F}_Q$ given by
\begin{equation}\label{loc-locus}
    \sigma = \sigma(z, \overline{z}), \qquad \phi_z = \phi_z (z, \overline{z}), \qquad
    \phi_{\overline{z}} = \big( \phi_{z} (z, \overline{z}) \big)^\dagger,
    \qquad A_\mu = D = F = \overline{F} =0,
\end{equation}
and obey the condition $[\phi_z, \sigma] = i D_z \sigma = 0$. Then the classical action
at the locus $\mathfrak{F}_Q$ is given by\footnote{If one could integrate out the scalar field $\sigma$,
the fixed action \eq{locus-action} would give rise to the action of two-dimensional NC $U(1)$ gauge theory.
However the scalar field must obey the condition $[\phi_z, \sigma] = 0$ and thus it is not
allowed to perform the Gaussian integration over the whole functional space of the $\sigma$ \cite{fukama13}.}
\begin{equation}\label{locus-action}
    \widetilde{S}_{5V}|_{\mathfrak{F}_Q} = \frac{1}{2 r^2 g_2^2} \int d^2 y
    \sigma \big( \sigma - 4ir [\phi_{\overline{z}}, \phi_z] \big),
\end{equation}
where $g_2 = \frac{g_5}{2 \pi r^{3/2}}$ is a two-dimensional gauge coupling constant.
After the matrix representation \eq{matrix-rep}, the classical action is mapped to the zero-dimensional
matrix model
\begin{equation}\label{matrix-locus}
    \widetilde{S}_{5V}|_{\mathfrak{F}_Q} = \frac{1}{2 r^2 g^2} \mathrm{Tr} \sigma \big( \sigma
    - 4ir [\phi^\dagger, \phi] \big),
\end{equation}
where $g = \frac{g_2}{\sqrt{2 \pi \alpha'}}$ is a coupling constant of the matrix model.
Recall that the matrix $\sigma$ must be subject to the condition $[\phi, \sigma] = 0$.

Now we compute the one-loop determinant coming from quadratic fluctuations of the fields about
the fixed points in \eq{loc-locus}. For that purpose, the gauge-fixing procedure is also necessary
for the computation of the path integral. We take the usual gauge-fixing term given by
\begin{equation}\label{gauge-fixing}
    \mathcal{L}_{5V}^{FP} = \overline{c} D_\mu D^\mu c + b D^\mu A_\mu.
\end{equation}
There remains the residual gauge symmetry that acts on
\begin{equation}\label{res-gauge}
    \sigma \to \sigma - i [\omega(z, \overline{z}), \sigma], \qquad
    \phi_z \to \phi_z - i [\omega(z, \overline{z}), \phi_z],
\end{equation}
where the gauge transformation parameter $\omega(z, \overline{z})$ is constant along the $\mathbb{S}^3$.
Thus the gauge symmetry is the redundancy of the background $\sigma$ and $\phi_z$, but not of the fluctuations.
Using this freedom, we can put the background as\footnote{Note that the action \eq{locus-action}
on-shell is in general divergent, which is the reason why the cyclic property \eq{cyc-int}
was not applied to the second term.
In order to regularize the on-shell action, one may replace the commutator $[\phi_{\overline{z}}, \phi_z]$ by $[\phi_{\overline{z}}, \phi_z] - \frac{1}{2\alpha'}$. Or one may relax the reality condition of the scalar
field $\sigma$ and complexify it. Then the solution to $[\phi_z, \sigma] = 0$ is enough to be $\sigma = \sigma (\overline{z})$, an anti-holomorphic function.}
\begin{equation}\label{gfix-back}
    \phi_z = - \frac{i}{2 \alpha'} \overline{z}, \qquad
    \sigma = \sigma_0 = \mathrm{constant}.
\end{equation}
For this gauge-fixing of the background fields, we will add another gauge-fixing term \cite{fukama13} given by
\begin{equation}\label{back-fixing}
    \mathcal{L}_{B}^{FP} = \overline{c} \partial_{\overline{z}} \partial_z c
    -2\alpha' b \big( [z, \phi_z] + i \big).
\end{equation}
The path integral of the ghost fields gives the one-loop determinant
\begin{equation}\label{back-det}
    \det(-r^2 \partial_{\overline{z}} \partial_z).
\end{equation}

Since we are interested in the large $t$ limit and perform the path integral over the fluctuations around
the fixed points defined by Eq. \eq{gfix-back}, let us expand the fields $\Phi$ about the saddle point
configuration in Eq. \eq{gfix-back} and rescale the fluctuation fields as
\begin{equation}\label{exp-saddle}
    \Phi = \Phi_0 + \frac{1}{\sqrt{t}} \delta \Phi
\end{equation}
and take the limit $t \to \infty$. $\Phi_0$ denotes the background at the fixed points in which
$\Phi_0 = - \frac{i}{2 \alpha'} \overline{z}, \; \frac{i}{2 \alpha'} z$, and $\sigma_0$
for $\phi_z, \; \phi_{\overline{z}}$, and $\sigma$, respectively, and $\Phi_0 = 0$, otherwise.
Taking $t$ to be large then allows us to keep only the quadratic terms in the Lagrangian \eq{q-action-5d}:
\begin{eqnarray}\label{quad-action-5d}
&& t \cdot \mathcal{L}_{5V}^Q = \frac{1}{g_5^2} \Big( \frac{1}{4} F_{\mu\nu} F^{\mu\nu}
    + \frac{1}{2} \partial^\mu \sigma \partial_\mu \sigma + 2 \partial_{\overline{z}} \sigma \partial_{z} \sigma
    + 2 (\partial^\mu \phi_{\overline{z}} - \partial_{\overline{z}} A^\mu)
    (\partial_\mu \phi_{z} - \partial_{z} A_\mu) \Big) \\
&& \hspace{1.5cm} + \frac{2ik_\mu}{g_5^2} \Big( \varepsilon^{\mu\nu\rho} (\partial_\nu \phi_{\overline{z}} - \partial_{\overline{z}} A_\nu)(\partial_\rho \phi_{z} - \partial_{z} A_\rho)
- (\partial^\mu \phi_{\overline{z}} - \partial_{\overline{z}} A^\mu) \partial_{z} \sigma
  + (\partial^\mu \phi_{z} - \partial_{z} A^\mu) \partial_{\overline{z}} \sigma \Big) \xx
&& \hspace{1.5cm} - 2\overline{\lambda} \Big( i \gamma^\mu \nabla_\mu - \frac{1}{2r} \Big) \lambda
    + 2 \overline{\psi} \Big( \gamma^\mu  \big( i \nabla_\mu + \frac{k_\mu}{r} \big) + \frac{1}{2r} \Big) \psi
    + 2 \overline{\psi} (1 - \gamma^\mu k_\mu) \partial_z \lambda
    + 4 \partial_{\overline{z}} \overline{\lambda} \psi, \nonumber
\end{eqnarray}
where all fields indicate the fluctuations $\delta \Phi$ in \eq{exp-saddle}.
Although the multiplication between fields has originally been defined by the star product \eq{star-prod},
we can ignore the star product for the quadratic terms
since all nontrivial star products in this case are total derivatives and thus can be dropped.
Hence we will regard the fluctuations  in the Lagrangian \eq{quad-action-5d}
as fields on $\mathbb{S}^3 \times \mathbb{C}$.
Integral over the auxiliary fields $D, \; F$ and $\overline{F}$
have already been performed, which contributes trivial constant terms to the partition function.
We will ignore an overall constant of the partition function.

In order to calculate the one-loop determinant of $U(1)$ gauge fields, we first proceed with separating
the gauge field into a divergenceless and pure divergent part:
\begin{equation}\label{decomp-u1}
    A_\mu = B_\mu - \partial_\mu \phi
\end{equation}
where $D^\mu B_\mu = 0$. Then the delta function constraint from Eq. \eq{gauge-fixing} becomes
$\delta( \square_0 \phi)$. One can see that the longitudinal mode in \eq{decomp-u1}
can be absorbed into the complex scalar field with the form
\begin{equation}\label{shift-long}
    \varphi_z = \phi_z + \partial_z \phi, \qquad \varphi_{\overline{z}}
    = \phi_{\overline{z}} + \partial_{\overline{z}} \phi,
\end{equation}
since $D_\mu \phi_z = \frac{1}{\sqrt{t}} \big(\partial_\mu \phi_z - \partial_z A_\mu \big)
+ \mathcal{O}(1/t)$ is invariant under the specified $U(1)$ gauge transformation and
so $D_\mu \phi_z = \frac{1}{\sqrt{t}} \big(\partial_\mu \varphi_z - \partial_z B_\mu \big)
+ \mathcal{O}(1/t)$.
Thus the longitudinal mode $A_\mu^L = - \partial_\mu \phi$ appears only in the gauge-fixing term
in Eq. \eq{gauge-fixing} and so we can integrate over $\phi$ using the delta function,
which picks up a Jacobian factor of $\det \square_0^{-1/2}$ \cite{kwy09}.\footnote{Note that
$\mathcal{D}A_\mu^L = \det \square_0^{1/2} \mathcal{D} \phi$ in the path integral measure.}
The integral of the ghosts in Eq. \eq{gauge-fixing}
contributes a factor of $\det \square_0$. Therefore the one-loop determinant from $b$ and $\phi$
as well as the ghosts $\overline{c}$ and $c$ contributes a factor
\begin{equation}\label{ghost-det}
\det \square_0^{\frac{1}{2}}.
\end{equation}

To evaluate the path integral of the bosonic fields, it is more convenient to use
a differential form notation. For that purpose, let us introduce the dreibeins $e^m = e^m_\mu dx^\mu
\in \Gamma(T^* \mathbb{S}^3)\; (m=1,2,3)$
on $\mathbb{S}^3$ obeying the structure equation $de^m = \frac{1}{r} \varepsilon^{mnp} e^n \wedge e^p$.
See appendix D for the differential geometry on $\mathbb{S}^3$.
The dual frame basis is given by $l_m = e_m^\mu \partial_\mu \in \Gamma(T \mathbb{S}^3)$ that
is left-invariant vector fields on $\mathbb{S}^3$ satisfying
the following commutation relation
\begin{equation}\label{comm-rel-vec}
    [l_m, l_n] = - \frac{2}{r} {\varepsilon_{mn}}^{p} l_p.
\end{equation}
Then the exterior differential operator $d: \Omega^p(\mathbb{S}^3) \to \Omega^{p+1}(\mathbb{S}^3)$
can be written as
\begin{equation}\label{ext-d}
    d = dx^\mu \partial_\mu = e^m l_m.
\end{equation}
Let us also define the three-dimensional Hodge duality $*: \Omega^p(\mathbb{S}^3)
\to \Omega^{3-p}(\mathbb{S}^3)$ as
\begin{equation} \label{hodge}
    *e^m = \frac{1}{2} \varepsilon^{mnp} e^n \wedge e^p, \qquad
    *(e^m \wedge e^n) = \varepsilon^{mnp} e^p
\end{equation}
and two operators acting on the dreibein by
\begin{equation}\label{int-s}
    \iota_{k} e^m = e^m_\mu k^\mu = k^m, \qquad S^m e^n = i \varepsilon^{mnp} e^p,
\end{equation}
where $k^m = \overline{\epsilon} \gamma^m \epsilon$ is the Killing vector field.
Using this form notation, the quadratic bosonic action on $\mathbb{S}^3$ can be written as
\begin{eqnarray}\label{bform-action}
&& t \cdot S_{VB}^Q = \frac{1}{g_5^2} \int d^2 y \int_{\mathbb{S}^3} \Big( \frac{1}{2} dB \wedge *dB +
 \frac{1}{2} d\sigma \wedge *d\sigma + 2 \partial_{\overline{z}} \sigma \partial_{z} \sigma *1 \xx
&& \hspace{3.0cm} + 2 (d \varphi_{\overline{z}} - \partial_{\overline{z}} B) \wedge
    * \big(1 - k \cdot S \big) (d \varphi_{z} - \partial_{z} B)  \xx
&& \hspace{3.0cm} + 2i \big( \partial_{\overline{z}} \sigma \iota_k (d \varphi_{z} - \partial_{z} B)
- \partial_{z} \sigma \iota_k (d \varphi_{\overline{z}} - \partial_{\overline{z}} B) \big) *1 \Big) \\
\label{bform-sq}
&& \hspace{1cm} = \frac{1}{g_5^2} \int d^2 y \int_{\mathbb{S}^3} \Big( \frac{1}{2} dB \wedge *dB +
 \frac{1}{2} d\sigma \wedge *d\sigma  \xx
&& \hspace{1.5cm} + 2 (d \varphi_{\overline{z}} - \partial_{\overline{z}} B
+ i k_m e^m \partial_{\overline{z}} \sigma) \wedge
    * \big(1 - k \cdot S \big) (d \varphi_{z} - \partial_{z} B - i k_m e^m \partial_{z} \sigma) \Big),
\end{eqnarray}
where $*1 = \sqrt{g} d^3 x$ is the volume form on $\mathbb{S}^3$ and $B = e^m B_m$ is a one-form connection
obeying the Lorenz gauge condition $d^\dagger B \equiv *d*B = l_m B_m = 0$.
Note that the exterior derivative \eq{ext-d} acts only on $\mathbb{S}^3$.
It is useful to use the fact that $dB \wedge *dB = d(B \wedge *dB) + B \wedge *(*d *dB)$
and the last line in Eq. \eq{bform-action} may be reduced to
$2i \big( \partial_{\overline{z}} \sigma \iota_k d \varphi_{z} - \partial_{z} \sigma \iota_k
d \varphi_{\overline{z}} \big)$ after integration by parts but it is convenient to keep the original form
to utilize the last expression in Eq. \eq{bform-sq}.

We expand a scalar field $\Phi(x, y)$ in the basis of the scalar spherical harmonics
$S_j^{\vec{m}} (x) \; (j=0, \frac{1}{2}, 1, \frac{3}{2}, \cdots; -j \leq \vec{m}= (m_1, m_2) \leq j)$
on $\mathbb{S}^3$ as
\begin{equation}\label{harm-exp}
  \Phi(x, y) = \sum_{j=0}^\infty \sum_{\vec{m}= -j}^j \Phi_j^{\vec{m}} (y)  S_j^{\vec{m}} (x).
\end{equation}
The harmonics $S_j^{\vec{m}} (x)$ belong to the $(j, j)$ representation of $SO(4) = SU(2)_L \times SU(2)_R$.
Let us use the ket notation for the spherical harmonics
\begin{equation}\label{ket-sharm}
    |j, m, m' \rangle \equiv S_j^{m, m'}.
\end{equation}
They obey the following properties:
\begin{eqnarray} \label{prop-shar}
  && -d^\dagger dS_j^{m, m'} = - *d*d S_j^{m, m'} = \frac{4 j(j+1)}{r^2} S_j^{m, m'}, \xx
  && \big( S_j^{m, m'} \big)^\dagger = (-1)^{m+m'} S_j^{-m, -m'}, \\
  && \int_{\mathbb{S}^3} (S_{k}^{n, n'})^\dagger S_j^{m, m'} \; *1
  = \delta_{kj} \delta^{nm} \delta^{n' m'}. \nonumber
\end{eqnarray}
In order to calculate the determinant of the operators in the quadratic Lagrangian \eq{bform-action},
we will take the same strategy as section 3.5 in Ref. \cite{kwy09}.
We identify $l_m = e^\mu_m \partial_\mu$ with $\frac{2i}{r} L_m$, where $L_m$ are operators
in the $su(2)$ Lie algebra. We also choose the Killing vector $k_\mu$ as $\delta_\mu^3$ \cite{fukama13,kwy09}
and thus $k = k_\mu \partial^\mu = \frac{2i}{r} L_3$.
Since we will use the $SU(2)_L$-invariant frame on $\mathbb{S}^3$ (when thinking of $\mathbb{S}^3$
as $SU(2)$ and letting it act on itself), the angular momentum operators $L_m$ are acting on the $SU(2)_L$
index $m$ in Eq. \eq{ket-sharm} and it is given by
\begin{equation}\label{s-harmonics}
\begin{array}{l}
L_\pm |j, m, m' \rangle = \sqrt{(j \mp m ) (j \pm m + 1 )}
    |j, m \pm 1, m' \rangle, \\
    L_3 |j, m, m' \rangle = m |j, m, m' \rangle, \qquad  - j \leq m \leq j,
\end{array}
\end{equation}
where $L_\pm = L_1 \pm i L_2$.

In order to have a similar expansion for the one-form $B = e^m B_m$, we need the other set of basis,
which can be constructed by considering a tensor product of the scalar spherical harmonics with
the dreibeins. In particular, the dreibeins $e^m$  on $\mathbb{S}^3$ are taken as the eigenstate of
the spin operators $\overrightarrow{S} \cdot \overrightarrow{S}$ and $S_3$ where
$(S^m)_{ln} = i \varepsilon^{lmn}$ is the spin-1 representation of $SU(2)$.
In terms of the spin-1 basis $|s=1, s_z\rangle \; (s_z = -1, 0, 1)$, they are given by
\begin{equation}\label{spin-1}
    e^\pm \equiv  \mp \frac{1}{\sqrt{2}} (e^1 \pm i e^2) = |s=1, \pm1 \rangle,
    \qquad e^0 \equiv e^3 = |s=1, 0 \rangle,
\end{equation}
and satisfy
\begin{eqnarray} \label{spin-1state}
  && S^\pm e^\mp =\sqrt{2} e^0, \qquad S^\pm e^\pm = 0, \qquad S^\pm e^0 = \sqrt{2} e^\pm, \xx
  && S_3 e^\pm = \pm e^\pm, \qquad S_3 e^0 = 0,
\end{eqnarray}
where $S^\pm = S^1 \pm i S^2$. The spin-1 basis $|s=1, s_z\rangle$ transforms as
the $(1,0)$ representation of $SU(2)_L \times SU(2)_R$.
Therefore the tensor product may be decomposed into the following irreducible representations,
\begin{equation}\label{irred-rep}
    (j,j) \otimes (1,0) = (j+1, j) \oplus (j-1, j) \oplus (j,j),
\end{equation}
where the last representation is a gradient of the scalar spherical harmonics.
A general vector field on $\mathbb{S}^3$ is then expanded as a combination of gradients of
the scalar spherical harmonics plus a set of vector spherical harmonics $V_{j\pm}^{\vec{m}} (x)$
which are in the representation $(j \pm 1, j)$ of $SU(2)_L \times SU(2)_R$ \cite{n=4s3,marino-rev}.
The vector spherical harmonics in the decomposition \eq{irred-rep} are then given by
\begin{equation}\label{vec-harmonic}
    V_{k,m;j, m_2} = \sum_{m_1=-j}^j \sum_{s=-1}^{+1} \langle j, m_1; 1, s|k,m \rangle\rangle
    S_j^{m_1,m_2} e^s,
\end{equation}
where $\langle j, m_1; 1, s|k,m \rangle\rangle$ are the Clebsch-Gordan coefficients of the spin-$j$
representation and the spin-1 representation of the $SU(2)$ group into the spin-$k$ representation,
with $k=(j- 1), \, j, \, (j+1)$. See appendix E for the Clebsch-Gordan coefficients.
Let us denote the vector spherical harmonics in \eq{vec-harmonic} by
$V_{k,j}^{\vec{m}} \equiv  V_{k,m;j, m_2}$. They have the following properties
\begin{equation} \label{vec-harm-ev}
\begin{array}{lll}
   * d V_{j+1,j}^{\vec{m}} = \frac{2(j+1)}{r} V_{j+1,j}^{\vec{m}},
   & d^\dagger  V_{j+1,j}^{\vec{m}} = 0, & j= 0, \frac{1}{2}, 1, \cdots, \\
  * d V_{j-1,j}^{\vec{m}} = - \frac{2j}{r} V_{j-1,j}^{\vec{m}}, & d^\dagger V_{j-1,j}^{\vec{m}} = 0,
  & j= 1, \frac{3}{2}, 2, \cdots, \\
  * d  V_{j,j}^{\vec{m}} = 0, &  V_{j,m;j,m_2} = -\frac{i}{2} \frac{r}{\sqrt{j(j+1)}} dS_j^{m_1, m_2},
  &  j= \frac{1}{2}, 1, \frac{3}{2}, \cdots,
\end{array}
\end{equation}
and form an orthonormal basis
\begin{equation}\label{vec-onbasis}
    \int_{\mathbb{S}^3}  {V_{k',j'}^{\vec{m}'}}^\dagger \wedge * V_{k,j}^{\vec{m}}
    = \delta_{k'k} \delta_{j' j} \delta^{\vec{m}' \vec{m}},
\end{equation}
where
\begin{equation}\label{vec-conj}
     {V_{k,j}^{\vec{m}}}^\dagger = \sum_{m_1=-j}^j \sum_{s=-1}^{+1} (-1)^{m_1 + m_2 + s}
     \langle \langle k,m | j, m_1; 1, s \rangle
    S_j^{-m_1, -m_2} e^{-s} = (-1)^{m+m_2} {V_{k,j}^{-\vec{m}}}.
\end{equation}
It is not difficult to derive the following formulae
\begin{eqnarray}\label{vec-formula1}
    && \int_{\mathbb{S}^3}  {S_{j'}^{\vec{m}'}}^\dagger \iota_k V_{k,j}^{\vec{m}} \; *1
    = \int_{\mathbb{S}^3}  {S_{j'}^{\vec{m}'}}^\dagger e^3 \wedge * V_{k,j}^{\vec{m}}
    = \delta_{j'j} \delta_{m_2' m_2} \langle j, m'_1; 1, 0|k,m \rangle\rangle, \xx
    && \hspace{3.2cm} = \delta_{j'j} \delta_{m_2' m_2} \delta_{m_1' m} \left\{
                                                     \begin{array}{ll}
                                                       \sqrt{\frac{(j+m+1)(j-m+1)}{(j+1)(2j+1)}},
& \hbox{for $k=j+1$;} \\
                                                       \frac{m}{\sqrt{j(j+1)}}, & \hbox{for $k=j$;} \\
                                                       - \sqrt{\frac{(j+m)(j-m)}{j(2j+1)}}, & \hbox{for $k=j-1$,}
                                                     \end{array}
                                                   \right. \\
    \label{vec-formula2}
    && \int_{\mathbb{S}^3}  {V_{k' j'}^{\vec{m}'}}^\dagger \wedge * \big[ (k \cdot S) V_{k,j}^{\vec{m}} \big]
    = \delta_{j'j} \delta_{m_2' m_2} \langle \langle k', m'|S_3| k,m \rangle\rangle.
\end{eqnarray}

We expand the transverse gauge field $B = B_+ + B_-$ in the basis \eq{vec-harmonic} as
\begin{eqnarray}\label{expan-b}
  && B_+ = \sum_{j = 0}^\infty \sum_{m=-(j+1)}^{j+1} \sum_{m' =-j}^j
  B^{\vec{m}}_{j+1, j} (z, \overline{z})  V_{j+1,j}^{\vec{m}} (x), \xx
  && B_- = \sum_{j = 1}^\infty \sum_{m=-(j-1)}^{j-1} \sum_{m' =-j}^j
  B^{\vec{m}}_{j-1, j}  (z, \overline{z}) V_{j-1,j}^{\vec{m}} (x),
\end{eqnarray}
where $B^{\vec{m}}_{k, j} (z, \overline{z}) = B_{k, m; j, m'} (z, \overline{z})$ are the transverse modes
depending on the coordinates on $\mathbb{C}$. Note that the longitudinal mode $A_L = - d \phi$ is expanded
in the basis $V_{j,j}^{\vec{m}} \propto dS_j^{\vec{m}}$ and decoupled from the physical fluctuations,
so we have already integrated it out in Eq. \eq{ghost-det}.
Since the action \eq{bform-action} is quadratic in fluctuations, it is straightforward to evaluate
the integral of fluctuation modes over $\mathbb{S}^3$.
The harmonic expansion of the bosonic action \eq{bform-action} can be
written as\footnote{\label{reality}It may be necessary
to keep track of the reality condition of the harmonic expansions such as Eqs. \eq{harm-exp} and \eq{expan-b}.
For example, for a real scalar field $\Phi(x,y)$, the reality condition in the harmonic expansion \eq{harm-exp}
is equivalent to the constraint $\big(\Phi_j^{m,m'}(x)\big)^\dagger = (-1)^{m+m'} \Phi_j^{-m, -m'}(x)$.
This kind of the reality condition has to be incorporated into the sum \eq{modeint-s3} and
the corresponding one-loop determinant. However, the reality condition may be easily implemented by
relaxing the condition and then taking the square-root of the final one-loop determinant.
We will take this strategy for simplicity since the one-loop determinant resulting from bosonic and
fermionic fluctuations will eventually be cancelled each other.}
\begin{equation}\label{modeint-s3}
     t \cdot S_{VB}^Q = \sum_{j=0}^\infty \sum_{\vec{m} = -j}^j \int d^2 y S_{B;j, \vec{m}}^Q (z, \overline{z}).
\end{equation}
Since the one-form gauge fields $B_\pm$ have a different allowed range for $j$ and $m$, special treatments
are needed when $m$ is close to $\pm j, \pm(j+1)$ and $j=0$.
The action $S_{B;j, \vec{m}}^Q$ of the modes for $j \geq 1, \; |m| \leq j-1, \; |m'| \leq j$, is given by
\begin{equation} \label{action1}
     \frac{(g_5 r)^2}{2} \cdot S_{B;j\geq 1, \vec{m}}^Q =
     \Xi_{j,\vec{m}}^\dagger \mathcal{M}_{j,m} \Xi_{j,\vec{m}}
\end{equation}
where $\Xi_{j,\vec{m}} = ( B^{\vec{m}}_{j-1, j}, B^{\vec{m}}_{j+1, j}, \varphi_{z, j}^{\vec{m}}, \varphi^{\vec{m}}_{\overline{z}, j}, \sigma^{\vec{m}}_{j})^T$ and
\begin{equation} \label{boson-matrix}
\begin{array}{c}
\mathcal{M}_{j,m} = \left(
                \begin{array}{cc}
                  A & B \\
                  C & D \\
                \end{array}
              \right),
\end{array}
\end{equation}
\begin{eqnarray} \label{m-abcd}
 && A = \left(
              \begin{array}{cc}
                j^2 - \frac{(j+m)}{j} r^2 \partial_{\overline{z}} \partial_z &  0 \\
               0  & (j+1)^2 - \frac{(j+1-m)}{j+1} r^2 \partial_{\overline{z}} \partial_z  \\
              \end{array}
            \right), \xx
 && B = ir \left(
              \begin{array}{ccc}
               (j+1)\sqrt{\frac{(j-m)(j+m)}{j(2j+1)}} \partial_{\overline{z}}
               & - (j+1)\sqrt{\frac{(j-m)(j+m)}{j(2j+1)}} \partial_{z}  &
               \sqrt{\frac{(j-m)(j+m)}{j(2j+1)}} r \partial_{\overline{z}} \partial_{z}  \\
                j \sqrt{\frac{(j-m+1)(j+m+1)}{(j+1)(2j+1)}} \partial_{\overline{z}}
                & - j \sqrt{\frac{(j-m+1)(j+m+1)}{(j+1)(2j+1)}} \partial_{z} &
                - \sqrt{\frac{(j-m+1)(j+m+1)}{(j+1)(2j+1)}} r \partial_{\overline{z}} \partial_{z} \\
              \end{array}
            \right), \xx
 && C =  ir \left(
              \begin{array}{cc}
               (j+1)\sqrt{\frac{(j-m)(j+m)}{j(2j+1)}} \partial_{z}
               & j \sqrt{\frac{(j-m+1)(j+m+1)}{(j+1)(2j+1)}} \partial_{z} \\
               - (j+1)\sqrt{\frac{(j-m)(j+m)}{j(2j+1)}} \partial_{\overline{z}}
                & - j \sqrt{\frac{(j-m+1)(j+m+1)}{(j+1)(2j+1)}} \partial_{\overline{z}} \\
                - \sqrt{\frac{(j-m)(j+m)}{j(2j+1)}} r \partial_{\overline{z}} \partial_{z} &
                 \sqrt{\frac{(j-m+1)(j+m+1)}{(j+1)(2j+1)}} r \partial_{\overline{z}} \partial_{z} \\
              \end{array}
            \right), \xx
 && D =  \left(
            \begin{array}{ccc}
            2 \big( j(j+1) - m \big) & 0 & - m r \partial_z \\
             0 & 2 \big( j(j+1) + m \big) & m r \partial_{\overline{z}} \\
             m r \partial_{\overline{z}} & - m r \partial_z
              &  j (j+1) - r^2 \partial_{\overline{z}} \partial_z \\
              \end{array}
               \right).
\end{eqnarray}
Note that one can put $m=0$ in the matrix $A$ in Eq. \eq{m-abcd} since the action \eq{modeint-s3}
is summing over $\vec{m}=(m, m')$ and thus any odd term under $\vec{m} \to -\vec{m}$ does not contribute
to the sum. To calculate the one-loop determinant for these modes, it is convenient to
shift the complex scalar field as
\begin{eqnarray} \label{shift-scalar}
&& \varphi_{z, j}^{\vec{m}} \to \varphi_{z, j}^{\vec{m}}
-  \frac{ir}{2 \big(j(j+1) - m \big)} \partial_{z} \Big( - \frac{j+1}{j} B^{\vec{m}}_{j-1, j} + \frac{j}{j+1}
B^{\vec{m}}_{j+1, j} + im  \sigma^{\vec{m}}_{j} \Big), \xx
&& \varphi_{\overline{z}, j}^{\vec{m}}  \to \varphi_{\overline{z}, j}^{\vec{m}}
+ \frac{i r}{2 \big(j(j+1) + m \big)} \partial_{\overline{z}} \Big( - \frac{j+1}{j} B^{\vec{m}}_{j-1, j} +
\frac{j}{j+1} B^{\vec{m}}_{j+1, j} + im \sigma^{\vec{m}}_{j} \Big)
\end{eqnarray}
after the scale transformation of gauge fields
\begin{eqnarray}\label{scale-b}
&& B^{\vec{m}}_{j-1, j} \to - \frac{1}{j} \sqrt{\frac{j(2j+1)}{(j-m)(j+m)}} B^{\vec{m}}_{j-1, j}, \xx
&& B^{\vec{m}}_{j+1, j} \to \frac{1}{j+1} \sqrt{\frac{(j+1)(2j+1)}{(j-m+1)(j+m+1)}}
B^{\vec{m}}_{j+1, j}.
\end{eqnarray}
Note that the shifting form \eq{shift-scalar} was suggested in the last line of Eq. \eq{bform-sq}.
The above scale transformation is just for convenience and it will be recovered later through
the Jacobian factor in the path integral measure.
The action \eq{action1} can then be written as
\begin{eqnarray} \label{mode1}
&& \frac{(g_5 r)^2}{2} \cdot S_{B;j\geq 1, \vec{m}}^Q =
\big (B_{j+1, j}^{\vec{m}} (z, \overline{z}) \big)^\dagger
b_{j+1, j}^{m} B_{j+1, j}^{\vec{m}} (z, \overline{z})
+ \big (B_{j-1, j}^{\vec{m}} (z, \overline{z}) \big)^\dagger
b_{j-1, j}^{m} B_{j-1, j}^{\vec{m}} (z, \overline{z}) \xx
&& \hspace{3.3cm} + 2 K_j^m |\varphi_{z, j}^{\vec{m}} (z, \overline{z})|^2
+ 2 K_j^{-m} |\varphi_{\overline{z}, j}^{\vec{m}} (z, \overline{z})|^2
+ \big (\sigma_{j}^{\vec{m}} (z, \overline{z}) \big)^\dagger
c_{j}^{m} \sigma_{j}^{\vec{m}} (z, \overline{z}) \xx
&& \hspace{3.3cm} - \frac{r^2}{2 K_j^m} |- \frac{j+1}{j} \partial_{z} B^{\vec{m}}_{j-1, j}(z, \overline{z})
+ \frac{j}{j+1} \partial_{z} B^{\vec{m}}_{j+1, j}(z, \overline{z})|^2 \xx
&& \hspace{3.3cm} - \frac{r^2}{2 K_j^{-m}} | - \frac{j+1}{j} \partial_{\overline{z}}
B^{\vec{m}}_{j-1, j} (z, \overline{z})
+ \frac{j}{j+1} \partial_{\overline{z}} B^{\vec{m}}_{j+1, j}(z, \overline{z})|^2 \\
&& \qquad + \frac{i r^2}{2}  \big(\sigma_{j}^{\vec{m}} (z, \overline{z}) \big)^\dagger \Big(
 ( N_j^m - N_j^{-m}) \partial_{\overline{z}} \partial_z B_{j+1,j}^{\vec{m}} (z, \overline{z})
+ ( \widetilde{N}_j^m  - \widetilde{N}_j^{-m}) \partial_{\overline{z}} \partial_z
B_{j-1,j}^{\vec{m}} (z, \overline{z})\Big) \xx
&& \qquad - \frac{i r^2}{2} \Big( \big(\partial_{\overline{z}} \partial_z B_{j+1,j}^{\vec{m}}
(z, \overline{z}) \big)^\dagger
 ( N_j^m - N_j^{-m})  + \big(\partial_{\overline{z}} \partial_z
 B_{j-1,j}^{\vec{m}} (z, \overline{z}) \big)^\dagger
( \widetilde{N}_j^m - \widetilde{N}_j^{-m}) \Big) \sigma_{j}^{\vec{m}} (z, \overline{z}), \nonumber
\end{eqnarray}
where
\begin{equation}\label{bc-cofop}
\begin{array}{l}
  K_j^m = j(j+1) - m, \quad N_j^m = \frac{j-m}{K_j^m}, \quad \widetilde{N}_j^m = \frac{j+m+1}{K_j^m}, \\
  b_{j+1, j}^{m} = \frac{(j+1)(2j+1)}{(j-m+1)(j+m+1)} \Big(1
  - \frac{1}{(j+1)^2} r^2 \partial_{\overline{z}} \partial_z \Big),  \\
   b_{j-1, j}^{m} = \frac{j(2j+1)}{(j-m)(j+m)} \Big(1
   - \frac{1}{j^2} r^2 \partial_{\overline{z}} \partial_z \Big),  \\
   c_{j}^{m} = j(j+1) - \frac{1}{2} \Big(\frac{(j-m)(j+m+1)}{K_j^m} + \frac{(j+m)(j-m+1)}{K_j^{-m}} \Big)
r^2 \partial_{\overline{z}} \partial_z.
\end{array}
\end{equation}
The crossing terms of $B^{\vec{m}}_{j\pm1, j}$ and $\sigma^{\vec{m}}_{j}$
can be diagonalized by shifting the scalar field $\sigma_j^{\vec{m}}$ as
$$ \sigma_j^{\vec{m}} \to \sigma_j^{\vec{m}} - \frac{i r^2}{2} \frac{1}{c_j^m} \Big(
 ( N_j^m - N_j^{-m}) \partial_{\overline{z}} \partial_z B_{j+1,j}^{\vec{m}} (z, \overline{z})
+ ( \widetilde{N}_j^m - \widetilde{N}_j^{-m}) \partial_{\overline{z}} \partial_z
B_{j-1,j}^{\vec{m}} (z, \overline{z})\Big). $$

The modes for the bosonic fluctuations after the above shifting may be arranged into the following form
\begin{eqnarray} \label{mode2}
&& \frac{(g_5 r)^2}{2} \cdot S_{B;j\geq 1, \vec{m}}^Q = 2 K_j^m |\varphi_{z, j}^{\vec{m}} (z, \overline{z})|^2
+ 2 K_j^{-m} |\varphi_{\overline{z}, j}^{\vec{m}} (z, \overline{z})|^2
+ \big (\sigma_{j}^{\vec{m}} (z, \overline{z}) \big)^\dagger
c_{j}^{m} \sigma_{j}^{\vec{m}} (z, \overline{z}) \xx
&& \hspace{3.3cm} + \big((B^{\vec{m}}_{j+1, j})^\dagger, (B^{\vec{m}}_{j-1, j})^\dagger \big)
   \left(
     \begin{array}{cc}
       a & b \\
       c & d \\
     \end{array}
   \right) \left(
             \begin{array}{c}
               B^{\vec{m}}_{j+1, j} \\
               B^{\vec{m}}_{j-1, j} \\
             \end{array}
           \right),
\end{eqnarray}
where
\begin{eqnarray*}
&& a = \frac{(j+1)(2j+1)}{(j-m+1)(j+m+1)} - \frac{1}{2} \Big( \frac{N_j^m}{j+m+1}
+ \frac{N_j^{-m}}{j-m+1} \Big)
 r^2 \partial_{\overline{z}} \partial_z \\
&& \qquad - ( N_j^m - N_j^{-m}) r^2 \partial_{\overline{z}} \partial_z \frac{1}{4 c_j^m}
( N_j^m - N_j^{-m}) r^2 \partial_{\overline{z}} \partial_z \equiv a_j^m, \\
&& b=c = - \frac{1}{2} \Big( \frac{1}{K_j^m} + \frac{1}{K_j^{-m}} \Big) r^2 \partial_{\overline{z}} \partial_z
- ( N_j^m - N_j^{-m}) r^2 \partial_{\overline{z}} \partial_z \frac{1}{4 c_j^m}
( \widetilde{N}_j^m - \widetilde{N}_j^{-m}) r^2 \partial_{\overline{z}} \partial_z, \\
&& d = \frac{j(2j+1)}{(j-m)(j+m)} - \frac{1}{2} \Big( \frac{\widetilde{N}_j^m}{j-m}
+ \frac{\widetilde{N}_j^{-m}}{j+m} \Big)
 r^2 \partial_{\overline{z}} \partial_z \\
&& \qquad - ( \widetilde{N}_j^m - \widetilde{N}_j^{-m}) r^2 \partial_{\overline{z}}\partial_z \frac{1}{4 c_j^m}
( \widetilde{N}_j^m - \widetilde{N}_j^{-m}) r^2 \partial_{\overline{z}} \partial_z.
\end{eqnarray*}
Then the one-loop determinant from these modes yields\footnote{Using the determinant formula for a block matrix
$$\det \left(
           \begin{array}{cc}
             A & B \\
             C & D \\
           \end{array}
         \right) = \det A \; \det \big(D - C A^{-1} B \big) = \det D \; \det \big(A - B D^{-1} C \big), $$
         one may directly calculate the determinant of the matrix $\mathcal{M}_{j,m}$. We checked it yields the same result.}
\begin{eqnarray}\label{det-vboson1}
    && \prod_{j=1}^\infty \prod_{m=-(j-1)}^{j-1} \prod_{m'=-j}^j \det \mathcal{M}^{-1}_{j,m} \xx
    &=& \prod_{j=1}^\infty \prod_{m=-(j-1)}^{j-1} \prod_{m'=-j}^j \frac{1}{4\big( j^2 (j+1)^2 -m^2 \big)}
\frac{1}{\det c_{j}^{m} \det \mathcal{B}_j^m},
\end{eqnarray}
where $\det D$ is the functional determinant of the operator $D$ since the expansion coefficients
in Eqs. \eq{harm-exp} and \eq{expan-b} are regarded as two-dimensional fields on $\mathbb{R}^2 \cong \mathbb{C}$
and
\begin{eqnarray*}
   \det \mathcal{B}_j^m &=& \frac{j(j+1)(j-m)(j+m)(j-m+1)(j+m+1)}{(2j+1)^2} \det \left(
     \begin{array}{cc}
       a & b \\
       c & d \\
     \end{array}
   \right) \\
   &=&  \frac{j^3 (j+1)^3 \det \Big((j(j+1)-m) - r^2 \partial_{\overline{z}} \partial_z \Big)\det \Big((j(j+1) + m) - r^2 \partial_{\overline{z}} \partial_z \Big)}{\big( j^2 (j+1)^2 -m^2 \big)\det c_{j}^{m}}.
\end{eqnarray*}
Here we have incorporated the Jacobian factor for the change of variable \eq{scale-b}.

After some similar algebra, the action $S_{B;j, \vec{m}}^Q$ of the modes for $j \geq 1/2, \; m = -j, \; |m'| \leq j$,
can be read as
\begin{eqnarray} \label{mode22}
 \frac{(g_5 r)^2}{2} \cdot S_{B;j\geq \frac{1}{2}, -j, m'}^Q &=&
\big (B_{j+1, j}^{-j, m'} (z, \overline{z}) \big)^\dagger a_j^{-j}
B_{j+1, j}^{-j, m'} (z, \overline{z}) \xx
&& + 2 j(j+2) |\varphi_{z, j}^{-j, m'} (z, \overline{z})|^2
+ 2 j^2 |\varphi_{\overline{z}, j}^{-j, m'} (z, \overline{z})|^2 \xx
&& + \big (\sigma_{j}^{-j, m'} (z, \overline{z}) \big)^\dagger
c_{j}^{-j} \sigma_{j}^{-j, m'} (z, \overline{z}),
\end{eqnarray}
and the one for $j \geq 1/2, \; m = j, \; |m'| \leq j$ as
\begin{eqnarray} \label{mode3}
 \frac{(g_5 r)^2}{2} \cdot S_{B;j\geq \frac{1}{2}, j, m'}^Q &=&
 \big (B_{j+1, j}^{j, m'} (z, \overline{z}) \big)^\dagger a_j^j
 B_{j+1, j}^{j, m'} (z, \overline{z}) \xx
&& + 2 j^2 |\varphi_{z, j}^{j, m'} (z, \overline{z})|^2
+ 2 j(j+2) |\varphi_{\overline{z}, j}^{j, m'} (z, \overline{z})|^2 \xx
&& + \big (\sigma_{j}^{j, m'} (z, \overline{z}) \big)^\dagger
c_{j}^{j} \sigma_{j}^{j, m'} (z, \overline{z}),
\end{eqnarray}
where
$$a_j^j = a_j^{-j} = \frac{(j+1)^2}{(j+2)c_j^j} \big(j(j+2)- r^2 \partial_{\overline{z}}\partial_z \big).$$
Therefore the one-loop determinant from these sector with $j \geq 1/2, \; m = \pm j, \; |m'| \leq j$
is given by
\begin{equation}\label{det-vboson2}
    \prod_{j=1/2}^\infty \prod_{m'=-j}^j \frac{1}{\big(j(j+1) \big)^6}
\frac{1}{\det \big( j(j+2)- r^2 \partial_{\overline{z}}\partial_z \big)^2}.
\end{equation}

Since the action $S_{B;j, \pm (j+1), m'}^Q$ of the modes with $j \geq 0$ is coming
from the contribution of $B_+$ only, we directly calculate the action without shifting
scalar fields and a scale transformation and it is then given by
\begin{eqnarray} \label{mode4}
 \frac{(g_5 r)^2}{2} \cdot S_{B;j\geq 0, \pm (j+1), m'}^Q &=&
\big (B_{j+1, j}^{-(j+1), m'} (z, \overline{z}) \big)^\dagger
\big( (j+1)^2 - r^2 \partial_{\overline{z}}\partial_z \big)
B_{j+1, j}^{-(j+1), m'} (z, \overline{z}) \xx
&& + \big (B_{j+1, j}^{j+1, m'} (z, \overline{z}) \big)^\dagger
\big( (j+1)^2 - r^2 \partial_{\overline{z}}\partial_z \big)
B_{j+1, j}^{j+1, m'} (z, \overline{z}),
\end{eqnarray}
which leads to the one-loop determinant
\begin{equation}\label{det-vboson3}
    \prod_{j=0}^\infty \prod_{m'=-j}^j \frac{1}{\det \big( (j+1)^2 - r^2 \partial_{\overline{z}}\partial_z \big)^2}.
\end{equation}
Finally, the action $S_{B;j=0, 0, 0}^Q$ contains remaining modes with $j=0$ and it is simple as
\begin{equation}\label{mode5}
\frac{(g_5 r)^2}{2} \cdot S_{B;j=0, 0, 0}^Q = \big (\sigma_{0}^{\vec{0}} (z, \overline{z})
-i B_{1, 0}^{0, 0} (z, \overline{z}) \big)^\dagger
(-r^2 \partial_{\overline{z}} \partial_z) \big( \sigma_{0}^{\vec{0}} (z, \overline{z})
- i B_{1, 0}^{0, 0} (z, \overline{z}) \big)
+ |B_{1, 0}^{0, 0} (z, \overline{z})|^2,
\end{equation}
and the one-loop determinant is given by $\det (-r^2 \partial_{\overline{z}} \partial_z)^{-1}$.

Note that the ghost factor \eq{ghost-det} in terms of the harmonic modes on $\mathbb{S}^3$
gives rise to the one-loop determinant
\begin{equation}\label{ghost-hdet}
\det \square_0^{\frac{1}{2}} = \prod_{j=\frac{1}{2}}^\infty \prod_{m=-j}^{j}
\prod_{m'=-j}^j j (j+1),
\end{equation}
which is the inverse of the contribution from the real scalar field $ \frac{1}{2} \int_{\mathbb{S}^3}
 d\sigma \wedge *d\sigma$.
See footnote \ref{reality} for our treatment of the reality condition.
Wrapping up all the contributions from the bosonic fluctuations and the ghost
contributions \eq{back-det} and \eq{ghost-hdet},
the one-loop determinant can be written as the form
\begin{equation}\label{det-vboson}
\Upsilon \prod_{j=\frac{1}{2}}^\infty \prod_{m=-j}^{j}
\prod_{m'=-j}^j \frac{1}{j^2 (j+1)^2 \det \Big(K_j^m - r^2 \partial_{\overline{z}} \partial_z \Big)
\det \Big(K_j^{-m} - r^2 \partial_{\overline{z}} \partial_z \Big)}
\end{equation}
with the factor
$$\Upsilon = \det (1 - 4 r^2 \partial_{\overline{z}} \partial_z)^2$$
up to an overall constant. This result is essentially the same as Ref. \cite{fukama13}
with the root $\alpha = 0$.

Next we consider the fermions in the quadratic Lagrangian \eq{quad-action-5d}.
Note that $P_- \equiv \frac{1}{2}(1 - \gamma^\mu k_\mu)$ in the fermionic Lagrangian is
a projection operator, i.e., $P^2_- = P_-$ and thus projects out some component of a spinor;
in our case with $k_\mu = \delta_\mu^3$, the upper component.
Fortunately the problem reduces to computing spin-orbit coupling in quantum mechanics
by identifying $l_m = e^\mu_m \partial_\mu$
and $\gamma^m = e^m_\mu \gamma^\mu$ with $\frac{2i}{r} L_m$ and $2 S^m$, respectively \cite{kwy09}.
Since $\nabla_\mu = \partial_\mu + \frac{i}{2r} \gamma_\mu$, we have the relation
\begin{equation}\label{dirac-ls}
    - i \gamma^\mu \nabla_\mu = - i \gamma^m l_m + \frac{3}{2r}
    = \frac{4}{r} \overrightarrow{L} \cdot \overrightarrow{S} + \frac{3}{2r}
= \frac{2}{r} \Big( (\overrightarrow{L} + \overrightarrow{S})^2 - \overrightarrow{L}
\cdot \overrightarrow{L} \Big).
\end{equation}
In order to get the harmonic expansion of the spinors $\lambda$ and $\psi$ on $\mathbb{S}^3$,
it is useful to introduce the eigenspinors $\theta_{k,m_s;j,m_2} (x) = \theta_{k,j}^{\vec{m}} (x)$ of the operator
$\mathbb{D} \equiv  \gamma^m \nabla_m$ by \cite{fukama13}
\begin{equation}\label{ham-fexp}
 \theta_{k,j}^{\vec{m}} (x) =  \sum_{m_1 = -j}^j \sum_{s=\pm (1/2)} \langle j, m_1;
\frac{1}{2}, s| k, m_s \rangle\rangle  S_j^{m_1,m_2} (x) \varsigma^s,
\end{equation}
where $\langle j, m_1; \frac{1}{2}, s| k, m_s \rangle\rangle $ are the Clebsch-Gordan coefficients of
the spin-$j$ representation and the spin-$\frac{1}{2}$ representation into the spin-$k$ representation,
with $k = j \pm \frac{1}{2}$ and the spinor $\varsigma^s$ satisfies the relation
$S^3 \varsigma^s = s \varsigma^s$ with $s = \pm \frac{1}{2}$. See appendix E for the Clebsch-Gordan coefficients.
The spinors $\theta_{k,j}^{\vec{m}} (x)$ obey the eigenvalue equations
\begin{equation}\label{spin-eigen}
    \begin{array}{l}
      \mathbb{D} \, \theta_{j + \frac{1}{2},m + \frac{1}{2};j,m'} = \frac{i}{r} \big( 2j + \frac{3}{2} \big)
      \theta_{j + \frac{1}{2},m + \frac{1}{2};j,m'}, \\
      \mathbb{D} \, \theta_{j - \frac{1}{2},m + \frac{1}{2};j,m'} = -\frac{i}{r} \big( 2j + \frac{1}{2} \big)
      \theta_{j - \frac{1}{2},m + \frac{1}{2};j,m'},
    \end{array}
\end{equation}
and form an orthonormal basis
\begin{equation}\label{spin-ortho}
    \int_{\mathbb{S}^3} \big(\theta_{k',j'}^{\vec{m}'} (x) \big)^\dagger \theta_{k,j}^{\vec{m}} (x) * 1 =
    \delta_{k'k} \delta_{j'j} \delta^{\vec{m}' \vec{m}},
\end{equation}
where the complex conjugate obeys the usual relation $\big( \theta_{k,j}^{\vec{m}} (x) \big)^\dagger
= (-1)^{m_s + m'} \theta_{k,j}^{-\vec{m}} (x)$. And it is easy to verify that
\begin{eqnarray}\label{formula-spin}
 \int_{\mathbb{S}^3} \big(\theta_{k',j'}^{\vec{m}'} (x) \big)^\dagger
(1-k_m \gamma^m) \theta_{k,j}^{\vec{m}} (x) *1
&=& \langle\langle k', m'| (1-2S^3)|k,m \rangle\rangle \delta_{j'j} \delta_{m_2' m_2}.
\end{eqnarray}

Using the spinor basis \eq{ham-fexp}, we expand the spinors $\lambda$ and $\psi$ as
\begin{eqnarray}\label{vspinor-exp}
    \lambda(x,y) &=& \sum_{j=0}^\infty \sum_{m=-(j+1)}^j \sum_{m'=-j}^j \lambda_{j + \frac{1}{2},m + \frac{1}{2};j,m'} (y) \theta_{j + \frac{1}{2},m + \frac{1}{2};j,m'} (x) \xx
    &&  + \sum_{j=1/2}^\infty \sum_{m=-j}^{j-1} \sum_{m'=-j}^j \lambda_{j - \frac{1}{2},m + \frac{1}{2};j,m'} (y)
    \theta_{j - \frac{1}{2},m + \frac{1}{2};j,m'} (x), \xx
    \psi(x,y) &=& \sum_{j=0}^\infty \sum_{m=-(j+1)}^j \sum_{m'=-j}^j \psi_{j + \frac{1}{2},m + \frac{1}{2};j,m'} (y) \theta_{j + \frac{1}{2},m + \frac{1}{2};j,m'} (x) \xx
    && + \sum_{j=1/2}^\infty \sum_{m=-j}^{j-1} \sum_{m'=-j}^j \psi_{j - \frac{1}{2},m + \frac{1}{2};j,m'} (y)
    \theta_{j - \frac{1}{2},m + \frac{1}{2};j,m'} (x).
\end{eqnarray}
We will use the shorthand notation for the coefficient modes such that
$\lambda_{j \pm \frac{1}{2},m + \frac{1}{2};j,m'} \to \lambda_{j \pm \frac{1}{2}}$ and
$\psi_{j \pm \frac{1}{2},m + \frac{1}{2};j,m'} \to \psi_{j \pm \frac{1}{2}}$ whenever such a notation is enough.
Since the Lagrangian \eq{quad-action-5d} is also quadratic in fermionic fields,
it is straightforward to evaluate the action for the mode expansion \eq{vspinor-exp}, which can be read as
\begin{equation}\label{fmodeint-s3}
     t \cdot S_{VF}^Q = \sum_{j=0}^\infty \sum_{\vec{m} = - (j+1)}^j \sum_{m'=-j}^j \int d^2 y
S_{F;j, \vec{m}}^Q (z, \overline{z}).
\end{equation}
In order to implement the Gaussian integration for the action \eq{fmodeint-s3},
we shift the spinor modes for $j \geq \frac{1}{2}, \; -j \leq m \leq (j-1), \; |m'| \leq j$ as
\begin{eqnarray*}
 && \lambda_{j + \frac{1}{2}}  \to \lambda_{j + \frac{1}{2}} + \frac{1}{j+1}
r \partial_{\overline{z}} \psi_{j + \frac{1}{2}}, \\
&& \overline{\lambda}_{j + \frac{1}{2}}  \to \overline{\lambda}_{j + \frac{1}{2}}
+ \frac{\sqrt{j-m}}{(j+1)(2j+1)} r \partial_{z} \Big( \sqrt{j-m} \overline{\psi}_{j + \frac{1}{2}}
+ \sqrt{j+m+1} \overline{\psi}_{j - \frac{1}{2}} \Big),\\
 && \lambda_{j - \frac{1}{2}}  \to \lambda_{j - \frac{1}{2}} - \frac{1}{j}
r \partial_{\overline{z}} \psi_{j - \frac{1}{2}}, \\
&& \overline{\lambda}_{j - \frac{1}{2}}  \to \overline{\lambda}_{j - \frac{1}{2}}
- \frac{\sqrt{j+m+1}}{j(2j+1)} r \partial_{z} \Big( \sqrt{j-m} \overline{\psi}_{j + \frac{1}{2}}
+ \sqrt{j+m+1} \overline{\psi}_{j - \frac{1}{2}} \Big).
\end{eqnarray*}
The corresponding action for these modes is then given by
\begin{eqnarray} \label{faction-1}
  S_{F;j \geq 1/2, \vec{m}}^Q &=& \frac{4}{r} \Big( (j+1) \overline{\lambda}_{j + \frac{1}{2}}
\lambda_{j + \frac{1}{2}} - j \overline{\lambda}_{j - \frac{1}{2}} \lambda_{j - \frac{1}{2}} \Big) \xx
 && - \frac{4}{r} ( \overline{\psi}_{j + \frac{1}{2}}, \overline{ \psi}_{j - \frac{1}{2}})
\left(
  \begin{array}{cc}
    \alpha & \beta \\
    \gamma & \delta \\
  \end{array}
\right)
\left(
  \begin{array}{c}
    \psi_{j + \frac{1}{2}} \\
    \psi_{j - \frac{1}{2}} \\
  \end{array}
\right)
\end{eqnarray}
with
\begin{eqnarray*}
&& \alpha = \frac{2j(j+1)-m}{2j+1} - \frac{j-m}{(j+1)(2j+1)} r^2
\partial_{\overline{z}} \partial_z \equiv \alpha_j^m, \\
&& \beta = \frac{\sqrt{(j-m)(j+m+1)}}{2j+1} \Big( 1 - \frac{1}{j+1}
r^2 \partial_{\overline{z}} \partial_z \Big), \\
&& \gamma = \frac{\sqrt{(j-m)(j+m+1)}}{2j+1} \Big( 1 + \frac{1}{j}
r^2 \partial_{\overline{z}} \partial_z \Big), \\
&& \delta = - \frac{2j(j+1)-m}{2j+1} + \frac{j+m+1}{j(2j+1)} r^2 \partial_{\overline{z}} \partial_z.
\end{eqnarray*}
Thus the above fermionic fluctuations lead to the one-loop determinant
\begin{equation}\label{det-vfermi1}
    \prod_{j=1/2}^\infty \prod_{m=-j}^{j-1} \prod_{m'=-j}^j  j^2 (j+1)^2 \det \Big((j(j+1)-m) -
    r^2 \partial_{\overline{z}} \partial_z \Big)^2.
\end{equation}
Here we have considered the fact that $\lambda_{j \pm \frac{1}{2}}$ and $\psi_{j \pm \frac{1}{2}}$
are complex spinors. Cf. footnote \ref{reality}.

For the remaining fermionic modes with $j \geq 0, \; m = -(j+1), \, j,\; |m'| \leq j$,
the corresponding action reads as
\begin{eqnarray} \label{faction-2}
 && S_{F;j \geq 0, m=-(j+1) \& j}^Q = \frac{4}{r} (j+1) \Big( \overline{\lambda}_{j + \frac{1}{2},
  -(j + \frac{1}{2});j, m'} \lambda_{j + \frac{1}{2}, - (j + \frac{1}{2});j, m'}
  + \overline{\lambda}_{j + \frac{1}{2}, j + \frac{1}{2};j, m'}
  \lambda_{j + \frac{1}{2}, j + \frac{1}{2};j, m'} \Big) \xx
  && \qquad  - \frac{4}{r} \Big( \overline{\psi}_{j + \frac{1}{2}, -(j + \frac{1}{2});j, m'}
  \alpha_{j}^{-(j+1)}  \psi_{j + \frac{1}{2}, -(j + \frac{1}{2});j, m'}
  + \overline{\psi}_{j + \frac{1}{2}, j + \frac{1}{2};j, m'} \alpha_{j}^{j}
  \psi_{j + \frac{1}{2}, j + \frac{1}{2};j, m'} \Big),
\end{eqnarray}
which leads to the one-loop determinant given by
\begin{equation}\label{det-vfermi2}
    \det ( 1 - r^2 \partial_{\overline{z}} \partial_z )^2
\prod_{j=1/2}^\infty \prod_{m'=-j}^j  j^2(j+1)^2
\det \big( (j+1)^2 -  r^2 \partial_{\overline{z}} \partial_z \big)^2.
\end{equation}
Combining all the contributions from the fermionic fluctuations yields the one-loop determinant
\begin{equation}\label{det-vfermi3}
  \frac{1}{\det ( 1 - 4 r^2 \partial_{\overline{z}} \partial_z )^2}
\prod_{j=1/2}^\infty \prod_{m=-j}^{j} \prod_{m'=-j}^j j^2 (j+1)^2 \det \Big((j(j+1)-m) -
    r^2 \partial_{\overline{z}} \partial_z \Big)^2.
\end{equation}

Note that the one-loop determinant \eq{det-vfermi3} from the fermionic fluctuations exactly cancels
the one \eq{det-vboson} from the bosonic fluctuations. This result is somewhat expected \cite{fukama13}
since our result corresponds to the five-dimensional $\mathcal{N}=2$ Yang-Mills theory
on $\mathbb{S}^3 \times \mathbb{R}^2$ since the noncommutativity can be ignored at the quadratic order.
However the classical action \eq{locus-action} at the localization locus needs not be quadratic
and thus the noncommutative structure between background fields must be kept.
The matrix representation of the background fields consequently gives rise to a zero-dimensional
matrix model with the action \eq{matrix-locus} subject to a perplexing constraint $[\phi, \sigma]=0$.
Thus we have explored the red arrows in Fig. \ref{flowchart} to derive a large $N$ matrix model
from a five-dimensional NC $U(1)$ gauge theory.

\subsection{Localization of hypermultiplet}

Using the same twisting as the vector multiplet, we can deduce the BRST transformations for the hypermultiplet
given by
\begin{eqnarray}\label{brst-hyper}
&& \delta_Q H_1 = 0, \quad \delta_Q H_2 = i g_5 \epsilon^T C_3 \chi,
\quad \delta_Q \overline{H}^1 = ig_5 \overline{\eta} \epsilon, \quad \delta_Q \overline{H}^2 = 0, \xx
&& \delta_Q \eta = - \frac{1}{g_5} \Big(\gamma^\mu D_\mu H_1 + [\sigma, H_1]
+ \frac{i}{r} H_1 \Big) \epsilon, \quad \delta_Q \chi = - \frac{2}{g_5} [\phi_z, H_1] \epsilon + F_1 \epsilon, \xx
&& \delta_Q \overline{\eta} = \frac{2}{g_5} \epsilon^T C_3 [\overline{H}^2, \phi_z]
+ \epsilon^T C_3 \overline{F}^2, \quad
\delta_Q \overline{\chi} = \frac{1}{g_5} \epsilon^T C_3 \Big( \gamma^\mu D_\mu \overline{H}^2
+ [\overline{H}^2, \sigma] - \frac{i}{r} \overline{H}^2 \Big), \\
&& \delta_Q F_1 = 0, \quad \delta_Q F_2 = i \epsilon^T C_3 \Big(\gamma^\mu D_\mu \eta
- 2[\phi_{\overline{z}}, \chi] - [\sigma, \eta] - \frac{i}{2r} \eta \Big)
- 2 \big([H_1, \epsilon^T C_3 \lambda] + [H_2, \overline{\psi}\epsilon] \big), \xx
&& \delta_Q \overline{F}^1 = - i \Big(D_\mu \overline{\chi} \gamma^\mu
+ 2[\phi_{\overline{z}}, \overline{\eta}] + [\sigma, \overline{\chi}] + \frac{i}{2r} \overline{\chi} \Big) \epsilon
- 2 \big([\overline{H}^1, \overline{\psi} \epsilon] - [\overline{H}^2, \epsilon^T C_3 \lambda] \big),
\quad \delta_Q \overline{F}^2 = 0. \nonumber
\end{eqnarray}
The above BRST transformations are nilpotent, i.e., $\delta^2_Q = 0$, as was expected.
After a tedious calculation, the BRST transformation of the action \eq{hyper-action} can be determined as
\begin{eqnarray} \label{brst-hvar}
\delta_Q S_{5H} &=& - \int d \upsilon \bigg[ \frac{1}{2g_5r} \Big( \overline{\eta} \gamma^\mu \epsilon D_\mu H_1
- (\epsilon^T C_3 \gamma^\mu \chi) D_\mu \overline{H}^2 + \overline{\eta} \epsilon [\sigma, H_1]
- \epsilon^T C_3 \chi [\overline{H}^2, \sigma] \Big)  \nonumber \\
&& \hspace{1.5cm} - \frac{1}{g_5 r} \Big(  \overline{\chi} \epsilon [\phi_z, H_1]
+ \epsilon^T C_3 \chi [\phi_z, \overline{H}^2] \Big)
+ \frac{1}{2 r} \Big( \overline{\chi} \epsilon F_1 + \epsilon^T C_3 \eta \overline{F}^2 \Big)  \nonumber \\
&& \hspace{1.5cm} + \frac{3i}{2 g_5 r^2} \Big( \overline{\eta} \epsilon H_1
+ \epsilon^T C_3 \chi \overline{H}^2 \Big) \bigg].
\end{eqnarray}
The above variation is exactly cancelled by the BRST transformation of the mass term \eq{hyper-mass}.
Thus the total action for the hypermultiplet is BRST invariant, i.e., $\delta_Q (S_{5H} + S^M_{5H}) = 0$.

Similarly we deform the action $S^{(\mathrm{inv})}_{5H} \equiv S_{5H} + S^M_{5H}$ by adding a BRST-exact term
\begin{eqnarray}\label{hyper-exact}
    S_{5H}^Q &=& \int d\upsilon  \delta_Q \Big[ (\delta_Q \eta)^\dagger \eta
    + \overline{\eta} (\delta_Q \overline{\eta})^\dagger + (\delta_Q \chi)^\dagger \chi
    + \overline{\chi} (\delta_Q \overline{\chi})^\dagger \Big] \xx
    &=& \int d\upsilon \mathcal{L}_{5H}^{QB} + \int d\upsilon  \mathcal{L}_{5H}^{QF}
\end{eqnarray}
where the bosonic part of the Lagrangian $\mathcal{L}_{5H}^Q$ is given by
\begin{eqnarray} \label{hyper-dbos}
\mathcal{L}_{5H}^{QB}  &=& \frac{1}{g_5^2} \Big( D^\mu \overline{H}^{\dot{\alpha}} D_\mu H_{\dot{\alpha}}
-[\sigma, \overline{H}^{\dot{\alpha}}] [\sigma, H_{\dot{\alpha}}]
+ \frac{1}{r^2} \overline{H}^{\dot{\alpha}} H_{\dot{\alpha}} \Big) \xx
&& + \Big( F_1 -  \frac{2}{g_5} [\phi_z, H_1] \Big)^\dagger
\Big( F_1 -  \frac{2}{g_5} [\phi_z, H_1] \Big)
+ \Big( F_2 +  \frac{2}{g_5} [\phi_{\overline{z}}, H_2] \Big)^\dagger
\Big( F_2 +  \frac{2}{g_5}[\phi_{\overline{z}}, H_2] \Big) \xx
   && + \frac{k_\mu}{g_5^2} \Big( \frac{1}{2} \epsilon^{\mu\nu\rho} F_{\nu\rho} + D^\mu \sigma \Big)
   \big( [\overline{H}^1, H_1] - [\overline{H}^2, H_2] \big)
\end{eqnarray}
up to total derivatives and the fermionic part by
\begin{eqnarray} \label{hyper-dfos}
\mathcal{L}_{5H}^{QF}  &=& \overline{\eta} [H_1, \lambda] - 2 [\overline{\lambda}, \overline{H}^1] \eta
+ 2 [\overline{\lambda}, H_2] C_3 \chi^* -[\overline{H}^2, \lambda^T] C_3 \chi   \xx
&& + k_\mu \Big( i \big(\overline{\eta} D^\mu \eta + \overline{\chi} D^\mu \chi \big)
+ \epsilon^{\mu\nu\rho} \big(\overline{\eta} \gamma_\nu D_\rho \eta -
\overline{\chi} \gamma_\nu D_\rho \chi \big)
+ i \big( [\sigma, \overline{\eta}] \gamma^\mu \eta - [\sigma, \overline{\chi}] \gamma^\mu \chi \big) \xx
   && - \overline{\eta} \gamma^\mu [H_1, \lambda] - [\overline{H}^2, \lambda^T] C_3 \gamma^\mu \chi
   + \frac{1}{2r} \big( \overline{\eta}\gamma^\mu \eta - 3 \overline{\chi} \gamma^\mu \chi \big) \Big).
\end{eqnarray}
Now the total classical action is defined by
\begin{equation}\label{th-action}
    \widetilde{S}_{5H} \equiv  S^{(\mathrm{inv})}_{5H} + t S_{5H}^Q.
\end{equation}
A fixed point of the BRST-exact action \eq{hyper-exact} is given by a solution to
$\delta_Q \eta = \delta_Q \chi = 0$ and $\delta_Q \overline{\eta} = \delta_Q \overline{\chi} = 0$
in addition to fermions $=0$. One gets $H_1 = F_1 = 0$ from the first condition and $H_2 = F_2 = 0$
from the second one. As a result, one finds no nontrivial background for the hypermultiplet and
thus $\widetilde{S}_{5H}|_{\mathfrak{F}_Q} = 0$.

Since we expand the fields in the hypermultiplet around the saddle point configuration
as Eq. \eq{exp-saddle} and take the limit $t \to \infty$,
it is enough to keep the quadratic order of the fluctuations.
After shifting the fields,
\begin{equation*}
    F_1 \to F_1 +  \frac{2}{g_5} [\phi_z, H_1], \qquad
    F_2 \to F_2 -  \frac{2}{g_5} [\phi_{\overline{z}}, H_2],
\end{equation*}
the bosonic Lagrangian \eq{hyper-dbos} reduces to a simple form:
\begin{equation}\label{hyper-bq2}
    t \cdot \mathcal{L}_{HB}^{Q}  = \frac{1}{g_5^2} \overline{H}^{\dot{\alpha}}\Big(-\partial^\mu \partial_\mu + \frac{1}{r^2} \Big) H_{\dot{\alpha}}.
\end{equation}
Note that we have already carried out the integration over the auxiliary fields $F_1$ and $F_2$
and their contributions to the partition function are simply an overall constant.
Similarly the fermionic Lagrangian \eq{hyper-dfos} also becomes a quadratic form:
\begin{eqnarray}\label{hyper-fq2}
    t \cdot \mathcal{L}_{HF}^{Q}  &=&  \overline{\eta} k_m \gamma^m \Big( i \gamma^n D_n \eta + \frac{1}{2r} \eta \Big)
+  \overline{\chi} k_m \gamma^n \Big( i \gamma^m D_n \chi -  \frac{3}{2r} \chi \Big) \xx
&=&  \overline{\eta} k_m \gamma^m \Big( i \gamma^n D_n \eta + \frac{1}{2r} \eta \Big)
+  \overline{\chi} k_m \gamma^m \Big( i \gamma^n D_n \chi +  \frac{1}{2r} \chi \Big) \xx
&& + \frac{4}{r} \overline{\chi}(S_+L_- - S_- L_+) \chi
\end{eqnarray}
where $S_\pm = \frac{1}{2}( \gamma^1 \pm i \gamma^2)$ acts on the spinors
$S_+ \varsigma^- = \varsigma^+, \; S_- \varsigma^+ = \varsigma^-$. One can show by a direct calculation
using the mode expansion \eq{harmonicex-hyper} below that the last term in Eq. \eq{hyper-fq2} vanishes.
A more easier way to see this is to consider a change of variable (cf. footnote \ref{japan-ref}),
$\chi \to C_3^{-1} \chi^*, \; \overline{\chi} \to \chi^T C_3$, under which
$\overline{\chi} k_m \gamma^n \big( i \gamma^m D_n \chi -  \frac{3}{2r} \chi \big)
\to \overline{\chi} k_m \gamma^m \big( i \gamma^n D_n \chi +  \frac{1}{2r} \chi \big)$
after integration by parts and using the relation $D_n k_m = - \frac{1}{r} \varepsilon_{mnp} k^p$.

The one-loop partition function for the hypermultiplet can be obtained in the same way as
the vector multiplet by employing the harmonic expansions of the fluctuations\footnote{Since the hypermultiplet
involves complex fields only, it is not necessary to care about the reality condition of
the harmonic expansions in  Eq. \eq{harmonicex-hyper}, unlike the vector multiplet
as we remarked in footnote \ref{reality}. Hence all the expansion coefficients
in \eq{harmonicex-hyper} will be regarded as independent.}
\begin{eqnarray} \label{harmonicex-hyper}
H_1 (x, y) &=& \sum_{j=0}^\infty \sum_{\vec{m}=-j}^j H_{1j}^{\vec{m}} (y) S_j^{\vec{m}} (x),
\qquad H_2 (x, y) = \sum_{j=0}^\infty \sum_{\vec{m}=-j}^j H_{2j}^{\vec{m}} (y) S_j^{\vec{m}} (x), \xx
    \eta(x,y) &=& \sum_{j=0}^\infty \sum_{m=-(j+1)}^j \sum_{m'=-j}^j \eta_{j + \frac{1}{2},m + \frac{1}{2};j,m'} (y) \theta_{j + \frac{1}{2},m + \frac{1}{2};j,m'} (x) \xx
    &&  + \sum_{j=1/2}^\infty \sum_{m=-j}^{j-1} \sum_{m'=-j}^j \eta_{j - \frac{1}{2},m + \frac{1}{2};j,m'} (y)
    \theta_{j - \frac{1}{2},m + \frac{1}{2};j,m'} (x), \xx
    \chi(x,y) &=& \sum_{j=0}^\infty \sum_{m=-(j+1)}^j \sum_{m'=-j}^j \chi_{j + \frac{1}{2},m + \frac{1}{2};j,m'} (y) \theta_{j + \frac{1}{2},m + \frac{1}{2};j,m'} (x) \xx
    && + \sum_{j=1/2}^\infty \sum_{m=-j}^{j-1} \sum_{m'=-j}^j \chi_{j - \frac{1}{2},m + \frac{1}{2};j,m'} (y)
    \theta_{j - \frac{1}{2},m + \frac{1}{2};j,m'} (x).
\end{eqnarray}
The one-loop determinant from the bosonic fluctuations can easily be obtained as
\begin{equation}\label{det-hbos}
\prod_{j=0}^\infty \prod_{m=-j}^{j} \prod_{m'=-j}^j \frac{1}{(2j+1)^4}.
\end{equation}
The above mode expansion for the fermionic Lagrangian \eq{hyper-fq2} leads to the action given by
\begin{eqnarray} \label{faction-hyper}
&& - \frac{1}{r} \sum_{j=0}^\infty \sum_{m'= - j}^j (2j+1)
\big( |\eta_{j + \frac{1}{2}, j + \frac{1}{2}; j, m'}|^2
- |\eta_{j + \frac{1}{2}, -(j + \frac{1}{2}); j, m'}|^2 \big) + ( \eta \to \chi) \xx
 && - \frac{1}{r} \sum_{j=1/2}^\infty \sum_{m= - j}^{j-1} \sum_{m'= - j}^j
( \overline{\eta}_{j + \frac{1}{2}}, \overline{\eta}_{j - \frac{1}{2}})
\left(
  \begin{array}{cc}
     \alpha & - \beta \\
    \beta & \alpha \\
  \end{array}
\right)
\left(
  \begin{array}{c}
    \eta_{j + \frac{1}{2}} \\
    \eta_{j - \frac{1}{2}} \\
  \end{array}
\right) + ( \eta \to \chi)
\end{eqnarray}
with
$$\alpha = 2m+1, \qquad \beta = 2 \sqrt{(j-m)(j+m+1)}.$$
Therefore the one-loop determinant from these fermionic modes yields a factor
\begin{equation}\label{det-hfer}
\prod_{j=0}^\infty \prod_{m=-j}^{j} \prod_{m'=-j}^j (2j+1)^4.
\end{equation}

Wrapping up the bosonic and fermionic contributions, one finds that the hypermultiplet contributes
just a constant to the total partition function.

\section{Localization of three-dimensional large $N$ gauge theory}

The relationship between a lower-dimensional large $N$ gauge theory and a higher-dimensional NC $U(1)$
gauge theory in Fig. \ref{flowchart} is an exact mathematical identity.
The identity in Fig. \ref{flowchart} is derived from the fact that the NC space \eq{nc-space} admits
a separable Hilbert space and NC $U(1)$ gauge fields become operators acting on the Hilbert space.
Using the matrix representation \eq{matrix-rep} of NC fields, we have obtained a three-dimensional large $N$
gauge theory described by the action \eq{3d-sym} for the vector multiplet and
the action \eq{mhyper-action} for the hypermultiplet.
Now we will explore the blue arrows in Fig. \ref{flowchart} to illuminate how to derive
the same large $N$ matrix model starting from a three-dimensional large $N$ gauge theory.
See Refs. \cite{local-qft7,local-qft8} and references therein for the localization
of three-dimensional quantum field theories.

The BRST invariant theory for the vector multiplet in the three-dimensional large $N$ gauge theory
is given by the action \eq{tv-action} by applying the isomorphic map \eq{imap-53}.
The localization locus $\mathfrak{F}_Q$ is defined by
\begin{eqnarray}\label{twvsusy-3d}
&& \delta_Q \lambda = \frac{1}{2} \Big( \frac{1}{2g_3} F_{\mu\nu} \gamma^{\mu\nu} + \frac{i}{g_3}
\gamma^\mu D_\mu \sigma  - i D \Big) \epsilon = 0, \xx
&& \delta_Q \psi = - \frac{i}{g_3} \Big( \gamma^\mu D_\mu \phi_z
- [\phi_z, \sigma] \Big) \epsilon = 0, \\
&& \delta_Q \overline{\psi} = i \overline{F} \epsilon^T C_3 = 0, \nonumber
\end{eqnarray}
where $\epsilon$ is a Killing spinor satisfying Eq. \eq{kill-spinor}.
The space of $\mathfrak{F}_Q$ consists of all possible solutions obeying the above conditions.
It may be characterized by the BPS equations given by
\begin{equation}\label{bps-locus}
\frac{1}{2} \varepsilon^{\mu\nu\rho} F_{\nu\rho} +  D^\mu \sigma = 0 \quad \&
\quad  D_\mu \phi_z = [\phi_z, \sigma] = 0
\end{equation}
and $D = F = \overline{F}=0$.
See, e.g., \cite{seok} for the localization at a Dirac monopole configuration.
But we will consider a more simplified set of solutions satisfying $A_\mu = \partial_\mu \sigma
= \partial_\mu \phi_z = 0$ and thus $\mathfrak{F}_Q$ is defined by the set of constants obeying
\begin{equation}\label{3d-locus}
    \mathfrak{F}_Q = \{ (\sigma, \phi_z)|  [\phi_z, \sigma] = 0 \}.
\end{equation}
Then the classical action at the locus $\mathfrak{F}_Q$ is given by the zero-dimensional
matrix model \eq{matrix-locus}.

The conventional choice of vacuum in the Coulomb branch of $U(N)$ gauge theory is given by
Eq. \eq{cc-vacuum}. It means that the locus $\mathfrak{F}_Q$ takes values in the Cartan subalgebra of
the Lie algebra of $\mathfrak{g}= U(N)$ such that
\begin{equation}\label{cartan}
    \sigma = \sum_{i=1}^N \sigma^i H_i, \qquad \phi_z = \sum_{i=1}^N \phi_z^i H_i.
\end{equation}
Here $H_i \; (i=1, \cdots, N)$ are generators of the Cartan subalgebra of $u(N)$ of rank $N$.
In this case the $U(N)$ gauge symmetry is broken to $U(1)^N$.
Note that the gauge group $\mathfrak{g}$ in our case is $U(N)$, in particular, in the limit $N \to \infty$.
In order to find all possible solutions defining the space of $\mathfrak{F}_Q$, it is important to notice
that the limit $N \to \infty$ opens a new phase of the Coulomb branch,
the so-called NC Coulomb branch \cite{q-emg,hea}.
For example it may be characterized by the vacuum \eq{nc-vacuum} satisfying the Moyal-Heisenberg algebra.
It should be emphasized that the NC Coulomb branch arises as a vacuum solution of the large $N$ gauge
theory \eq{3d-sym} and it saves the NC nature of matrices while the conventional vacuum \eq{cartan}
dismisses the property.

To be specific, the locus $\mathfrak{F}_Q$ in the NC Coulomb branch is given by
\begin{equation}\label{vac-sigma}
    \langle \sigma \rangle_{\mathrm{vac}} = \sigma_0 \mathbb{I}_{N \times N}
\end{equation}
and
\begin{eqnarray}\label{vac-phi}
\begin{array}{l}
\langle \phi_z \rangle_{\mathrm{vac}} = - \frac{i}{\sqrt{2 \alpha'}} \left(
                                                                     \begin{array}{ccccc}
                                                                       0 & 0 & 0 & \cdots & 0 \\
                                                                       1 & 0 & 0 & \cdots & 0 \\
                                                                       0 & \sqrt{2} & 0 & \cdots & 0 \\
                                                                       0 & 0 & \sqrt{3} & \cdots & 0 \\
                                                                       \vdots & \vdots & \vdots & \vdots & \vdots \\
                                                                     \end{array}
                                                                   \right)
\equiv - \frac{i}{2 \alpha'} \overline{z}, \\
\langle \phi_{\overline{z}} \rangle_{\mathrm{vac}} = \frac{i}{\sqrt{2 \alpha'}} \left(
                                               \begin{array}{cccccc}
                                                0 & 1 & 0 & 0 & \cdots & 0 \\
                                                0 & 0 & \sqrt{2} & 0 & \cdots & 0 \\
                                                0 & 0 & 0 & \sqrt{3} & \cdots & 0 \\
                                                \vdots & \vdots & \vdots & \vdots & \vdots & \vdots \\
                                                                     \end{array}
                                                                   \right)
\equiv \frac{i}{2 \alpha'} z.
\end{array}
\end{eqnarray}
It is then obvious that $\langle [\phi_z, \sigma] \rangle_{\mathrm{vac}} = 0$ and
$\langle [\phi_{\overline{z}}, \phi_z] \rangle_{\mathrm{vac}} = \frac{1}{2\alpha'} \mathbb{I}_{N \times N}$.
Of course, we have to take the limit $N \to \infty$ to make sense of this NC vacuum.\footnote{It might
be remarked that $\sqrt{2\alpha'}\phi_{\overline{z}}$ corresponds to $a = \frac{x+ip}{\sqrt{2\hbar}}$
and $\sqrt{2\alpha'}\phi_{z}$ to $a^\dagger = \frac{x-ip}{\sqrt{2\hbar}}$ and the operators $x$ and $p$ obey
the Heisenberg algebra $[x, p] = i\hbar$. As is well-known from quantum mechanics, the representation space
of the Heisenberg algebra is the infinite-dimensional Fock space $\mathcal{H}$ and so the representation
$\phi_{\overline{z}}$ and $\phi_{z}$ on the Hilbert space $\mathcal{H}$ requires infinite-dimensional matrices.}
One may observe that the NC vacuum \eq{vac-phi} can be represented by the root system $\mathfrak{r}$ of
the Lie algebra $su(N)$ as
\begin{equation}\label{phi-root}
    \langle \phi_z \rangle_{\mathrm{vac}} = - \frac{i}{\sqrt{2 \alpha'}} \sum_{\alpha \in \mathfrak{r}}
\phi_z^\alpha E_\alpha.
\end{equation}
Therefore the NC Coulomb branch is in sharp contrast to the conventional vacuum \eq{cartan}
which takes values in the Cartan subalgebra of $u(N)$ only.

The localization of a large $N$ gauge theory at the conventional Coulomb branch \eq{cartan}
has been discussed by many authors \cite{kwy09,seok,jaff10,hhl11,imamu,hoyo12}.
See also a review \cite{marino-rev}. So we will focus on the localization at the NC Coulomb branch.
Let us represent all possible deformations of the vacuum $\mathfrak{F}_Q$ by
\begin{equation}\label{mat-fluct}
\begin{array}{l}
   \sigma (x, z, \overline{z}) = \sigma_0 \mathbb{I}_{N \times N} + \frac{1}{\sqrt{t}}
\delta \sigma (x, z, \overline{z}), \\
  \phi_z (x, z, \overline{z}) = - \frac{i}{2 \alpha'} \overline{z} + \frac{1}{\sqrt{t}}
\delta \phi_z (x, z, \overline{z}), \\
  \phi_{\overline{z}} (x, z, \overline{z}) = \frac{i}{2 \alpha'} z + \frac{1}{\sqrt{t}}
\delta \phi_{\overline{z}} (x, z, \overline{z}), \\
\Phi (x, z, \overline{z}) = \frac{1}{\sqrt{t}}
\delta \Phi (x, z, \overline{z}),
\end{array}
\end{equation}
where $\Phi (x, z, \overline{z})$ collectively represents remaining fields with the vanishing
vacuum expectation value at $\mathfrak{F}_Q$.
The notation in Eq. \eq{mat-fluct} means the large $N$ matrices such that, for example,
\begin{equation}\label{exp-matrix}
  \Phi (x, z, \overline{z}) \cong \sum_{i=1}^N  \Phi^i (x) H_i +
\sum_{\alpha \in \mathfrak{r}} \Phi^\alpha (x) E_\alpha.
\end{equation}
In other words, the matrix representation of $\Phi (x, z, \overline{z})$ on the Fock space $\mathcal{H}$
can be expanded in the Chevalley basis $(H_i, E_{\pm \alpha})$ of the Lie algebra $u(N)$
in the limit $N \to \infty$.
With this notation, it is obvious that the localization of the three-dimensional large $N$ gauge theory around
the locus \eq{3d-locus} is exactly parallel to the five-dimensional case and thus arrives at the results
\eq{det-vboson} and \eq{det-vfermi3} for the one-loop determinant from the bosonic and fermionic fluctuations
 described by large $N$ matrices in Eq. \eq{mat-fluct}. So we confirm the duality
in Fig. \ref{flowchart} to illustrate how to use a five-dimensional NC $U(1)$ gauge theory for the localization
of the vector multiplet in the three-dimensional large $N$ gauge theory.

We can apply the same idea to the hypermultiplet in a three-dimensional large $N$ gauge theory.
The BRST invariant theory for the hypermultiplet in the three-dimensional large $N$ gauge theory
is given by the action \eq{th-action} by applying the isomorphic map \eq{imap-53}.
The localization locus $\mathfrak{F}_Q$ for the hypermultiplet is defined by
\begin{eqnarray} \label{locus-mh}
&& \delta_Q \eta = - \frac{1}{g_3} \Big(\gamma^\mu D_\mu H_1 + [\sigma, H_1]
+ \frac{i}{r} H_1 \Big) \epsilon = 0, \xx
&& \delta_Q \chi = - \frac{2}{g_3} [\phi_z, H_1] \epsilon + F_1 \epsilon = 0, \xx
&& \delta_Q \overline{\eta} = \frac{2}{g_3} \epsilon^T C_3 [\overline{H}^2, \phi_z]
+ \epsilon^T C_3 \overline{F}^2 = 0, \xx
&& \delta_Q \overline{\chi} = \frac{1}{g_3} \epsilon^T C_3 \Big( \gamma^\mu D_\mu \overline{H}^2
+ [\overline{H}^2, \sigma] - \frac{i}{r} \overline{H}^2 \Big) = 0.
\end{eqnarray}
Given the locus $\mathfrak{F}_Q$ of the vector multiplet, the solution for the above equations
is trivial and it is given by $H_1 = H_2 = F_1 = F_2 = 0$.
The fluctuations around the locus $\mathfrak{F}_Q$ are described by
\begin{equation}\label{3d-hfluct}
\Phi (x, z, \overline{z}) = \frac{1}{\sqrt{t}}
\delta \Phi (x, z, \overline{z}),
\end{equation}
which we assume the same expansion \eq{exp-matrix} in terms of the Chevalley basis $(H_i, E_{\pm \alpha})$
of the Lie algebra $u(N)$. So we will eventually arrive at the result \eq{det-hbos} and \eq{det-hfer}
for the one-loop determinant for the hypermultiplet described by the three-dimensional large $N$ gauge theory.
This result also confirms the duality in Fig. \ref{flowchart} between a five-dimensional NC $U(1)$ gauge theory
and a three-dimensional large $N$ gauge theory.

\section{Discussion}

We emphasize that a NC space realizes a remarkable duality between a higher-dimensional NC $U(1)$
gauge theory and a lower-dimensional large $N$ gauge theory  \cite{q-emg,mine15}.
This duality is simply derived from a very elementary fact that the NC space \eq{nc-space} denoted by
$\mathbb{R}^{2n}_\theta$ is equivalent to the Heisenberg algebra of an $n$-dimensional
harmonic oscillator. A well-known property from quantum mechanics is that the NC space $\mathbb{R}^{2n}_\theta$
admits a separable Hilbert space and NC $U(1)$ gauge fields become operators acting on the Hilbert space.
The matrix representation of dynamical operators on the Hilbert space immediately leads to the picture
depicted in Fig. \ref{flowchart}. Therefore the relationship between a lower-dimensional large $N$ gauge theory
and a higher-dimensional NC $U(1)$ gauge theory in the figure is an exact mathematical identity.
In this correspondence, the dynamical variables in the lower-dimensional large $N$ gauge theory take values in
$\mathcal{A}_N (\mathbb{S}^3) = C^\infty (\mathbb{S}^3) \otimes \mathcal{A}_N$ and
those in the higher-dimensional NC $U(1)$ gauge theory take values in
$\mathcal{A}_\theta (\mathbb{S}^3) = C^\infty (\mathbb{S}^3) \otimes \mathcal{A}_\theta$.
We have applied the localization technique to this correspondence.
The result reveals a rich duality between NC $U(1)$ gauge theories and large $N$ matrix models
in various dimensions, as clearly summarized in Fig. \ref{flowchart}.

We note that both $\mathcal{A}_N (\mathbb{S}^3)$ and $\mathcal{A}_\theta (\mathbb{S}^3)$ are associative
NC algebras and $\{\mathcal{A}_N (\mathbb{S}^3), \; [\bullet, \bullet] \}$ and
$\{\mathcal{A}_\theta (\mathbb{S}^3), \; [-, -] \}$ form a Lie algebra under their bracket operation.
Moreover, given a Hilbert space $\mathcal{H}$, a matrix in Eq. \eq{matrix-rep} is an element of
a linear map $\rho: \mathcal{H} \to \mathcal{H}$, i.e., $\rho = \mathrm{End}(\mathcal{H})$
and a linear representation $\rho: \mathcal{A}_\theta (\mathbb{S}^3) \to \mathcal{A}_N (\mathbb{S}^3)$
in $\mathcal{H}$ is a Lie algebra homomorphism.
However there is another important lesson that we have learned from quantum mechanics.
For example, the momentum (position) operator in the Heisenberg algebra \eq{nc-phase} can be represented
by a differential operator in position (momentum) space, i.e., $p_i = -i\hbar \frac{\partial}{\partial x^i}$
or $x^i = i\hbar \frac{\partial}{\partial p_i}$. Recall that we have used the left-invariant vector fields in
Eq. \eq{ang-op} to represent the $SU(2)$ Lie algebra.
More generally a NC algebra $\mathcal{A}_\theta$ always has a representation in terms
of a differential (graded) Lie algebra $\mathfrak{D}$ and the map $\mathcal{A}_\theta \to \mathfrak{D}$
is also a Lie algebra homomorphism \cite{rev-mncg}. To be specific, let us apply the Lie algebra homomorphism
$\mathcal{A}_\theta \to \mathfrak{D}$ to the dynamical variables in Fig. \ref{flowchart}.
We will get a set of differential operators derived from the five-dimensional NC $U(1)$ gauge theory
on $\mathbb{S}^3 \times \mathbb{R}_\theta^2$, which is isomorphically mapped to a three-dimensional
large $N$ gauge theory through the matrix representation as shown in Fig. \ref{flowchart}.
An interesting problem is to identify the theory described by the set of differential operators.
It turns out \cite{q-emg} that the theory in a classical limit describes
a five-dimensional gravity whose asymptotic (vacuum) geometry corresponds to $\mathbb{S}^3 \times \mathbb{R}^2$
and the relationship between the five-dimensional gravity and the three-dimensional large $N$ gauge theory
is the well-known gauge/gravity duality or large $N$ duality.

Therefore the localization of a higher-dimensional NC $U(1)$ gauge theory and a lower-dimensional
large $N$ gauge theory in Fig. \ref{flowchart} can be interpreted as a localization of a five-dimensional
gravity emergent from the gauge theory. A configuration at the localization locus $\mathfrak{F}_Q$ will
be mapped to a BPS geometry according to the correspondence $\mathcal{A}_\theta \to \mathfrak{D}$.
This means that there exists an isomorphic map from the NC $U(1)$ gauge
theory to the Einstein gravity which completes the large $N$ duality \cite{mine15}.
In our case, Eq. \eq{3d-locus} corresponds to a vacuum geometry $\mathbb{S}^3 \times \mathbb{R}^2$.
As we pointed out in section 4, the locus is characterized by the BPS equations \eq{bps-locus}
whose solution is, in general, nontrivial, e.g. $U(N)$ instantons on $\mathbb{S}^3$ such as Nahm monopoles
and $U(1)^N$ monopoles in $\mathbb{R} \times \mathbb{S}^2$ \cite{seok}.
Of course, putting instantons on a compact space is highly nontrivial.
Nevertheless solutions exist, e.g., \cite{adhm}. It is known \cite{hsy-prl} that NC $U(1)$ instantons
on $\mathbb{R}_\theta^{2n}$ are equivalent to $n$-dimensional Calabi-Yau manifolds in a commutative limit.
Therefore it will be interesting to consider a nontrivial locus such as BPS solutions
and study their emergent geometry around the locus from the geometric point of view.

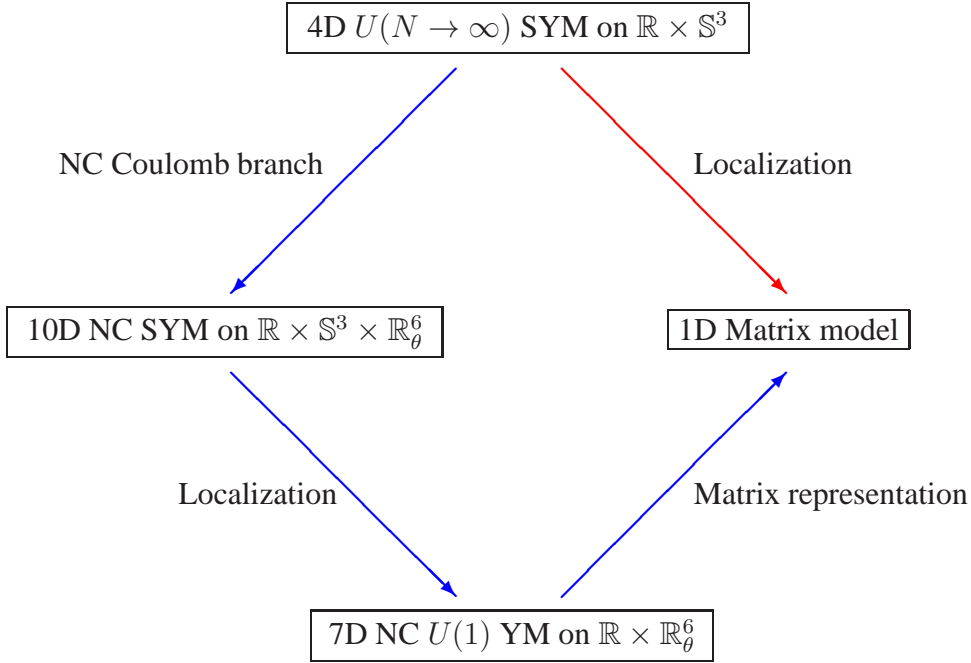
\begin{figure}
\centering
\begin{picture}(400,250)

%%%%%%%%%%%%%%%%%%%%%%%%%%%%%%%%%%%%%%%%%%%%%%%%%%%%%%%%%%%%%%%%%%%%%%%%%%
% \put(initial point of x,y coord. where the origin is SW corner){Text}  %
% \put(coord.){\vector(x,y direction of vector){Length}}                 %
%%%%%%%%%%%%%%%%%%%%%%%%%%%%%%%%%%%%%%%%%%%%%%%%%%%%%%%%%%%%%%%%%%%%%%%%%%

% Top
\put(116,242){\framebox[1.1\width]{4D $U(N \rightarrow \infty)$ SYM on $\mathbb{R} \times \mathbb{S}^3$}}
% Left
\put(10,127){\framebox[1.1\width]{10D NC SYM on $\mathbb{R} \times \mathbb{S}^{3} \times \mathbb{R}_\theta^6$}}
% Right
\put(260,127){\framebox[1.1\width]{1D Matrix model}}
% Bottom
\put(125,12){\framebox[1.1\width]{7D NC $U(1)$ YM on $\mathbb{R} \times \mathbb{R}_{\theta}^{6}$}}

% NE
\put(270,190){Localization}
\thicklines
\textcolor[rgb]{0.00,0.00,1.00}{\put(180,230){\vector(-1,-1){85}}}

% NW
\put(30,190){NC Coulomb branch}

\thicklines
\textcolor[rgb]{0.98,0.00,0.00}{\put(220,230){\vector(1,-1){85}}}

% SE
\put(270,65){Matrix representation}
\thicklines
\textcolor[rgb]{0.00,0.00,1.00}{\put(95,115){\vector(1,-1){85}}}

% SW
\put(75,65){Localization}
\thicklines
\textcolor[rgb]{0.00,0.00,1.00}{\put(220,30){\vector(1,1){85}}}
\end{picture}
\caption{Localization for large $N$ duality}
\label{fig-duality}
\end{figure}

Our localization scheme outlined in Fig. \ref{flowchart} may be directly applied to a localization problem
in the AdS/CFT correspondence \cite{local-qft12}.
The $AdS_5$ space has a boundary $\mathbb{R} \times \mathbb{S}^3$
in global coordinates. Hence one may consider the $\mathcal{N}=4$ $U(N)$ SYM theory
on $\mathbb{R} \times \mathbb{S}^3$ \cite{n=4s3,n=4s3ads} to study the $AdS_5/CFT_4$ duality.
The localization technique provides us a powerful tool for a nonperturbative analysis
of the large $N$ duality \cite{pestun}. The $\mathcal{N}=4$ SYM theory has six adjoint scalar fields and
the AdS/CFT duality typically considers the $N \to \infty$ limit of $U(N)$ gauge group.
Therefore one can consider a vacuum in the NC Coulomb branch by turning on
vacuum expectation values of the adjoint scalar fields such that the vacuum moduli
obey the Heisenberg algebra \eq{nc-vacuum} with $\mathrm{rank} (B) = 6$.
As we illustrated in section 4, the fluctuations around the NC Coulomb branch \eq{nc-vacuum} are
described by a ten-dimensional $\mathcal{N}=1$ supersymmetric NC $U(1)$ gauge theory \cite{hea}.
Although these two theories are defined in different dimensions with different gauge groups,
they are mathematically equivalent to each other as depicted in Fig. \ref{fig-duality}.
In this paper we have shown that the localization of a large $N$ gauge theory at the
NC Coulomb branch is equivalent to the localization of a higher-dimensional NC $U(1)$ gauge theory.
The corresponding picture for the AdS/CFT correspondence has been summarized in Fig. \ref{fig-duality}.
As we remarked in section 1, the NC field theory representation of
a lower-dimensional large $N$ gauge theory in the NC Coulomb branch
will provide us a powerful machinery to identify gravitational variables dual to
large $N$ matrices \cite{q-emg}. Hence one may study a nonperturbative aspect of the AdS/CFT correspondence
using the localization technique along the flowchart in Fig. \ref{fig-duality}.
We think that Fig. \ref{fig-duality} will be a straightforward generalization
of Fig. \ref{flowchart}. We hope to address this interesting problem in the near future.

\section*{Acknowledgments}

We thank Teruhiko Kawano for a kind help on some calculational details.
B.H.L. was supported by National Research Foundation of Korea (NRF) grant funded
by the Korea government (MSIP) (No.2014R1A2A1A01002306).
D.H. was supported by the Korea Ministry of Education, Science and Technology, Gyeongsangbuk-Do and Pohang City.
H.S.Y. was supported by the National Research Foundation of Korea (NRF) grant funded
by the Korea government (MOE) (No. NRF-2015R1D1A1A01059710).

\appendix

\section{Notation, conventions and useful formulae}

\subsection{Gamma matrices}

The five-dimensional gamma matrices $\Gamma^{M}, \; M=1,\cdots,5$, are given by
\begin{equation}
\Gamma^{m} = \gamma^{m} \otimes \sigma^{2}, \quad \Gamma^{4}  = \mathbf{1} \otimes \sigma^{1}, \quad
\Gamma^{5} = \mathbf{1} \otimes \sigma^{3},
\end{equation}
and the three-dimensional gamma matrices are defined by
\begin{equation}
\gamma^{m}  = \sigma^{m}, \qquad m=1,2,3
\end{equation}
where $\sigma^{m}$ are the Pauli matrices.
The gamma matrices satisfy the Dirac algebra
\begin{eqnarray}
&& \{\Gamma^{M}, \Gamma^{N} \} = 2 \delta^{MN}, \\
&& \{\gamma^{m}, \gamma^{n} \} = 2 \delta^{mn}
\end{eqnarray}
and the Lorentz generators are defined by
\begin{equation}
J^{MN} = \frac{1}{2} \Gamma^{MN} = \frac{1}{4}[\Gamma^{M}, \Gamma^{N}], \qquad
J^{mn} = \frac{1}{2} \gamma^{mn} = \frac{1}{4}[\gamma^{m}, \gamma^{n}]
= \frac{i}{2} {\varepsilon^{mn}}_p \gamma^p.
\end{equation}
Useful (anti-)commutation relations are
\begin{equation}\label{comm-anti}
\{\Gamma^{MN}, \Gamma^{L} \} = 2 \Gamma^{MNL},
\qquad  [\Gamma^{MN}, \Gamma^{L}] = 2 (\Gamma^{M}\delta^{NL} - \Gamma^{N}\delta^{ML}).
\end{equation}
Similar relations hold for the three-dimensional gamma matrices $\gamma^m$.

\subsection{Charge conjugation matrices}

The five-dimensional charge conjugation matrix $C_5$ obeys
\begin{equation}
(\Gamma^{M})^{T} = C_{5} \Gamma^{M} C_{5}^{-1}, \qquad (C_{5})^{T}  = -C_{5}
\end{equation}
where $T$ denotes the transpose of a matrix.
It is related to the three-dimensional charge conjugation matrix $C_3 = i \sigma^2$ by
\begin{equation}
C_{5} = C_{3} \otimes \mathbf{1}
\end{equation}
and thus $C_3$ satisfies the relation
\begin{equation}
(\gamma^{m})^{T} = - C_{3} \gamma^{m} C_{3}^{-1}, \qquad (C_{3})^{T} = -C_{3}.
\end{equation}

\subsection{Fermion bilinears}

Symplectic Majorana spinors satisfy the following transposition property of fermion bilinears:
\begin{eqnarray} \label{bi-spinor}
&& \overline{\psi}_{\dot{\alpha}} \chi^{\dot{\alpha}}
= - \overline{\chi}_{\dot{\alpha}} \psi^{\dot{\alpha}}, \xx
&& \overline{\psi}_{\dot{\alpha}} \Gamma^{M} \chi^{\dot{\alpha}}
= - \overline{\chi}_{\dot{\alpha}} \Gamma^{M}
\psi^{\dot{\alpha}},  \\
&& \overline{\psi}_{\dot{\alpha}} \Gamma^{MN} \chi^{\dot{\alpha}}
= \overline{\chi}_{\dot{\alpha}} \Gamma^{MN} \psi^{\dot{\alpha}}. \nonumber
\end{eqnarray}
The raising and lowering of $SU(2)_R$ indices are defined by
\begin{equation} \label{rl-spinor}
 \chi_{\dot{\alpha}} \equiv  \chi^{\dot{\beta}} \varepsilon_{\dot{\beta}\dot{\alpha}}, \qquad
 \psi^{\dot{\alpha}} \equiv  \psi_{\dot{\beta}} \varepsilon^{\dot{\beta}\dot{\alpha}}.
\end{equation}
Then the following relation is deduced:
\begin{equation}\label{rl-flip}
 \overline{\psi}_{\dot{\alpha}} \chi^{\dot{\alpha}}
= - \overline{\psi}^{\dot{\alpha}} \chi_{\dot{\alpha}}
\end{equation}
which should not be confused with the first one in \eq{bi-spinor}.
Our convention for the $SU(2)_R$ invariant tensors, $\varepsilon_{\dot{\alpha}\dot{\beta}}$ and $\varepsilon^{\dot{\alpha}\dot{\beta}}$, is given by
\begin{equation}
\varepsilon_{12}  = \varepsilon^{21}  =  1, \qquad \varepsilon_{21}  = \varepsilon^{12} =  -1
\end{equation}
and thus $\varepsilon_{\dot{\alpha}\dot{\gamma}} \varepsilon^{\dot{\gamma}\dot{\beta}} = \delta_{\dot{\alpha}}^{~\dot{\beta}}$.

\subsection{Lie algebra $\mathfrak{g}$}

The gauge group for NC $U(1)$ gauge theories is $U(1)_\star$ whose element is given by
\begin{equation}\label{nc-u1}
    e^{i\lambda}_\star = \sum_{k=0}^\infty \frac{i^k}{k!} \overbrace{\lambda \star \cdots
    \star \lambda}^{k \; \mathrm{times}} \in U(1)_\star
\end{equation}
where $\lambda = \lambda(X) \in \mathcal{A}_{\theta}$. We consider the matrix representation \eq{matrix-rep}
of the NC gauge parameter $\lambda \in \mathcal{A}_{\theta}$ which leads to a gauge transformation parameter
in $U(N \to \infty)$ gauge theory. In this way, we get the gauge group $U(N)$ for large $N$ gauge theories with
the limit $N \to \infty$, i.e., $U(1)_\star \to U(N)$ by $e^{i\lambda}_\star \mapsto  e^{i\Lambda}$ where
$\Lambda(x) = \sum_{a=1}^{N^2} \lambda^a (x) T^a$.
The Lie algebra generators in $u(N)$ are split into $su(N)$ generators $T^a \; (a = 1, \cdots, N^2 - 1)$
and a $u(1)$ generator $T^{N^2} = \frac{1}{\sqrt{N}} \mathbb{I}$. The $su(N)$ generators are normalized as $\mathrm{Tr}(T^a T^b) = \delta^{ab}$ and obey the commutation relation
\begin{equation}\label{lie-alg}
    [T^a, T^b] = i f^{abc} T^c.
\end{equation}
It is convenient to introduce the Chevalley basis $(H_i, E_{\pm \alpha})$ for a simple Lie algebra,
i.e. $su(N)$, obeying the relations
\begin{equation}\label{chevalley}
    [H_i, E_\alpha ] = \alpha_i E_\alpha, \qquad
    [E_\alpha, E_{-\alpha}] = \alpha_i H_i \equiv \alpha \cdot H
\end{equation}
where $i = 1, \cdots, N-1$ and $\alpha \in \mathfrak{r}$ is an element of the root system $\mathfrak{r}$
of the Lie algebra $su(N)$.

\subsection{Integral on NC space}

For the star product \eq{star-prod}, the integral
\begin{equation}\label{nc-int}
    \int d^5 X f_1 (X) \star f_2 (X) \star  \cdots \cdots \star f_n (X)
\end{equation}
is invariant under cyclic permutations of the smooth functions $f_i$ \cite{ncft-rev}.
In particular, the following useful relations are deduced from this property:
\begin{equation} \label{com-int}
 \int d^5 X f (X) \star g (X) = \int d^5 X g (X) \star f (X),
\end{equation}
\begin{equation}\label{cyc-int}
\int d^5 X f_1 (X) \star f_2 (X) \star f_3 (X) =
    \int d^5 X f_2 (X) \star f_3 (X) \star f_1 (X) =
    \int d^5 X f_3 (X) \star f_1 (X) \star f_2 (X).
\end{equation}
Note that the above cyclic permutations have been derived from the assumption that the functions $f_i$
appropriately behave, i.e., rapidly decay, at asymptotic infinities so that total derivative terms
can be dropped. Thus one may worry about the first term in \eq{5d-action} since
$\widehat{\mathcal{F}}_{MN}$ does not decay to zero but approaches to a constant value at infinity.
Fortunately, constant terms do not introduce any trouble for the cyclic permutation of the integral
because they are immune from the star product and so they can be placed outside the integral.
For example, if one of $f_i$'s in Eq. \eq{cyc-int} is constant, Eq. \eq{cyc-int} reduces to
Eq. \eq{com-int}. Consequently the constant terms in the star product do not threaten
the cyclic property \eq{cyc-int} unless the integral is divergent.
In this case the cyclic permutation of the integral \eq{nc-int} can be implemented with impunity.

This property can also be understood using the matrix representation \eq{matrix-rep}.
In the matrix representation, the integral \eq{nc-int} is transformed into the trace over matrices, i.e.,
\begin{equation}\label{int=tr}
    \int d^5 X = (2 \pi \theta) \int d^3 x \mathrm{Tr}.
\end{equation}
Therefore the cyclic property  of the integral \eq{nc-int} corresponds to the cyclic permutation of
the matrix trace, e.g.,
\begin{equation}\label{cyc-mat}
 \mathrm{Tr} f_1 (x) f_2 (x) f_3 (x) = \mathrm{Tr} f_2 (x) f_3 (x) f_1 (x) =
 \mathrm{Tr} f_3 (x) f_1 (x) f_2 (x)
\end{equation}
for $N \times N$ matrices $f_{1,2,3}(x)$ over $\mathbb{R}^3$ or $\mathbb{S}^3$.
Note that the background $B$-field is mapped to the identity matrix (see $\mathbf{A.4}$) and
so it can freely escape from the trace. Hence the previous argument is confirmed.

\section{Vanishing cubic terms in supersymmetric transformations}

This appendix is to check the supersymmetric invariance of five-dimensional NC $U(1)$ gauge theory,
in particular, the vanishing of the fermionic cubic terms in supersymmetric variations \cite{ni-ra02}.

As in the commutative case, after cancellation of all the quadratic terms,
we are left with the cubic terms of $\Psi$ field:
\begin{equation}\label{f-cubic}
\overline{\Psi}_{\dot{\alpha}} \Gamma^{M} [\overline{\Sigma}_{\dot{\beta}} \Gamma_{M} \Psi^{\dot{\beta}},
\Psi^{\dot{\alpha}}] - \overline{\Psi}_{\dot{\alpha}} [\overline{\Sigma}_{\dot{\beta}} \Psi^{\dot{\beta}},
\Psi^{\dot{\alpha}}].
\end{equation}
In order to show the vanishing of the cubic terms in (\ref{f-cubic}),
we need the Fierz identity for gamma matrices
\begin{equation}\label{fierz-gamma}
    \delta_{\alpha\gamma} \delta_{\delta\beta} = \frac{1}{4} \delta_{\alpha\beta} \delta_{\delta\gamma}
    + \frac{1}{4} (\Gamma^M)_{\alpha\beta} (\Gamma_M)_{\delta\gamma}
    - \frac{1}{8} (\Gamma^{MN})_{\alpha\beta} (\Gamma_{MN})_{\delta\gamma}
\end{equation}
which leads to the identity
\begin{equation}\label{fierz-id}
    (\overline{\phi}\eta) (\overline{\epsilon}\zeta) = - \frac{1}{4} (\overline{\eta}\zeta)
    (\overline{\epsilon}\phi)
    - \frac{1}{4} (\overline{\eta}\Gamma^M \zeta)(\overline{\epsilon} \Gamma_M \phi)
    + \frac{1}{8} (\overline{\eta}\Gamma^{MN} \zeta)(\overline{\epsilon} \Gamma_{MN} \phi)
\end{equation}
for arbitrary symplectic Majorana spinors  $\epsilon, \zeta, \eta$ and $\phi$.
First note that, using the identity $\overline{\Psi}_{\dot{\alpha}} (\overline{\Sigma}_{\dot{\beta}}
\Psi^{\dot{\beta}}) \Psi^{\dot{\alpha}} : = - (\Psi^{\dot{\alpha}})^T (\overline{\Psi}_{\dot{\alpha}})^T
(\overline{\Sigma}_{\dot{\beta}} \Psi^{\dot{\beta}}) = - \overline{\Psi}_{\dot{\alpha}} \Psi^{\dot{\alpha}}
(\overline{\Sigma}_{\dot{\beta}} \Psi^{\dot{\beta}})$, we get
\begin{equation}\label{0-add}
  \overline{\Psi}_{\dot{\alpha}} [\overline{\Sigma}_{\dot{\beta}} \Psi^{\dot{\beta}},
\Psi^{\dot{\alpha}}] := - 2 \overline{\Psi}_{\dot{\alpha}} \Psi^{\dot{\alpha}}
\overline{\Sigma}_{\dot{\beta}} \Psi^{\dot{\beta}}
\end{equation}
where $:=$ means the equality under the integral $\int_M d^5 X$, i.e., up to total derivative terms.
Similarly, we have
\begin{equation}\label{1-add}
 \overline{\Psi}_{\dot{\alpha}} \Gamma^M [\overline{\Sigma}_{\dot{\beta}} \Gamma_M \Psi^{\dot{\beta}},
\Psi^{\dot{\alpha}}] := - 2 \overline{\Psi}_{\dot{\alpha}} \Gamma^M \Psi^{\dot{\alpha}}
\overline{\Sigma}_{\dot{\beta}} \Gamma_M \Psi^{\dot{\beta}}.
\end{equation}
But, one can show by the same calculation that
\begin{equation}\label{2-add}
 \overline{\Psi}_{\dot{\alpha}} \Gamma^{MN} [\overline{\Sigma}_{\dot{\beta}} \Gamma_{MN} \Psi^{\dot{\beta}},
\Psi^{\dot{\alpha}}] := 0.
\end{equation}
Using this result, let us rewrite Eq. (\ref{f-cubic}) as the following form
\begin{eqnarray} \label{f-cubic-re}
&& \overline{\Psi}_{\dot{\alpha}} [\overline{\Sigma}_{\dot{\beta}} \Psi^{\dot{\beta}},
\Psi^{\dot{\alpha}}] + \overline{\Psi}_{\dot{\alpha}} \Gamma^{M} [\overline{\Sigma}_{\dot{\beta}} \Gamma_{M} \Psi^{\dot{\beta}}, \Psi^{\dot{\alpha}}] - \frac{1}{2} \overline{\Psi}_{\dot{\alpha}} \Gamma^{MN} [\overline{\Sigma}_{\dot{\beta}} \Gamma_{MN} \Psi^{\dot{\beta}}, \Psi^{\dot{\alpha}}]
- 2\overline{\Psi}_{\dot{\alpha}} [\overline{\Sigma}_{\dot{\beta}} \Psi^{\dot{\beta}},
\Psi^{\dot{\alpha}}] \nonumber \\
&=& - 4 \overline{\Psi}^{\dot{\beta}} [\overline{\Sigma}_{\dot{\beta}} \Psi^{\dot{\alpha}},
\Psi_{\dot{\alpha}}] - 2\overline{\Psi}_{\dot{\alpha}}
[\overline{\Sigma}_{\dot{\beta}} \Psi^{\dot{\beta}}, \Psi^{\dot{\alpha}}]
\end{eqnarray}
where we applied the Fierz identity (\ref{fierz-gamma}) to the first three terms
and used the identity $\overline{\Psi}_{\dot{\alpha}} \Psi^{\dot{\beta}} \overline{\Sigma}_{\dot{\beta}}
 \Psi^{\dot{\alpha}} : = \overline{\Psi}^{\dot{\beta}} (\overline{\Sigma}_{\dot{\beta}} \Psi^{\dot{\alpha}})  \Psi_{\dot{\alpha}}$. Note that the first term in \eq{f-cubic-re} can be written as
\begin{eqnarray}\label{1f-re}
 4 \overline{\Psi}^{\dot{\beta}} [\overline{\Sigma}_{\dot{\beta}} \Psi_{\dot{\alpha}},
\Psi^{\dot{\alpha}}] &:=& 4 \overline{\Sigma}_{\dot{\beta}}\Psi_{\dot{\alpha}}
(\overline{\Psi}^{\dot{\alpha}}{\Psi}^{\dot{\beta}}
- \overline{\Psi}^{\dot{\beta}}{\Psi}^{\dot{\alpha}}) \nonumber\\
&=& - 4 \overline{\Psi}_{\dot{\alpha}} \Psi^{\dot{\alpha}} \overline{\Sigma}_{\dot{\beta}} \Psi^{\dot{\beta}},
\end{eqnarray}
using the identities $\overline{\Psi}^{\dot{\beta}} (\overline{\Sigma}_{\dot{\beta}} \Psi_{\dot{\alpha}})
\Psi^{\dot{\alpha}} := - (\Psi^{\dot{\alpha}})^T (\overline{\Psi}^{\dot{\beta}})^T
(\overline{\Sigma}_{\dot{\beta}} \Psi_{\dot{\alpha}})= \overline{\Psi}^{\dot{\alpha}}
\Psi^{\dot{\beta}}(\overline{\Sigma}_{\dot{\beta}} \Psi_{\dot{\alpha}})$
and $(\overline{\Psi}^{\dot{\alpha}}{\Psi}^{\dot{\beta}}
- \overline{\Psi}^{\dot{\beta}}{\Psi}^{\dot{\alpha}}) = - \varepsilon^{\dot{\alpha}\dot{\beta}}
\overline{\Psi}_{\dot{\gamma}}{\Psi}^{\dot{\gamma}}$. After using Eqs. (\ref{0-add}) and (\ref{1f-re}),
one can finally see that the two terms in (\ref{f-cubic-re}) exactly cancel each other.
This completes the proof of the supersymmetric invariance of five-dimensional NC $U(1)$ gauge theory.

\section{Closed supersymmetric algebra}

In this appendix, we present a detailed result for the closedness of the supersymmetry algebra
on $\mathbb{S}^3 \times \mathbb{R}_{\theta}^2$ for the vector multiplet.
The modified supersymmetry transformations generated by the spinor $\epsilon$
obeying Eq. \eq{kill-spinor} will be denoted by $\Delta_\epsilon = \delta_\epsilon + \delta'_\epsilon$
where $\delta_\epsilon$-transformations are given by Eq. \eq{susytr-3d}
with the replacement $g_3 \to g_5$. The result on $\mathbb{S}^3$ is exactly the same
as the five-dimensional case if $g_5$ is replaced by $g_3$.

First, the vector multiplet satisfies the following closed algebra given by
\begin{eqnarray} \label{c-susy1}
 &&  [\Delta_\eta, \Delta_\epsilon] A_\mu = - i \zeta^\nu F_{\nu\mu} + i \zeta D_\mu \sigma, \\
\label{c-susy2}
 &&  [\Delta_\eta, \Delta_\epsilon] \phi_z = - i \zeta^\mu D_{\mu} \phi_z - \zeta [\sigma, \phi_z], \\
\label{c-susy3}
 &&  [\Delta_\eta, \Delta_\epsilon] \sigma = - i \zeta^\mu D_{\mu} \sigma, \\
\label{c-susy4}
 &&  [\Delta_\eta, \Delta_\epsilon] \lambda = - i \zeta^\mu \big( D_{\mu} \lambda
 + \frac{i}{2r} \gamma_\mu \lambda \big) - \zeta \big( [\sigma, \lambda] + \frac{1}{r} \lambda \big), \\
\label{c-susy5}
 &&  [\Delta_\eta, \Delta_\epsilon] \psi = - i \zeta^\mu \big( D_{\mu} \psi
 + \frac{i}{2r} \gamma_\mu \psi \big) - \zeta \big( [\sigma, \psi] + \frac{1}{r} \psi \big), \\
\label{c-susy6}
 &&  [\Delta_\eta, \Delta_\epsilon] F = - i \zeta^\mu D_{\mu} F - \zeta \big( [\sigma, F]
 + \frac{2}{r} F \big), \\
\label{c-susy7}
 &&  [\Delta_\eta, \Delta_\epsilon] D = - i \zeta^\mu D_{\mu} D - \zeta [\sigma, D],
\end{eqnarray}
where
\begin{equation}\label{2-para}
    \zeta^\mu = \overline{\epsilon} \gamma^\mu \eta - \overline{\eta} \gamma^\mu \epsilon,
    \qquad \zeta = \overline{\epsilon} \eta - \overline{\eta} \epsilon
\end{equation}
and the covariant derivative $D_\mu$ contains gauge and spin connections.
In order to get the above result, we have used at several places the three-dimensional Fierz identity,
\begin{equation}\label{3d-fierz}
    (\overline{\epsilon} \psi)  \phi =  - \frac{1}{2} \psi (\overline{\epsilon}\phi)
    - \frac{1}{2} \gamma_\mu \psi (\overline{\epsilon} \gamma^\mu \phi)
\end{equation}
for complex spinors $\overline{\epsilon}, \psi$ and $\phi$.
It is useful to recall that the transformation parameters $\epsilon$ and $\eta$ do not depend on
the NC coordinates $y^a \in \mathbb{R}^2_{\theta}$, so they are immune from the star product that
is implicitly assumed for all multiplications. It is easy to see that $[\Delta_\eta, \Delta_\epsilon]$
acts as an even symmetry of the theory since it can be written as a sum of a translation generated
by the parameters $\zeta^\mu$, a gauge transformation by $\rho = \zeta^\mu A_\mu + \zeta \sigma$,
a Lorentz transformation by $\kappa_{\mu\nu} = \frac{1}{r} \varepsilon_{\mu\nu\lambda} \zeta^\lambda$
and a $U(1)$ transformation in $SU(2)_R$ symmetry by $\upsilon = \frac{\zeta}{r}$ \cite{hst-5dsusy}.
Thus it verifies that the modified supersymmetry transformations $\Delta_\epsilon$ form
a closed algebra even off-shell.

\section{Harmonic analysis on $\mathbb{S}^3$}

Any element of $SU(2)$ can be written in the form
\begin{equation}\label{su2-gel}
    g = \left(
          \begin{array}{cc}
            \alpha & \beta \\
            -\overline{\beta} & \overline{\alpha} \\
          \end{array}
        \right), \qquad |\alpha|^2 + |\beta|^2 =1.
\end{equation}
The Maurer-Cartan (MC) forms $\omega^m$ on the $SU(2)$ group manifold are defined by
$$ g^{-1} dg = \sum_{m=1}^3 \tau_m \omega^m,$$
and they satisfy
\begin{equation}\label{su2-steq}
    d\omega^m - \frac{1}{2} \varepsilon^{mnp} \omega^n \wedge \omega^p = 0.
\end{equation}
The basis of the $su(2)$ Lie algebra obeys the relation
$$ [ \tau_m, \tau_n] = - \varepsilon^{mnp} \tau_p. $$
See the appendix in Ref. \cite{marino-rev} for their explicit coordinate representations.

We can use the MC forms to analyze the differential geometry of $\mathbb{S}^3$.
The dreibein of $\mathbb{S}^3$ is proportional to $\omega^m$, and we write
\begin{equation}\label{dreibein}
    e^m = \frac{r}{2} \omega^m_\mu dx^\mu = \frac{r}{2} \omega^m.
\end{equation}
In terms of the dreibeins, the metric on $\mathbb{S}^3$ is given by
\begin{equation}\label{su2-metric}
    g_{\mu\nu} = e^m_\mu e^n_\nu \delta_{mn}.
\end{equation}
The inverse dreibein is defined by
$$ E_m^\mu = g^{\mu\nu} e_{\nu m},$$
which can be used to define left-invariant vector fields
\begin{equation}\label{left-vec}
    l_m = E_m^\mu \partial_\mu.
\end{equation}
They satisfy $e^m (l_n) = \delta^m_n$ and the commutation relations
\begin{equation}\label{vec-lie}
    [l_m, l_n] = - \frac{2}{r} \varepsilon_{mnp} l_p.
\end{equation}
If we introduce the operators $L_m$ through
\begin{equation}\label{ang-op}
    l_m = \frac{2i}{r} L_m,
\end{equation}
they obey the standard commutation relations of the $SU(2)$ angular momentum operators:
\begin{equation}\label{su2-ang}
    [L_m, L_n] = i \varepsilon_{mnp} L_p.
\end{equation}

The spin connection ${\omega^m}_n$ is introduced via the torsion-free condition
$$ de^m + {\omega^m}_n \wedge e^n = 0.$$
In our case this condition can be solved by
\begin{equation}\label{su2-spin-conn}
    {\omega^m}_n = \frac{1}{r} \varepsilon_{mnp} e^p
\end{equation}
using Eq. \eq{su2-steq}.
The torsion-free condition also leads to the explicit expression,
$${\omega^m}_{n \mu} = E^\nu_n (\partial_\mu e_\nu^m - \Gamma^\lambda_{\mu\nu} e_\lambda^m )$$
or, equivalently,
$$ \partial_\mu e_\nu^m = \Gamma^\lambda_{\mu\nu} e_\lambda^m - e^n_\nu {\omega^m}_{n \mu},
\qquad \partial_\mu E_m^\nu = E^\nu_n {\omega^n}_{m \mu} - \Gamma^\nu_{\mu\lambda} E^\lambda_m.$$
The curvature tensor is given by
$${R^m}_n = d {\omega^m}_{n} + {\omega^m}_{p} \wedge {\omega^p}_{n}
= \frac{1}{r^2} e^m \wedge e^n,$$
or, equivalently, $R_{mnmn} = \frac{1}{r^2}$ (no sum). Thus the Ricci tensor and the Ricci scalar
are given by $R_{mn} = \frac{2}{r^2} \delta_{mn}$ and $R= \frac{6}{r^2}$, respectively.

The scalar Laplacian on $\mathbb{S}^3$ can be written in local coordinates as
\begin{equation}\label{scalar-lap1}
    \square_0 \phi = - \frac{1}{\sqrt{\det g}} \sum_{\mu, \nu} \frac{\partial}{\partial x^\mu}
    \Big( \sqrt{\det g} g^{\mu\nu} \frac{\partial \phi}{\partial x^\nu} \Big)
\end{equation}
or equivalently
\begin{equation}\label{scalar-lap2}
\square_0  = -  g^{\mu\nu} \partial_\mu \partial_\nu +  g^{\mu\nu} \Gamma^\lambda_{\mu\nu}
\partial_\lambda.
\end{equation}
It can be written, in terms of left-invariant vector fields, as
\begin{eqnarray} \label{lap-left}
- \square_0 &=& \sum_{m=1}^3 l_m^2 \xx
  &=&  E^{\mu}_m E^{\nu}_m \partial_\mu \partial_\nu +  E^{\mu}_m \frac{\partial E_m^\nu}{\partial x^\mu}
\frac{\partial}{\partial x^\nu}.
\end{eqnarray}
The Peter-Weyl theorem says that any square-integrable function on $\mathbb{S}^3 \cong SU(2)$ can be written as
a linear combination of the spherical harmonics in Eq. \eq{ket-sharm} as was illustrated in \eq{harm-exp}.

The dreibeins $e^m$ are taken as the eigenstate of the spin operators
$\overrightarrow{S} \cdot \overrightarrow{S}$ and $S_3$ where $(S^m)_{ln} = i \varepsilon^{lmn}$ is
the spin-1 representation of $SU(2)$. The vector spherical harmonics on $\mathbb{S}^3$ are
then constructed by considering a tensor product of the scalar spherical harmonics
with the spin-1 basis $|s=1, s_z\rangle \; (s_z = -1, 0, 1)$.
The space of one-forms on $\mathbb{S}^3$ can be decomposed using
the vector spherical harmonics in Eq. \eq{vec-harmonic}.
The vector Laplacian $\square_1 \equiv *d*d $ acts on a one-form $B = e^m B_m$ obeying
the Lorentz gauge condition $d^\dagger B \equiv *d*B = l_m B_m = 0$ as follows:
\begin{equation}\label{bl-op}
 *d*d B = \Big( - l_n l_n B_m + \frac{2}{r} {\epsilon_m}^{np} l_n B_p + \frac{4}{r^2} B_m \Big) e^m.
\end{equation}

Using the dreibein, we can define locally inertial gamma matrices as $\gamma_m = E_m^\mu \gamma_\mu$
which satisfy the relations
\begin{equation*}
\{\gamma_m, \gamma_n \} = 2 \delta_{mn}, \qquad [\gamma_m, \gamma_n] = 2i \varepsilon_{mnp} \gamma_p.
\end{equation*}
The covariant derivative acting on a spinor is defined by
\begin{eqnarray} \label{cov-spin}
  \nabla_\mu  &=& \partial_\mu + \frac{1}{4} \omega_\mu^{mn} \gamma_{mn}
  = \partial_\mu + \frac{i}{2} e_\mu^{m} \gamma_{m} \xx
  &=&  \partial_\mu + \frac{i}{2} \gamma_{\mu}.
\end{eqnarray}
It follows that the Dirac operator is
\begin{equation}\label{dirac-op-app}
    - i \slashed{D} = - i \gamma^\mu \partial_\mu + \frac{3}{2r}
    = - i \gamma^m l_m + \frac{3}{2r}.
\end{equation}
The Laplacian for the Dirac operator obeys the relation
\begin{eqnarray}\label{dirac-sq}
- \slashed{D}^2 &=& - \gamma^\mu \gamma^\nu \nabla_\mu \nabla_\nu
= - ( g^{\mu\nu} + \gamma^{\mu\nu} ) \nabla_\mu \nabla_\nu  \xx
&=& \square_{\frac{1}{2}} - \frac{1}{2} \gamma^{mn} R_{mn} =
\square_{\frac{1}{2}} + \frac{1}{4} R.
\end{eqnarray}

If we introduce the spin operators $S_m = \frac{1}{2} \gamma_m$ satisfying the $su(2)$ algebra
$[S_m, S_n ] = i \varepsilon_{mnp} S_p$, the Dirac operator reads as
\begin{equation}\label{dirac-oprep}
 - i \slashed{D} = \frac{4}{r} \Big( \overrightarrow{L} \cdot \overrightarrow{S} + \frac{3}{8} \Big).
\end{equation}
Since the total angular momentum is defined by $\overrightarrow{J} = \overrightarrow{L} + \overrightarrow{S}$
such that $\overrightarrow{L} \cdot \overrightarrow{S} = \frac{1}{2} (\overrightarrow{J}^2
- \overrightarrow{L}^2 - \overrightarrow{S}^2)$ and $\overrightarrow{S}$ corresponds to spin $s=\frac{1}{2}$
and $\overrightarrow{L}$ to $j$, the possible eigenvalues of $\overrightarrow{J}$ are $j \pm 1/2$.
Thus the eigenvalues of the Dirac operator \eq{dirac-op-app} are equal to
\begin{equation}\label{dirac-eigenval}
    \frac{2}{r} \Big( \big(j \pm \frac{1}{2} \big) \big(j  \pm \frac{1}{2} + 1 \big) - j(j+1) \Big)
    = \left\{
        \begin{array}{ll}
          \frac{1}{r} \big( 2j + \frac{3}{2} \big) & \hbox{for $+$;} \\
          - \frac{1}{r} \big( 2j + \frac{1}{2} \big) & \hbox{for $-$,}
        \end{array}
      \right.
\end{equation}
with degeneracies
\begin{equation}\label{dirac-eigendeg}
    d_{j \pm \frac{1}{2}} = \Big( 2 \big(j \pm \frac{1}{2} \big) + 1 \Big) \big(2j + 1 \big)
    = \left\{
        \begin{array}{ll}
          2 (j+1) (2j + 1) & \hbox{for $+$;} \\
           2 j ( 2j + 1) & \hbox{for $-$.}
        \end{array}
      \right.
\end{equation}
The eigenvectors of the Dirac operator are given by the spinor spherical harmonics introduced
in Eq. \eq{ham-fexp}.

\section{Clebsch-Gordan coefficients}

We reproduce the Clebsch-Gordan coefficients for the products $j \otimes 1$
and $j \otimes \frac{1}{2}$ in Ref. \cite{fukama13} for reader's convenience.

$\bullet$ The spin $k = j+1$ representation

\begin{eqnarray*}
   && | k=j+1, m \rangle\rangle = \frac{1}{\sqrt{2(j+1)(2j+1)}}
\Big( \sqrt{(j+m)(j+m+1)}|j, m-1 \rangle |1, 1 \rangle \xx
&& + \sqrt{2(j+m+1)(j-m+1)}|j, m \rangle |1, 0 \rangle
+ \sqrt{(j-m)(j-m+1)}|j, m+1 \rangle |1, -1 \rangle \Big).
\end{eqnarray*}

$\bullet$ The spin $k = j$ representation

\begin{eqnarray*}
   && | k=j, m \rangle\rangle = \frac{1}{\sqrt{2j(j+1)}}
\Big( - \sqrt{(j+m)(j-m+1)}|j, m-1 \rangle |1, 1 \rangle \xx
&& + \sqrt{2} m |j, m \rangle |1, 0 \rangle
+ \sqrt{(j-m)(j+m+1)}|j, m+1 \rangle |1, -1 \rangle \Big).
\end{eqnarray*}

$\bullet$ The spin $k = j-1$ representation

\begin{eqnarray*}
   && | k=j-1, m \rangle\rangle = \frac{1}{\sqrt{2j(2j+1)}}
\Big( \sqrt{(j-m)(j-m+1)}|j, m-1 \rangle |1, 1 \rangle \xx
&& - \sqrt{2(j+m)(j-m)}|j, m \rangle |1, 0 \rangle
+ \sqrt{(j+m)(j+m+1)}|j, m+1 \rangle |1, -1 \rangle \Big).
\end{eqnarray*}

The spin operator $S_3$ acts on the state $|k, m \rangle\rangle$ as

\begin{equation*}
   S_3 |j+1, m \rangle\rangle = \frac{m}{j+1} |j+1, m \rangle\rangle
- \sqrt{\frac{j(j-m +1)(j+m+1)}{(j+1)^2 (2j+1)}}|j, m \rangle \rangle.
\end{equation*}

\begin{eqnarray*}
   S_3 |j, m \rangle\rangle &=& \frac{1}{\sqrt{j(j+1)}} \Big(
- \sqrt{\frac{j^2 (j-m +1)(j+m+1)}{(j+1) (2j+1)}}|j+1, m \rangle \rangle \xx
&& + \sqrt{\frac{m^2}{j(j+1)}}|j, m \rangle \rangle
- \sqrt{\frac{(j+1)^2 (j-m )(j+m)}{j (2j+1)}}|j-1, m \rangle \rangle \Big).
\end{eqnarray*}

\begin{equation*}
   S_3 |j-1, m \rangle\rangle = - \frac{m}{j} |j-1, m \rangle\rangle
- \sqrt{\frac{(j+1)(j-m)(j+m)}{j^2 (2j+1)}}|j, m \rangle \rangle.
\end{equation*}

$\bullet$ The spin $k = j+\frac{1}{2}$ representation $(m = -(j+1), -j, \cdots, j-1, j)$

\begin{equation*}
  | k=j+\frac{1}{2}, m + \frac{1}{2} \rangle\rangle = \sqrt{\frac{j-m}{2j+1}}
|j, m+1 \rangle |\frac{1}{2}, - \frac{1}{2} \rangle
+ \sqrt{\frac{j+m+1}{2j+1}}
|j, m \rangle |\frac{1}{2}, \frac{1}{2} \rangle.
\end{equation*}

$\bullet$ The spin $k = j-\frac{1}{2}$ representation $(m = -j, \cdots, j-1)$

\begin{equation*}
  | k=j-\frac{1}{2}, m + \frac{1}{2} \rangle\rangle = \sqrt{\frac{j+m+1}{2j+1}}
|j, m+1 \rangle |\frac{1}{2}, - \frac{1}{2} \rangle
+ \sqrt{\frac{j-m}{2j+1}}
|j, m \rangle |\frac{1}{2}, \frac{1}{2} \rangle.
\end{equation*}

The spin operator $S_3$ acts on the state $|k, m \rangle\rangle$ as

\begin{equation*}
 S_3 |j+\frac{1}{2}, m + \frac{1}{2} \rangle\rangle = \frac{2m+1}{2(2j+1)}
|j+\frac{1}{2}, m+\frac{1}{2} \rangle \rangle
- \frac{\sqrt{(j-m)(j+m+1)}}{2j+1}
|j - \frac{1}{2}, m + \frac{1}{2} \rangle \rangle.
\end{equation*}

\begin{equation*}
 S_3 |j-\frac{1}{2}, m + \frac{1}{2} \rangle\rangle =
- \frac{\sqrt{(j-m)(j+m+1)}}{2j+1}
|j + \frac{1}{2}, m + \frac{1}{2} \rangle \rangle
- \frac{2m+1}{2(2j+1)}
|j-\frac{1}{2}, m+\frac{1}{2} \rangle \rangle.
\end{equation*}

\end{document}